\documentclass[vecphys,multphys]{svmult}

\usepackage{makeidx}     % allows index generation
\usepackage{graphicx}    % standard LaTeX graphics tool
\usepackage{epstopdf}    % to handle eps figures in pdflatex       
\usepackage{multicol}    % used for the two-column index
\usepackage[bottom]{footmisc}% places footnotes at page bottom
\usepackage{amsmath,amssymb,bbm,bm}
\usepackage{psfrag}

\usepackage{enumerate}   % for custom numbered lists like a), b) 
\calctocindent 
\smartqed

\newcommand{\assign}{:=}
\newcommand{\bignone}{\,}
\newcommand{\emdash}{ -- }
\newcommand{\mathpi}{\pi}
\newcommand{\mathd}{\D}
\newcommand{\mathe}{\eul}
\newcommand{\mathi}{\imag}
\newcommand{\tmem}[1]{\emph{#1}}
\newcommand{\tmop}[1]{\ensuremath{\operatorname{#1}}}
\newcommand{\tmmathbf}[1]{\vec{#1}}
\newcommand{\tmstrong}[1]{\emph{#1}}
\newcommand{\tmtextbf}[1]{\textbf{#1}}
\newcommand{\tmtextit}[1]{\textit{#1}}
\newcommand{\um}{-}

\newenvironment{itemizedot}{\begin{itemize}}{\end{itemize}}
\newcommand{\TeXmacs}{T\kern-.1667em\lower.5ex\hbox{E}\kern-.125emX\kern-.1em\lower.5ex\hbox{\textsc{m\kern-.05ema\kern-.125emc\kern-.05ems}}}

\makeatletter
\def\tableofcontents{\section*{\contentsname\markboth{{\contentsname}}%
                                                    {{\contentsname}}}
 \def\authcount##1{\setcounter{auco}{##1}\setcounter{@auth}{1}}
 \def\lastand{\ifnum\value{auco}=2\relax
                 \unskip{} \andname\
              \else
                 \unskip \lastandname\
              \fi}%
 \def\and{\stepcounter{@auth}\relax
          \ifnum\value{@auth}=\value{auco}%
             \lastand
          \else
             \unskip,
          \fi}%
 \@starttoc{toc}\if@restonecol\twocolumn\fi}
\makeatother

\begin{document}
\mainmatter

\setcounter{page}{223}
\setcounter{chapter}{4}
\title{Introduction to Decoherence Theory}
\author{Klaus Hornberger}
\institute{Arnold Sommerfeld Center for Theoretical Physics,
Ludwig--Maximilians--Universit\"at M\"unchen, Theresienstra{\ss}e 37, 80333 Munich, Germany
}

\maketitle
\section{The Concept of Decoherence}

This introduction to the theory of decoherence is aimed
at readers with an interest in the science of quantum information. In that field, one
is usually content with simple, abstract descriptions of non-unitary ``quantum
channels'' to account for imperfections in quantum processing tasks.
However, in order to justify such models of non-unitary evolution and to
understand their limits of applicability it is important to know their
physical basis. I will therefore emphasize the dynamic and microscopic
origins of the phenomenon of decoherence, and will relate it to concepts from
quantum information where applicable, in particular to the theory of quantum
measurement.
\renewcommand{\thefootnote}{}
\footnote{\textsf{This text corresponds to a chapter in: 
\emph{A. Buchleitner, C. Viviescas, and M. Tiersch (Eds.),
Entanglement and Decoherence. Foundations and Modern Trends,
Lecture Notes in Physics, Vol.~{768}, Springer, Berlin (2009).}
\\Short reference: K. Hornberger, Lect. Notes Phys. \textbf{768}, 223-278 (2009)}}
\renewcommand{\thefootnote}{\arabic{footnote}}
\setcounter{footnote}{0}

The study of decoherence, though based at the heart of quantum theory, is a
relatively young subject. It was initiated in the 1970's and 1980's with the
work of H.~D.~Zeh and W.~Zurek on the emergence of classicality in the quantum
framework. Until that time the orthodox interpretation of quantum mechanics
dominated, with its strict distinction between the classical macroscopic world
and the microscopic quantum realm. The mainstream attitude concerning the
boundary between the quantum and the classical was that this was a purely
philosophical problem, intangible by any physical analysis. This changed with
the understanding that there is no need for denying quantum mechanics to hold
even macroscopically, if one is only able to understand within the framework
of quantum mechanics why the macro-world {\tmem{appears}} to be classical. For
instance, macroscopic objects are found in approximate position eigenstates of
their center-of-mass, but never in superpositions of macroscopically distinct
positions. The original motivation for the study of decoherence was to explain
these effective {\tmem{super-selection rules}} and the apparent emergence of
classicality within quantum theory by appreciating the crucial role played by
the environment of a quantum system.

Hence, the relevant theoretical framework for the study of decoherence is the
theory of {\tmem{open quantum systems}}, which treats the effects of an
uncontrollable environment on the quantum evolution. Originally developed to
incorporate the phenomena of friction and thermalization in the quantum
formulation, it has of course a much longer history than decoherence theory.
However, we will see that the intuition and approximations developed in the
traditional treatments of open quantum systems are not necessarily appropriate
to yield a correct description of decoherence effects, which may take place on
a time scale much shorter than typical relaxation phenomena. In a sense, one
may say that while the traditional treatments of open quantum systems focus on
how an environmental ``bath'' affects the system, the emphasis in decoherence
is more on the contrary question, namely how the system affects and disturbs
environmental degrees of freedom, thereby revealing information about its
state.

The physics of decoherence became very popular in the last decade, mainly due
to advances in experimental technology. In a number of experiments the gradual
emergence of classical properties in a quantum system could be observed, in
agreement with the predictions of decoherence theory. Needles to say, a second
important reason for the popularity of decoherence is its relevance for
quantum information processing tasks, where the coherence of a relatively
large quantum system has to be maintained over a long time.

Parts of these lecture notes are based on the books on decoherence by E.~Joos
et al. {\cite{Joos2003a}} and on open quantum systems by H.-P.~Breuer \&
F.~Petruccione {\cite{Breuer2002a}}, and on the lecture notes of W.~Strunz
{\cite{Strunz2002a}}. Interpretational aspects, which are not covered here,
are discussed in {\cite{Bacciagaluppi2005a,Schlosshauer2005a}}, and useful
reviews by W.~H.~Zurek and J.~P.~Paz can be found in
{\cite{Zurek2003a,Paz2001a}}. This chapter deals exclusively with
conventional, i.e. environmental decoherence, as opposed to spontaneous
reduction theories {\cite{Bassi2003a}}, which aim at \ ``solving the
measurement problem of quantum mechanics'' by modifying the Schr\"odinger
equation. These models are conceptually very different from environmental
decoherence, though their predictions of super-selection rules are often
qualitatively similar.

\subsection{Decoherence in a Nutshell}
\label{KHsec:nutshell}

Let us start by discussing the basic decoherence effect in a rather general
framework. As just mentioned, we need to account for the unavoidable coupling
of the quantum system to its environment. Although these environmental degrees
of freedom are to be treated quantum mechanically, their state must be taken
unobservable for all practical purposes, be it due to their large number or
uncontrollable nature. In general, the detailed temporal dynamics induced by
the environmental interaction will be very complicated, but one can get an
idea of the basic effects by assuming that the interaction is sufficiently
short-ranged to admit a description in terms of scattering theory. In this
case, only the map between the asymptotically free states before and after the
interaction needs to be discussed, thus avoiding a temporal description of the
collision dynamics.

Let the quantum state of a system be described by the density operator
$\rho$ on the Hilbert space $\mathcal{H}$. We take the system to interact with
a single environmental degree of freedom at a time -- think of a phonon, a
polaron, or a gas particle scattering off your favorite implementation of a
quantum register. Moreover, let us assume, for the time being, that this
environmental ``particle'' is in a pure state $\rho_{\text{E}} = |
\psi_{\tmop{in}} \rangle \langle \psi_{\tmop{in}} |_{\text{E}}$, with $|
\psi_{\tmop{in}} \rangle_{\text{E}} \in \mathcal{H}_{\text{E}}$. The
scattering operator $\mathsf{S}_{\tmop{tot}}$ maps between the in- and
out-asymptotes in the total Hilbert space $\mathcal{H}_{\tmop{tot}}
=\mathcal{H} \otimes \mathcal{H}_E$, and for sufficiently short-ranged
interaction potentials we may identify those with the states before and after
the collision. The initially uncorrelated system and environment turn into a
joint state,
\begin{eqnarray}
  \text{[before collision]} \bignone &  & \rho_{\tmop{tot}} = \rho \otimes |
  \psi_{\tmop{in}} \rangle \langle \psi_{\tmop{in}} |_{\text{E}} \;, \\
  \text{[after collision]}  &  & \rho'_{\tmop{tot}} = \mathsf{S} _{\tmop{tot}}
  [\rho \otimes | \psi_{\tmop{in}} \rangle \langle \psi_{\tmop{in}}
  |_{\text{E}}] \mathsf{S}_{\tmop{tot}}^{\dag} \;.  \label{KHeq:rhop}
\end{eqnarray}
Now let us assume, in addition, that the interaction is {\tmem{non-invasive}}
\index{interaction!-- non-invasive} with respect to a certain system property.
This means that there is a number of distinct system states, such that the
environmental scattering off these states causes no transitions in the system.
For instance, if these distinguished states correspond to the system being
localized at well-defined sites then the environmental particle should induce
no hopping between the locations. In the case of elastic scattering, on the
other hand, they will be the energy eigenstates. Denoting the set of these
mutually orthogonal system states by $\left\{ |n \rangle \right\} \in
\mathcal{H}$, the requirement of non-invasiveness means that
$\mathsf{S}_{\tmop{tot}}$ commutes with those states, that is, it has the form
\begin{equation}
  \mathsf{S}_{\tmop{tot}} = \sum_n |n \rangle \langle n| \bignone \otimes
  \mathsf{S}_n \;,
\end{equation}
where the $\mathsf{S}_n$ are scattering operators acting in the environmental
Hilbert space. The insertion into (\ref{KHeq:rhop}) yields
\begin{eqnarray}
  \rho'_{\tmop{tot}} & = & \sum_{m, n} \langle m| \rho |n \rangle |m \rangle
  \langle n| \otimes \mathsf{S}_m | \psi_{\tmop{in}} \rangle \langle
  \psi_{\tmop{in}} |_{\text{E}}  \mathsf{S}_n^{\dag} \nonumber \\
  & \equiv & \sum_{m, n} \rho_{mn} |m \rangle \langle n| \otimes |
  \psi_{\tmop{out}}^{\left( m \right)} \rangle \langle
  \psi_{\tmop{out}}^{\left( n \right)} |_{\text{E}} \;,
\end{eqnarray}
and disregarding the environmental state by performing a partial trace we get
the {\tmem{system}} state after the interaction:
\begin{equation}
  \rho' = \tmop{tr}_{\text{E}}  \left( \rho_{\tmop{tot}}' \right) = \sum_{m,
  n} |m \rangle \langle n| \rho_{mn}  \underset{\langle
  \psi_{\text{out}}^{\left( n \right)} | \psi_{\text{out}}^{\left( m \right)}
  \rangle_{\text{E}}}{\underbrace{\langle \psi_{\tmop{in}} | \bignone
  \mathsf{S}_n^{\dag}  \mathsf{S}_m | \psi_{\tmop{in}} \rangle_{\text{E}}}} \;.
  \label{KHtrace1}
\end{equation}
Since the $\mathsf{S}_n$ are unitary the diagonal elements, or
{\tmem{populations}}\index{state!-- populations}, are indeed unaffected,
\begin{eqnarray}
  &  & \rho'_{m m} = \rho_{m m} \;, 
\end{eqnarray}
while the off-diagonal elements, or {\tmem{coherences}}\index{state!-- coherences}, get multiplied by the overlap of the environmental states
scattered off the system states $m$ and $n$,
\begin{equation}
  \rho_{m n}' = \rho_{m n} \langle \psi_{\text{out}}^{\left( n \right)} |
  \psi_{\text{out}}^{\left( m \right)} \rangle \label{KHScoh} \;.
\end{equation}
This factor has a modulus of less than one so that the coherences, which
characterize the ability of the system state to display a superposition
between $|m \rangle$ and $|n \rangle$, get suppressed.\footnote{The value $|
    \rho_{m n} |$ determines the maximal fringe visibility in a general
    interference experiment involving the states $|m \rangle$ and $|n
    \rangle$, as described by the projection on a general superposition {$|
      \psi_{\theta, \varphi} \rangle_{} = \cos \left( \theta \right) \text{$|m
        \rangle$+$\mathe^{\mathi \varphi} $sin$\left( \theta \right) |n
        \rangle$}$}.} It is important to note that this loss of coherence
occurs in a {\tmem{special basis}}, which is determined only by the scattering
operator, i.e. by the type of environmental interaction, and to a degree that
is determined by both the environmental state and the interaction.

This loss of the ability to show quantum behavior due to the interaction with
an environmental quantum degree of freedom is the basic effect of decoherence.
One may view it as due to the arising correlation between the system with the
environment. After the interaction the joint quantum state of system and
environment is no longer separable, and part of the coherence initially
located in the system now resides in the non-local correlation between system
and the environmental particle; it is lost once the environment is
disregarded. A complementary point of view argues that the interaction
constitutes an information transfer from the system to the environment. The
more the overlap in (\ref{KHScoh}) differs in magnitude from unity, the more
an observer could in principle learn about the system state by measuring the
environmental particle. Even though this measurement is never made, the
complementarity principle then explains that the wave-like interference
phenomenon characterized by the coherences vanishes as more information
discriminating the distinct, ``particle-like'' system states is revealed.

To finish the introduction, here is a collection of other characteristics and
popular statements about the decoherence phenomenon. One often hears that
decoherence ({\tmem{i}}) can be extremely fast as compared to all other
relevant time scales, ({\tmem{ii}}) that it can be interpreted as an indirect
measurement process, a monitoring of the system by the environment,
({\tmem{iii}}) that it creates dynamically a set of preferred states (``robust
states'' or ``pointer states'')\index{pointer states}\index{state!-- pointer|see{pointer states}}\index{pointer!-- states}\index{pointer states!-- robust states}
which seemingly do not obey the superposition principle, thus providing the
microscopic mechanism for the emergence of effective super-selection rules,
and ({\tmem{iv}}) that it helps to understand the emergence of classicality in
a quantum framework. These points will be illustrated in the following, though
({\tmem{iii}}) has been demonstrated only for very simple model systems and
({\tmem{iv}}) depends to a fair extent on your favored interpretation of
quantum mechanics.

\subsection{General Scattering Interaction}
\label{KHsec:scattering}
\index{interaction!-- scattering}

In the above demonstration of the decoherence effect the choice of the
interaction and the environmental state was rather special. Let us therefore
now take $\mathsf{S}_{\tmop{tot}} \tmop{and} \rho_{\text{E}}$ to be arbitrary
and carry out the same analysis. Performing the trace in (\ref{KHtrace1}) in
the eigenbasis of the environmental state, $\rho_{\text{E}} = \sum_{\ell}
p_{\ell} | \psi_{\ell} \rangle \langle \psi_{\ell} | \bignone_{\text{E}}$, we
have
\begin{eqnarray}
   \rho' = \tmop{tr}_{\text{E}}  \left( \mathsf{S}_{\tmop{tot}} [\rho_{}
   \otimes \rho_{\text{E}}] \mathsf{S}_{\tmop{tot}}^{^{\dag}} \right) & = &
   \sum_{j, \ell} \bignone p_{\ell} \langle \psi_j | \mathsf{S}_{\tmop{tot}} |
   \psi_{\ell} \rangle_{\text{E}} \rho_{} \langle \psi_{\ell} |
   \mathsf{S}^{\dag}_{\tmop{tot}} | \psi_j \rangle_{\text{E}} \nonumber\\
   & = & \sum_k \mathsf{W}_k \bignone \rho_{}  \mathsf{W}_k^{\dag} \;, 
   \label{KHops}
\end{eqnarray}
where the $\langle \psi_j | \mathsf{S}_{\tmop{tot}} | \psi_{\ell} \rangle$ are
operators in $\mathcal{H}$. After subsuming the two indices $j, \ell$ into a
single one, we get the second line with the {\tmem{Kraus operators}}\index{Kraus!-- operators}
$\mathsf{W}_k \bignone $ given by
\begin{eqnarray}
   \mathsf{W}_k & = & \sqrt{p_{\ell_k}} \langle \psi_{j_k} |
   \mathsf{S}_{\tmop{tot}} | \psi_{\ell_k} \rangle \;. \label{KHKrausop}
\end{eqnarray}
It follows from the unitarity of $\mathsf{S}_{\tmop{tot}}$ that they satisfy
\begin{equation}
   \sum_k \mathsf{W}_k^{\dag}  \mathsf{W}_k \bignone =\mathbbm{I} \;.
\end{equation}
This implies that (\ref{KHops}) is the operator-sum representation of a
completely positive map $\Phi : \rho \mapsto \rho'$ (see Sect.
\ref{KHsemigroups}). In other words, the scattering transformation has the
form of the most general evolution of a quantum state that is compatible with
the rules of quantum theory. In the operational formulation of quantum
mechanics this transformation is usually called a {\tmem{quantum operation}}
\index{quantum!-- operation} {\cite{Kraus1983a}}, the quantum information
community likes to call it a {\tmem{quantum channel}}\index{quantum!-- channel}. Conversely, given an arbitrary quantum channel, one can also
construct a scattering operator $\mathsf{S}_{\tmop{tot}}$ and an environmental
state $\rho_{\text{E}}$ giving rise to the transformation, though it is
usually not very helpful from a physical point of view to picture the action
of a general, dissipative quantum channel as due to a single scattering event.

\subsection{Decoherence as an Environmental Monitoring
 Process}\index{continuous monitoring}\index{continuous monitoring!-- and decoherence}\index{decoherence!-- as continuous monitoring }
\label{KHsec:monitoring}

We are now in a position to relate the decoherence of a quantum system to the
information it reveals to the environment. Since the formulation is based on
the notion of an {\tmem{indirect measurement}} it is necessary to first collect some aspects of
measurement theory {\cite{Busch1991a,Holevo2001a}}.

\subsubsection{Elements of General Measurement Theory}

\paragraph{Projective Measurements}
\index{measurement!-- projective}

This is the type of measurement discussed in standard textbooks of quantum
mechanics. A projective operator $| \alpha \rangle \langle \alpha | \equiv
\mathsf{P}_{\alpha} = \mathsf{P}_{\alpha}^2 = \mathsf{P}_{\alpha}^{\dag}$ is
attributed to each possible outcome $\alpha$ of an idealized measurement
apparatus. The probability of the outcome $\alpha$ is obtained by the Born
rule
\begin{eqnarray}
  \tmop{Prob} (\alpha | \rho) & = & \tmop{tr} \left( \mathsf{P}_{\alpha} \rho
  \right) = \langle \alpha | \rho | \alpha \rangle \;, 
\end{eqnarray}
and after the measurement of $\alpha$ the state of the quantum system is given
by the normalized projection
\begin{eqnarray}
  \mathcal{M}:\;\rho & \mapsto  & \mathcal{M} \left( \rho | \alpha
  \right) = \frac{\mathsf{P}_{\alpha} \rho \mathsf{P}_{\alpha}}{\tmop{tr}
  \left( \mathsf{P}_{\alpha} \rho \right)} \;.  \label{KHeq:projectivem}
\end{eqnarray}
The basic requirement that the projectors form a resolution of the identity
operator,
\begin{eqnarray}
  \sum_{\alpha} \mathsf{P}_{\alpha} & = & \mathbbm{I} \;,  \label{KHPres}
\end{eqnarray}
ensures the normalization of the corresponding probability distribution \linebreak
$\tmop{Prob} \left( \alpha | \rho \right)$.

If the measured system property corresponds to a self-adjoint operator
$\mathsf{A}$ the $\mathsf{P}_{\alpha}$ are the projectors into its
eigenspaces, so that its expectation value is
\begin{eqnarray}
  \langle \mathsf{A} \rangle & = & \tmop{tr} \left( \mathsf{A} \rho \right) \;.
  \nonumber
\end{eqnarray}
If $\mathsf{A}$ has a continuous spectrum the outcomes are characterized by
intervals of a real parameter, and the sum in (\ref{KHPres}) should be
replaced by a projector-valued Stieltjes integral {$\int \mathd \mathsf{P}
 (\alpha) \bignone \bignone$=$\mathbbm{I}$}, or equivalently by a Lebesgue
integral over a {\tmem{projector-valued measure}} (PVM)\index{projector-value measure (PVM)} {\cite{Busch1991a,Holevo2001a}}.

It is important to note that projective measurements are not the most general
type of measurement compatible with the rules of quantum mechanics. In fact,
non-destructive measurements\index{measurement!-- non-destructive} of a quantum system are usually not of the projective
kind.

\paragraph{Generalized Measurements}

In the most general measurement situation, a positive (and therefore
hermitian) operator $\mathsf{F}_{\alpha} > 0$ is attributed to each outcome
$\alpha$. Again, the collection of operators corresponding to all possible
outcomes must form a resolution of the identity operator,
\begin{eqnarray}
   \sum_{\alpha} \mathsf{F}_{\alpha} \bignone & = & \mathbbm{I}. 
\end{eqnarray}
In particular, one speaks of a {\tmem{positive operator-valued measure}}
(POVM)\index{positive operator-valued measure (POVM)} in the case of a continous outcome parameter, $\int \mathd \mathsf{F}
(\alpha) =\mathbbm{I} \bignone$, and the probability (or probability density
in the continuous case) of outcome $\alpha$ is given by
\begin{eqnarray}
   \tmop{Prob} (\alpha | \rho) & = & \tmop{tr} \left( \mathsf{F}_{\alpha} \rho
   \right) . 
\end{eqnarray}
The effect on the system state of a generalized measurement is described by a
nonlinear transformation
\begin{eqnarray}
    \mathcal{M}:\;\rho & \mapsto &  \mathcal{M} \left( \rho | \alpha
   \right) = \frac{\sum_k \mathsf{M}_{\alpha, k} \rho \bignone 
   \mathsf{M}^{\dagger}_{\alpha, k}}{\tmop{tr} \left( \mathsf{F}_{\alpha} \rho
   \right)}  \label{KHeq:transformation-state-measurement}
\end{eqnarray}
involving a norm-decreasing completely positive map in the numerator (see
Sect. \ref{KHsemigroups}), and a normalization which is subject to the
{\tmem{consistency requirement }}
\begin{eqnarray}
   \sum_k \mathsf{M}_{\alpha, k}^{\dag}  \mathsf{M}_{\alpha, k} & = &
   \mathsf{F}_{\alpha} \bignone .  \label{KHeq:consireq}
\end{eqnarray}
The operators $\mathsf{M}_{\alpha, k}$ appearing in
(\ref{KHeq:transformation-state-measurement}) are called {\tmem{measurement
   operators}}\index{operator!-- measurement}\index{measurement!-- operator}, and they serve to characterize
the measurement process completely. The $\mathsf{F}_{\alpha}$ are sometimes
called ``effects'' or ``measurement elements''. Note that different
measurement operators $\mathsf{M}_{\alpha, k}$ can lead to the same
measurement element $\mathsf{F}_{\alpha}$.

A simple class of generalized measurements are {\tmem{unsharp
   measurements}}\index{measurement!-- unsharp}, where a number of projective
operators are lumped together with probabilistic weights in order to account
for the finite resolution of a measurement device or for classical noise in
its signal processing. However, generalized measurements schemes may also
perform tasks which are seemingly impossible with a projective measurement,
such as the error-free discrimination of two non-orthogonal states
{\cite{Helstrom1976a,Chefles2000a}}.

\paragraph{Efficient Measurements}

A generalized measurement is called {\tmem{efficient}}\index{measurement!-- efficient} if there is only a
single summand in (\ref{KHeq:transformation-state-measurement}) for each
outcome $\alpha$,
\begin{eqnarray}
  \mathcal{M} \left( \rho | \alpha \right) & = & \frac{\mathsf{M}_{\alpha}
  \rho \bignone  \mathsf{M}_{\alpha}^{\dag}}{\tmop{tr} \left(
  \mathsf{M}_{\alpha}^{\dag} \mathsf{M}_{\alpha} \rho \right)}\;, 
  \label{KHeq:efficientm}
\end{eqnarray}
implying that pure states are mapped to pure states. In a sense, these are
measurements where no unnecessary, that is no classical, uncertainty is
introduced during the measurement process, see below. By means of a
(left) polar decomposition and the consistency requirement
(\ref{KHeq:consireq}) efficient measurement operators have the form
\begin{equation}
  \mathsf{M}_{\alpha} = \mathsf{U}_{\alpha}  \sqrt{\mathsf{F}_{\alpha}}\;,
\end{equation}
with an unitary operator $\mathsf{U}_{\alpha}$. This way the state after
efficient measurement can be expressed in a form which decomposes the
transformation into a ``raw measurement''\index{measurement!-- raw} described
by the $\mathsf{F}_{\alpha}$ and a ``measurement
back-action''\index{measurement!-- back-action} given by the
$\mathsf{U}_{\alpha}$:
\begin{equation}
  \mathcal{M} \left( \rho | \alpha \right) =
  \underset{\text{back-action}}{\underbrace{\mathsf{U}_{\alpha}}}
  \underset{\parbox{.7in}{\centering\scriptsize raw \\ measurement}}
{\underbrace{\frac{\sqrt{\mathsf{F}_{\alpha}} \rho
  \sqrt{\mathsf{F}_{\alpha}}}{\tmop{tr} \left( \mathsf{F}_{\alpha} \rho
  \right)}}} 
  \underset{\text{back-action}}{\underbrace{\mathsf{U}_{\alpha}^{\dag}}} \;.
  \label{KHeq:effectivemeastrafo}
\end{equation}

In this transformation the positive operators $\sqrt{\mathsf{F}_{\alpha}}$
``squeeze'' the state along the measured property and expand it along the
other, complementary ones, similar to what a projector would do, while the
back-action operators $\mathsf{U}_{\alpha}$ ``kick'' the state by transforming
it in a way that is reversible, in principle, provided the outcome $\alpha$
is known. Note that the projective measurements (\ref{KHeq:projectivem}) are a
sub-class in the set of back-action-free efficient measurements.

\subsubsection{Indirect Measurements}\index{measurement!-- indirect}

In an {\tmem{indirect measurement}} one tries to obtain information about the
system in a way that disturbs it as little as possible. This is done by
letting a well-prepared microscopic quantum probe interact with the system.
This probe is then measured by projection, i.e. destructively, so that one
can infer properties of the system without having it brought into contact with
a macroscopic measurement device. Let $\rho_{\text{probe}}$ be the prepared
state of the probe, $\mathsf{S}_{\tmop{tot}}$ describe the interaction between
system and probe, and $\mathsf{P}_{\alpha}$ be the projectors corresponding to
the various outcomes of the probe measurement. The probability of measuring
$\alpha$ is determined by the reduced state of the probe after interaction,
i.e.
\begin{equation}
  \tmop{Prob} (\alpha | \rho) = \tmop{tr}_{\text{probe}} \left(
  \mathsf{P}_{\alpha} \rho'_{\text{probe}} \right) = \tmop{tr}_{\text{probe}}
  \left( \mathsf{P}_{\alpha} \tmop{tr}_{\text{sys}} (
  \mathsf{S}_{\tmop{tot}} [\rho \otimes \rho_{\text{probe}}]
  \mathsf{S}_{\tmop{tot}}^{\dag}) \right) \;.
\end{equation}
By pulling out the system trace (extending the projectors to
${\mathcal{H}_{\tmop{tot}} = {\mathcal{H} \otimes \mathcal{H}_{\text{p}}}}$)
and using the cyclic permutability of operators under the trace we have
\begin{equation}
  \tmop{Prob} (\alpha | \rho) = \tmop{tr} \left(
  \mathsf{S}_{\tmop{tot}}^{\dag} [\mathbbm{I} \otimes \mathsf{P}_{\alpha}]
  \mathsf{S}_{\tmop{tot}}  \left[ \rho \otimes \rho_{\text{probe}} \right]
  \right) = \tmop{tr} \left( \mathsf{F}_{\alpha} \rho \right) \;,
\end{equation}
with microscopically defined measurement elements
\begin{eqnarray}
  \mathsf{F}_{\alpha} & = & \tmop{tr}_{\text{probe}} \left(
  \mathsf{S}_{\tmop{tot}}^{\dag} [\mathbbm{I} \otimes \mathsf{P}_{\alpha}]
  \mathsf{S}_{\tmop{tot}} [ \mathbbm{I} \otimes \rho_{\text{probe}}] \right) >
  0 
\end{eqnarray}
satisfying $\sum_{\alpha} \mathsf{F}_{\alpha} =\mathbbm{I} \bignone .$ Since
the probe measurement is projective, we can also specify the new system state
conditioned on the click at $\alpha$ of the probe detector,
\begin{eqnarray}
  \mathcal{M} \left( \rho | \alpha \right) & = & \tmop{tr}_{\text{probe}} 
  \left( \mathcal{M}_{\tmop{tot}} \left( \rho_{\tmop{tot}} | \alpha \right)
  \right) \nonumber\\
  & = & \tmop{tr}_{\text{probe}}  \left( \frac{[\mathbbm{I} \otimes
  \mathsf{P}_{\alpha}] \mathsf{S}_{\tmop{tot}}  \left[ \rho \otimes
  \rho_{\text{probe}} \right]  \mathsf{S}_{\tmop{tot}}^{\dag} [\mathbbm{I}
  \otimes \mathsf{P}_{\alpha}]}{\tmop{tr} \left( \mathsf{F}_{\alpha} \rho
  \right)} \right) \nonumber\\
  & = & \sum_k \frac{\mathsf{M}_{\alpha, k} \rho \mathsf{M}_{\alpha,
  k}^{\dag}}{\tmop{tr} \left( \mathsf{F}_{\alpha} \rho \right)} \bignone \;. 
\end{eqnarray}
In the last step a convex decomposition of the initial probe state into pure
states was inserted, {$\rho_{\text{probe}} = \sum_k w_k | \psi_k \rangle
 \langle \psi_k | \bignone$}. Taking $\mathsf{P}_{\alpha} = | \alpha \rangle
\langle \alpha |$ we thus get a microscopic description also of the
measurement operators,
\begin{equation}
  \mathsf{M}_{\alpha, k} = \sqrt{w_k} \langle \alpha | \mathsf{S}_{\tmop{tot}}
  | \psi_k \rangle \;.
\end{equation}
This shows that an indirect measurement is efficient\index{measurement!-- efficient} (as defined above) if the probe is initially in a pure state,
i.e. if there is no uncertainty introduced in the measurement process, apart
from the one imposed by the uncertainty relations on $\rho_{\text{probe}}$.

If we know that an indirect measurement has taken place, but do not know its
outcome $\alpha$ we have to resort to a probabilistic (Bayesian) description
of the new system state. It is given by the sum over all possible outcomes
weighted by their respective probabilities,
\begin{equation}
  \rho' = \sum_{\alpha} \tmop{Prob} (\alpha | \rho) \mathcal{M} \left( \rho |
  \alpha \right) = \sum_{\alpha, k} \mathsf{M}_{\alpha, k} \rho
  \mathsf{M}_{\alpha, k}^{\dag} \;.
\end{equation}
This form is the same as above in (\ref{KHops}) and
(\ref{KHKrausop}), where the basic effect of decoherence has been described.
This indicates that the decoherence process can be legitimately viewed as a
consequence of the information transfer from the system to the environment.
The complementarity principle can then be invoked to understand which
particular system properties lose their quantum behavior, namely those
complementary to the ones revealed to the environment. This ``monitoring
interpretation'' of the decoherence process will help us below to derive
microscopic master equations.

\subsection{A Few Words on Nomenclature}

Since decoherence phenomena show up in quite different sub-communities of
physics, a certain confusion and lack of uniformity developed in the
terminology. This is exacerbated by the fact that decoherence often reveals
itself as a loss of fringe visibility in interference experiments {\emdash} a
phenomenon, though, which may have other causes than decoherence proper. Here
is an attempt of clarification:
\begin{itemizedot}
\item {\tmstrong{decoherence:}}\index{decoherence} In the original sense, an
 environmental quantum effect affecting macroscopically distinct states. The
 term is nowadays applied to mesoscopically different states as well, and
 even for microscopic states, as long as it refers to the {\tmem{quantum}}
 effect of environmental, i.e. in practice unobservable, degrees of freedom.

 However, the term is often (ab-)used for any other process reducing the
 purity of a micro-state.

\item {\tmstrong{dephasing:}}\index{dephasing} In a narrow sense, this
 describes the phenomenon that coherences, i.e., the off-diagonal elements of
 the density matrix, get reduced in a particular basis, namely the energy
 eigenbasis of the system.  It is a statement about the effect, and not the
 cause. In particular, dephasing may be {\tmem{reversible}} if it is not due
 to decoherence, as revealed e.g. in spin-echo experiments.

 This phrase should be treated with great care since it is used differently
 in various sub-communities. It is taken as a synonym to
 ``{\tmem{dispersion}}'' in molecular physics and in nonlinear optics, as a
 synonym to ``{\tmem{decoherence}}'' in condensed matter, and often as a
 synonym to ``{\tmem{phase averaging}}'' in matter wave optics. It is also
 called a {\tmem{$T_2$-process}} in NMR and in condensed matter physics (see
 below).

\item {\tmstrong{phase averaging:}}\index{phase!-- averaging} A classical noise
 phenomenon entering through the dependence of the unitary system evolution
 on external control parameters which fluctuate (parametric average over
 unitary evolutions). A typical example are the vibrations of an
 interferometer grating or the fluctuations of the classical magnetic field
 in an electron interferometer due to technical noise. Empirically, phase
 averaging is often hard to distinguish from decoherence proper.

\item {\tmstrong{dispersion:}}\index{dispersion} Coherent broadening of wave
 packets during the unitary evolution, e.g. due to a velocity dependent group
 velocity or non-harmonic energy spacings. This unitary effect may lead to a
 reduction of signal oscillations, for instance, in molecular pump-probe
 experiments.

\item {\tmstrong{dissipation:}}\index{dissipation} Energy exchange with the
 environment leading to thermalization. Usually accompanied by decoherence,
 but see Sect.~\ref{KHsec:examples} for a counter-example.
\end{itemizedot}

\section{Case Study: Dephasing of Qubits}
\label{KHsec:qubit}
\index{qubit!-- dephasing}\index{dephasing!-- of qubits}

So far, the discussion of the temporal dynamics of the decoherence process was
circumvented by using a scattering description. Before going to the general
treatment of open quantum systems in Sect.~\ref{KHsec:Markov}, it is helpful
to take a closer look on the time evolution of a special system where the
interaction with a model environment can be treated exactly
{\cite{Palma1996a,Breuer2002a}}.

\subsection{An Exactly Solvable Model}\label{KHsec:qubitmodel}

Let us take a two-level system, or qubit, described by the Pauli spin operator
$\sigma_z$, and model the environment as a collection of bosonic field modes.
In practice, such fields can yield an appropriate effective description even
if the actual environment looks quite differently, in particular if the
environmental coupling is a sum of many small contributions.{\footnote{A
    counter-example would be the presence of a degenerate environmental degree
    of freedom, such as a bistable fluctuator.}} What is fairly non-generic
in the present model is the type of coupling between system and environment,
which is taken to {{commute}} with the system Hamiltonian.

The total Hamiltonian thus reads
\begin{equation}
  \mathsf{H}_{\tmop{tot}} = \underbrace{\frac{\hbar \omega}{2} \sigma_z +
  \sum_k \hbar \omega_k  \mathsf{b}_k^{\dag} \mathsf{b}_k}_{\mathsf{H}_0}
  \bignone + \underset{\mathsf{H}_{\text{int}}}{\underbrace{\sigma_z  \sum_k
  \left( g_k  \mathsf{b}_k^{\dag} + g_k^{\ast}  \mathsf{b}_k  \right)
  \bignone}} \;,
\end{equation}
with the usual commutation relation for the mode operators of the bosonic
field modes, {$[ \mathsf{b}_i, \mathsf{b}^{\dag}_k] = \delta_{i k}$}, and
coupling constants $g_k$. The fact that the system Hamiltonian commutes with
the interaction, guarantees that there is no energy exchange between system
and environment so that we expect pure dephasing.

By going into the interaction picture one transfers the trivial time evolution
generated by $\mathsf{H}_{0\bignone}$ to the operators (and indicates this
with a tilde). In particular,
\begin{equation}
  \widetilde{\mathsf{H}}_{\tmop{int}} \left( t \right) = \mathe^{\mathi
  \mathsf{H} _0 t / \hbar}  \mathsf{H}_{\tmop{int}} \mathe^{\um \mathi
  \mathsf{H} _0 t / \hbar} = \sigma_z  \sum_k (g_k \mathe^{\mathi \omega_k t} 
  \mathsf{b}_k^{\dag} + g_k^{\ast} \mathe^{- \mathi \omega_k t}  \mathsf{b}_k)
  \bignone \;, \label{KHintpic}
\end{equation}
where the second equality is granted by the commutation $[\sigma_z,
\mathsf{H}_{\tmop{int}}] = 0$. The time evolution due to this Hamiltonian can
be formally expressed as a Dyson series,\index{Dyson series}
\begin{eqnarray}
  \widetilde{\mathsf{U}} (t) & = & \mathcal{T}_{\leftarrow} \exp \left( \um
  \frac{\mathi}{\hbar}  \int_0^t \mathd t' \widetilde{
  \mathsf{H}}_{\tmop{int}} (t') \bignone \right) \nonumber\\
  & = & \sum_{n = 0}^{\infty} \bignone \frac{1}{n!} \left( \frac{1}{\mathi
  \hbar} \right)^n \int_0^t \mathd t_1 \cdots \bignone \mathd t_n
  \mathcal{T}_{\leftarrow} \left[ \widetilde{ \mathsf{H}}_{\tmop{int}} (t_1)
  \cdots \widetilde{ \mathsf{H}}_{\tmop{int}} (t_n) \right] \;, 
\end{eqnarray}
where $\mathcal{T}_{\leftarrow}$ is the time ordering operator\index{operator! -- time ordering}\index{time ordering operator} (putting the operators with larger time arguments to the left).
Due to this time ordering requirement the series usually cannot be evaluated
exactly (if it converges at all). However, in the present case the commutator
of $\widetilde{ \mathsf{H}}_{\tmop{int}}$ at different times is not an
operator, but just a c-number,
\begin{equation}
  [ \widetilde{ \mathsf{H}}_{\tmop{int}} \left( t \right), \widetilde{
  \mathsf{H}}_{\tmop{int}} (t')] = 2 \mathi \sum_k \left| g_k \right|^2 \sin
  \left( \bignone \omega_k (t' - t) \right) \;.
\end{equation}
As a consequence, the time evolution differs only by a time-dependent phase
from the one obtained by casting the operators in their natural
order,{\footnote{To obtain the time evolution $\widetilde{\mathsf{U}} \left( t
    \right)$ for the case $[ \widetilde{\mathsf{H}} \left( t \right),
    \widetilde{\mathsf{H}} \left( t' \right)] = c \mathbbm{I}$ define the
    operators
 \begin{eqnarray}
   \mathsf{\Phi} \left( t \right) & = & \frac{1}{\hbar} \int_0^t \mathd t'
   \bignone \widetilde{\mathsf{H}} \left( t' \right) \nonumber
 \end{eqnarray}
 and $\overline{\mathsf{U}} \left( t \right) = \exp [\mathi \mathsf{\Phi}
 \left( t \right)] \widetilde{\mathsf{U}} \left( t \right)$. This way
 $\overline{\mathsf{U}} \left( t \right)$ describes the ``additional'' motion
 due to the time ordering requirement. It satisfies
 \begin{eqnarray}
   \partial_t \overline{\mathsf{U}} \left( t \right) & = & \left( \left[
   \frac{\mathd}{\mathd t} \mathe^{\mathi \mathsf{\Phi} \left( t \right)}
   \right] \mathe^{- \mathi \mathsf{\Phi} \left( t \right)} + \frac{1}{\mathi
   \hbar} \mathe^{\mathi \mathsf{\Phi} \left( t \right)} \widetilde{\mathsf{H}}
   \left( t \right) \mathe^{\um \mathi \mathsf{\Phi} \left( t \right)} \right)
   \overline{\mathsf{U}} \left( t \right) \;. \nonumber
 \end{eqnarray}
 The derivative in square brackets has to be evaluated with care since the
 $\widetilde{\mathsf{H}} \left( t \right)$ do not commute at different times.
 By first showing that $[ \mathsf{A}, \partial_t \mathsf{A}] = c \in
 \mathbbm{C}$ implies $\partial_t \mathsf{A}^n = n \mathsf{A}^{n - 1}
 \partial_t \mathsf{A} - \frac{1}{2} n \left( n - 1 \right) c \mathsf{A}^{n -
 2}$ one finds
 \begin{eqnarray}
   \frac{\mathd}{\mathd t} \mathe^{\mathi \mathsf{\Phi} \left( t \right)} & = &
   \frac{- 1}{\mathi \hbar} \mathe^{\mathi \mathsf{\Phi} \left( t \right)}
   \widetilde{\mathsf{H}} \left( t \right) + \frac{1}{2 \hbar^2} \mathe^{\mathi
   \mathsf{\Phi} \left( t \right)} \left[ \int_0^t \mathd t'
   \widetilde{\mathsf{H}} \left( t' \right), \widetilde{\mathsf{H}} \left( t
   \right) \bignone \right] \;. \nonumber
 \end{eqnarray}
 Therefore, we have $\partial_t \overline{\mathsf{U}} \left( t \right) = \left(
 2 \hbar^2 \right)^{- 1} \int_0^t \mathd t' [ \widetilde{\mathsf{H}} \left( t'
 \right), \widetilde{\mathsf{H}} \left( t \right)] \bignone
 \overline{\mathsf{U}} \left( t \right)$, which can be integrated to yield
 finally
 \begin{eqnarray}
   \widetilde{\mathsf{U}} \left( t \right) & = & \exp \left( - \frac{1}{2
   \hbar^2} \int_0^t \mathd t_1 \int_0^{t_1} \mathd t_2 \left[
   \widetilde{\mathsf{H}} \left( t_1 \right), \widetilde{\mathsf{H}} \left( t_2
   \right) \right] \bignone \right) \mathe^{- \mathi \mathsf{\Phi} \left( t
   \right)} \;. \nonumber
 \end{eqnarray}}}
\begin{equation}
   \widetilde{\mathsf{U}} (t) = \mathe^{\mathi \varphi (t)} \exp \left( \um
   \frac{\mathi}{\hbar}  \int_0^t \mathd t'  \widetilde{
   \mathsf{H}}_{\tmop{int}} (t') \right) \;,
\end{equation}
where the phase is given by
\begin{eqnarray}
  \varphi \left( t \right) & = & \frac{\mathi}{2 \hbar^2} \int_0^t \mathd t_1
  \int_0^t \mathd t_2 \Theta \left( t_1 - t_2 \right) \left[
  \widetilde{\mathsf{H}}_{\tmop{int}} \bignone \left( t_1 \right),
  \widetilde{\mathsf{H}}_{\tmop{int}} \bignone \left( t_2 \right) \right] \;. 
\end{eqnarray}
One can now perform the integral over the interaction Hamiltonian to get
\begin{equation}
  \widetilde{\mathsf{U}} (t) = \mathe^{\mathi \varphi (t)} \exp \left(
  \frac{1}{2} \sigma_z  \sum_k \left( \alpha_k (t) \mathsf{b}_k^{\dag} -
  \alpha_k^{\ast} (t) \mathsf{b}_k \right) \bignone \right) \;,
\end{equation}
with complex, time-dependent functions
\begin{equation}
  \alpha_k (t) \assign 2 g_k  \frac{1 - \mathe^{\mathi \omega_k t}}{\hbar
  \omega_k} \;. \label{KHalphat}
\end{equation}
The operator $\widetilde{\mathsf{U}} (t)$ is diagonal in the eigenbasis of the
system, and it describes how the environmental dynamics depends on the state
of the system. In particular, if the system is initially in the upper level,
$| \psi \rangle = | \uparrow \rangle$, one has
\begin{equation}
  \widetilde{\mathsf{U}} (t) | \uparrow \rangle | \xi_0 \rangle_{\text{E}} =
  \mathe^{\mathi \varphi (t)} | \uparrow \rangle \prod_k \mathsf{D}_k \left(
  \frac{\alpha_k (t)}{2} \right) | \xi_0 \rangle \hspace{0.6em} =:
  \mathe^{\mathi \varphi (t)} | \uparrow \rangle | \xi_{\uparrow} \left( t
  \right) \rangle_{\text{E}} \;, \label{KHeq:D1}
\end{equation}
and for the lower state
\begin{equation}
  \widetilde{\mathsf{U}} (t) | \downarrow \rangle | \xi_0 \rangle_{\text{E}} =
  \mathe^{\mathi \varphi (t)} | \downarrow \rangle \prod_k \mathsf{D}_k \left(
  - \frac{\alpha_k (t)}{2} \right) | \xi_0 \rangle \hspace{0.6em} =:
  \hspace{0.6em} \mathe^{\mathi \varphi (t)} | \downarrow \rangle |
  \xi_{\downarrow} \left( t \right) \rangle_{\text{E}} \;. \label{KHeq:D2}
\end{equation}
Here we introduced the unitary {\tmem{displacement operators}}\index{operator!-- displacement}\index{displacement operator} for the $k$-th field mode,
\begin{eqnarray}\label{KHeq:displacementoperator}
  \text{$\mathsf{D}_k (\alpha)$} & = & \exp (\alpha \mathsf{b}^{\dag}_k -
  \alpha^{\ast}  \mathsf{b}_k) \;, 
\end{eqnarray}
which effect a translation of the field state in its attributed phase space.
In particular, the coherent state\index{coherent states!-- definition} $| \alpha
\rangle_k$ of the field mode $k$ is obtained from its ground state $|0
\rangle_k$ by ${| \alpha \rangle_k \assign \mathsf{D}_k (\alpha) |0
 \rangle_k}$ {\cite{Walls1994a}}.

The equations (\ref{KHeq:D1}) and (\ref{KHeq:D2}) show that the collective
state of the field modes gets displaced by the interaction with the system and
that the sense of the displacement is determined by the system state.

Assuming that the states of system and environment are initially uncorrelated,
$\rho_{\tmop{tot}} \left( 0 \right) = \rho \otimes \rho_{\text{E}},$ the
time-evolved system state reads{\footnote{In fact, the assumption
   {$\rho_{\tmop{tot}} \left( 0 \right) = \rho \otimes \rho_{\text{E}}$} is
   quite unrealistic if the coupling is strong, as discussed below.
   Nonetheless, it certainly represents a valid initial state.}}
\begin{equation}
  \tilde{\rho} (t) = \tmop{tr}_{\text{E}}  \left( \widetilde{\mathsf{U}} (t)
  [\rho \otimes \rho_{\text{E}}] \widetilde{\mathsf{U}}^{\dag} (t) \right) \;.
\end{equation}
It follows from (\ref{KHeq:D1}) and (\ref{KHeq:D2}) that the populations are
unaffected,
\begin{eqnarray*}
  \langle \uparrow | \tilde{\rho} (t) | \uparrow \rangle & = & \langle
  \uparrow | \tilde{\rho} (0) | \uparrow \rangle \;,\\
  \langle \downarrow | \tilde{\rho} (t) | \downarrow \rangle & = & \langle
  \downarrow | \tilde{\rho} (0) | \downarrow \rangle\;,
\end{eqnarray*}
while the coherences are suppressed by a factor which is given by the trace
over the displaced initial field state,
\begin{equation}
  \langle \uparrow | \tilde{\rho} (t) | \downarrow \rangle = \langle \uparrow
  | \tilde{\rho} (0) | \downarrow \rangle \underbrace{\tmop{tr}_{\text{E}} 
  \left( \prod_k \mathsf{D}_k (\alpha_k (t)) \rho_{\text{E}} \bignone
  \right)}_{\chi \left( t \right)} \;.
\end{equation}
Incidentally, the complex suppression factor $\chi \left( t \right)$ is equal
to the {\tmem{Wigner characteristic function}}\index{Wigner!-- characteristic
 function} of the original environmental state at the points $\alpha_k (t)$,
i.e. it is given by the Fourier transform of its Wigner function
{\cite{Hillery1984a}}.

\subsubsection{Initial Vacuum State}

If the environment is initially in its vacuum state, $\rho_{\text{E}} =
\bigotimes_k \bignone |0 \rangle \langle 0|_k$, the $| \xi_{\uparrow} \left( t
\right) \rangle_{\text{E}}$ \ and $| \xi_{\downarrow} (t) \rangle_{\text{E}}$
defined in (\ref{KHeq:D1}), (\ref{KHeq:D2}) turn into multi-mode coherent
states, and the suppression factor can be calculated immediately to yield:
\begin{eqnarray}
  \chi_{\tmop{vac}} (t) & = & \prod_k \langle 0| \mathsf{D}_k (\alpha_k (t))
  |0 \rangle_k \bignone = \prod_k \exp \left( \um \frac{\left| \alpha_k (t)
  \right|^2}{2} \right) \nonumber\\
  & = & \exp \left( \um \sum_k 4 \left| g_k \right|^2  \frac{1 - \cos \left(
  \omega_k t \right)}{\hbar^2 \omega_k^2} \bignone \right)  \bignone \;.
  \label{KHchivac}
\end{eqnarray}
For times that are short compared to the field dynamics, $t \ll \omega_k^{\um
 1}$, one observes a Gaussian decay of the coherences. Modifications to this
become relevant at $\omega_k t \cong 1$, provided $\chi_{\tmop{vac}} \left( t
\right)$ is then still appreciable, i.e. for $4 \left| g_k \right|^2 /
\hbar^2 \omega_k^2 \ll 1$. Being a sum over periodic functions,
$\chi_{\tmop{vac}} (t)$ is quasi-periodic, that is, it will come back
arbitrarily close to unity after a large period (which increases exponentially
with the number of modes). These somewhat artificial Poincar\'e
   recurrences vanish if we replace the sum over the discrete modes by an
integral over a continuum with mode density $\mu$,
\begin{eqnarray}
  \sum_k f (\omega_k) & \longrightarrow & \int_0^{\infty} \mathd \omega \mu
  (\omega) f (\omega) \;,  \label{KHcontlim}
\end{eqnarray}
for any function $f$. This way the coupling constants $g_k$ get replaced by
the \tmem{spectral density} of the environment,
\begin{eqnarray}
  J (\omega) & = & 4 \mu (\omega) \left| g (\omega) \right|^2 \;. 
  \label{KHeq:J}
\end{eqnarray}
This function characterizes the environment by telling how effective the
coupling is at a certain frequency.

\subsubsection{Thermal State}\index{thermal!-- state}\index{state!-- thermal}

If the environment is in a thermal state with temperature $T$,
\begin{equation}
  \rho_{\text{E}} = \rho_{\tmop{th}} = \frac{\mathe^{- \mathsf{H}_{\text{E}} /
  k_B T}}{\tmop{tr} \left( \mathe^{- \mathsf{H}_{\text{E}} / k_B T} \right)} =
  \bigotimes_k \underbrace{\left( 1 - \mathe^{- \hbar \omega_k / k_B T} \right)
  \bignone  \sum_{n = 0}^{\infty} \mathe^{- \hbar \omega_k n / k_B T} \bignone
  |n \rangle \langle n|_k}_{= \rho_{\tmop{th}}^{(k)}} \;,
\end{equation}
the suppression factor reads{\footnote{This can be found in a small exercise
    by using the Baker--Hausdorff relation with $\exp \left( \alpha
      \mathsf{b}^{\dag} - \alpha^{\ast} \mathsf{b} \right) = \exp (\um |
    \alpha |^2 \text{/2}) \exp \left( \alpha \mathsf{b}^{\dag} \right) \exp
    \left( \um \alpha^{\ast} \mathsf{b} \right)$, and the fact that coherent
    states\index{coherent states!-- definition} satisfy the eigenvalue equation
    $\mathsf{b} | \beta \rangle = \beta | \beta \rangle$, have the number
    representation\index{coherent states!-- number representation} ${\langle n|
      \beta \rangle = \exp (\um | \beta |^2 / 2) \beta^n / \sqrt{n!}}$, and
    form an over-complete set\index{coherent states!-- overcompleteness} with
    {$\mathbbm{I}= \pi^{- 1} \int \mathd^2 \beta | \beta \bignone \rangle
      \langle \beta |$}.}}
 \begin{equation}
   \chi (t) = \prod_k \tmop{tr} \left( \mathsf{D}_k (\alpha_k (t))
   \rho_{\tmop{th}}^{(k)} \right) \bignone = \prod_k \exp \left( \um
   \frac{\left| \alpha_k (t) \right|^2}{2} \coth \left( \frac{\hbar \omega_k}{2
   k_B T} \right) \right) \bignone \label{KHchi} \;.
 \end{equation}
 This factor can be separated into its vacuum component (\ref{KHchivac}) and a
 thermal component, $\chi (t) = \mathe^{- F_{\tmop{vac}} (t)} \mathe^{-
 F_{\tmop{th}} (t)}$, with the following definitions of the vacuum and the
 thermal decay functions:
 \begin{eqnarray}
   F_{\tmop{vac}} (t) & \assign & \sum_k 4 \left| g_k \right| ^2 \frac{1 - \cos
   \left( \omega_k t \right)}{\hbar^2 \omega_k^2} \;, \\
   F_{\tmop{th}} (t) & \assign & \sum_k 4 \left| g_k \right| ^2 \frac{1 - \cos
   \left( \omega_k t \right)}{\hbar^2 \omega_k^2}  \left( \coth \left(
   \frac{\hbar \omega_k}{2 k_B T} \right) - 1 \right) \;.
 \end{eqnarray}

\subsection{The Continuum Limit}

Assuming that the field modes are sufficiently dense we replace their sum by
an integration. Noting (\ref{KHcontlim}), (\ref{KHeq:J}) we have
\begin{eqnarray}
  F_{\tmop{vac}} (t) & \longrightarrow & \int_0^{\infty} \mathd \omega J
  (\omega) \bignone  \frac{1 - \cos \left( \omega_{} t \right)}{\hbar^2
  \omega^2}  \label{KHGva}\\
  F_{\tmop{th}} (t) & \longrightarrow & \int_0^{\infty} \mathd \omega J
  (\omega) \bignone  \frac{\left. 1 - \cos (\omega t \right)}{\hbar^2
  \omega^2}  \left( \coth \left( \frac{\hbar \omega}{2 k_B T} \right) - 1
  \right) .  \label{KHGth}
\end{eqnarray}
So far, the treatment was exact. To continue we have to specify the spectral
density in the continuum limit. A typical model takes $g \propto
\sqrt{\omega}$, so that the spectral density of a $d$-dimensional field can be
written as {\cite{Weiss1999a}}
\begin{equation}
  J (\omega) = a \omega \left( \frac{\omega}{\omega_c} \right)^{d - 1}
  \mathe^{\um \omega / \omega_c} \label{KHeq:Jdim}
\end{equation}
with ``damping strength'' $a > 0$. Here, $\omega_c$ is a characteristic
frequency ``cutoff'' where the coupling decreases rapidly, such as the Debye
frequency in the case of phonons.

\subsubsection{Ohmic Coupling}\index{ohmic coupling}

For $d = 1$ the spectral density (\ref{KHeq:Jdim}) increases linearly at small
$\omega$ (``{\tmem{Ohmic coupling}}''). One finds
\begin{equation}
  F_{\tmop{vac}} (t) = \frac{a}{2 \hbar^2} \log (1 + \omega_c^2 t^2) \;,
  \label{KHGAvac}
\end{equation}
which bears a strong $\omega_c$ dependence. Evaluating the second integral
requires to assume that the cutoff $\omega_c$ is large compared to the thermal
energy, $kT \ll \hbar \omega_c$:
\begin{equation}
  F_{\tmop{th}} (t) \simeq \frac{a}{\hbar^2} \log \left( \frac{\sinh \left( t
  / t_{\text{T}} \right)}{t / t_{\text{T}}} \right) \;. \label{KHGath}
\end{equation}
Here $t_T = \hbar / \left( \pi k_B T \right)$ is a thermal quantum time scale.
The corresponding frequency $\omega_1 = 2 / t_T$ is called the (first)
{\tmem{Matsubara frequency}}\index{Matsubara frequency}, which also shows up
if imaginary time path integral techniques are used to treat the influence of
bosonic field couplings {\cite{Weiss1999a}}. For large times the decay
function $F_{\tmop{th}} (t)$ shows the asymptotic behavior
\begin{eqnarray}
  F_{\tmop{th}} (t) & \sim & \frac{a}{\hbar^2}  \frac{t}{t_T} \hspace{1em}
  \text{[as $t \rightarrow \infty$]} \;. 
\end{eqnarray}

It follows that the decay of coherence is characterized by rather different
regimes. In the short time regime ($t < \omega_c^{\um 1}$) we have the
perturbative behavior
\begin{eqnarray}
  F (t) & \simeq & \frac{a}{2 \hbar^2} \omega_c^2 t^2 \hspace{2em} \text{[for
  $t \ll \omega_c^{- 1}$]\;,} 
\end{eqnarray}
which can also be obtained from the short-time expansion of the time evolution
operator. Note that the decay is here determined by the overall width
$\omega_c$ of the spectral density. The intermediate region, {$\omega_c^{\um
   1} < t < \omega_1^{- 1}$}, is dominated by $F_{\tmop{vac}} (t)$ and called
the {\tmem{vacuum regime}}\index{vacuum regime},
\begin{eqnarray*}
  F (t) & \simeq & \frac{a}{\hbar^2} \log \left( \omega_c t \right)
  \hspace{2em} \text{[for $\omega_c^{- 1} \ll t \ll \omega_1^{- 1}$]} \;.
\end{eqnarray*}
Beyond that, for large times the decay is dominated by the thermal suppression
factor,
\begin{eqnarray}
  F (t) & \simeq & \frac{a}{\hbar^2}  \frac{t}{t_T} \hspace{2em} \text{[for
  $\omega_1^{- 1} \ll t$]} \;. \label{KHeq:thermalregime} 
\end{eqnarray}
In this {\tmem{thermal regime}}\index{thermal!-- regime} the decay shows the
exponential behavior typical for the Markovian master equations discussed
below. Note that the decay rate for this long time behavior is determined by
the {\tmem{low frequency behavior}} of the spectral density, characterized by
the damping strength $a$ in (\ref{KHeq:Jdim}), and is proportional to the
temperature $T$.

\subsubsection{Super-Ohmic Coupling}

For $d = 3$, the case of a ``super-Ohmic'' bath, the integrals (\ref{KHGva}),
(\ref{KHGth}) can be calculated without approximation. We note only the
long-time behavior of the decay,
\begin{equation}
  \lim_{t \rightarrow \infty} F (t) = 2 a \left( \frac{k_B T}{\hbar \omega_c}
  \right)^2 \psi' \left( 1 + \frac{k_B T}{\hbar \omega_c} \right) < \infty \;.
  \label{KHeq:Fasymp}
\end{equation}
Here $\psi \left( z \right)$ stands for the Digamma function, the logarithmic
derivative of the gamma function. Somewhat surprisingly, the coherences do not
get completely reduced as $t \rightarrow \infty$, even at a finite
temperature. This is due to the suppressed influence of the important
low-frequency contributions to the spectral density in three dimensions (as
compared to lower dimensions). While such a suppression of decoherence is
plausible for intermediate times, the limiting behavior (\ref{KHeq:Fasymp}) is
clearly a result of our simplified model assumptions. It will be absent if
there is a small anharmonic coupling between the bath modes
{\cite{Machnikowski2005a}} or if there is a small admixture of different
couplings to $ \mathsf{H}_{\tmop{int}}$.

\subsubsection*{Decoherence by ``Vacuum Fluctuations''?}
\index{decoherence!-- by vacuum fluctuations?}

The foregoing discussion seems to indicate that the ``vacuum fluctuations'' attributed to the quantized field modes are responsible for a general decoherence process, which occurs at short time scales even if the field is in its ground state. This ground state is non-degenerate and the only way to change it is to increase the energy of the field. But in our model the interaction Hamiltonian $H_{\rm int}$ commutes with the system Hamiltonian, so that it cannot describe energy exchange between qubit and field. One would therefore expect that after the interaction the field has the same energy as before, so that an initial vacuum state remains unchanged and decoherence cannot take place.

This puzzle is resolved by noting that the initial state  $\rho_{\text{E}} =
\bigotimes_k \bignone |0 \rangle \langle 0|_k$ is an eigenstate only
in the absence of the coupling $H_{\rm int}$, but not of the total
Hamiltonian. By starting with the product state $\rho_{\rm
  tot}(0)=\rho\otimes\rho_E$  we do not account for this possibly
strong coupling. At an infinitesimally small time later, system and
field thus suddenly feel that they are coupled to each other, which
leads to  a renormalization of their energies (as described by the Lamb shift discussed in Sect.~\ref{KHsec:weakcoup}). The factor $\chi_{\rm vac}$ in Eq.~(\ref{KHchivac}) describes the `initial jolt' produced by this sudden switching on of the coupling.

It follows that the above treatment of the short time dynamics, though formally correct, does  not give a physically reasonable picture if the system state is prepared in the presence of the coupling. In this case, one should rather work with the eigenstates of the total Hamiltonian, often denoted as ``dressed states''. If we start with a superposition of those two dressed states, which correspond in the limit of vanishing coupling to the two system states and the vacuum field, the resulting dynamics will show no further loss of coherence. This is consistent with the above notion that at zero temperature elastic processes cannot lead to decoherence  \cite{Imry2002a}.

\subsection{Dephasing of $N$ Qubits}

Let us now discuss the generalization to the case of $N$ qubits which do not
interact directly among each other. Each qubit may have a different coupling
to the bath modes. The system Hamiltonian is then 
\begin{equation}
  \mathsf{H}_{\tmop{tot}} = \bignone \underbrace{\sum_{j = 0}^{N - 1}
  \frac{\hbar \omega_j}{2} \sigma_z^{(j)} + \sum_k \hbar \omega_k 
  \mathsf{b}_k^{\dag} \mathsf{b}_k}_{\mathsf{H}_0} \bignone + \sum_{j = 0}^{N
  - 1} \bignone \sigma^{(j)}_z  \sum_k \left( g^{(j)}_k  \mathsf{b}_k^{\dag} +
  [g^{(j)}_k]^{\ast} \mathsf{b}_k  \right) \bignone \;.
\end{equation}
Similar to above, the time evolution in the interaction picture reads
\begin{equation*}
\widetilde{\mathsf{U}} (t) = \mathe^{\mathi \varphi \left( t \right)} \exp
   \left( \frac{1}{2}  \sum_{j = 0}^{N - 1} \sigma_z^{(j)}  \sum_k \bignone
   \bignone \left( \alpha^{(j)}_k  \mathsf{b}_k^{\dag} -
   [\alpha^{(j)}_k]^{\ast} \mathsf{b}_k  \right) \bignone \right)\;,
\end{equation*}
where the displacement of the field modes now depends on the $N$-qubit state.

As an example, we take $N = 2$ qubits and only a single vacuum mode. For the
inital qubit states
\begin{eqnarray}
  | \phi \rangle = c_{11} | \uparrow \uparrow \rangle + c_{00} | \downarrow
  \downarrow \rangle &  &  \label{KHn2phi}
\end{eqnarray}
and
\begin{eqnarray}
  | \psi \rangle = c_{10} | \uparrow \downarrow \rangle + c_{01} | \downarrow
  \uparrow \rangle &  & \;, \label{KHn2psi}
\end{eqnarray}
we obtain, respectively,
\begin{equation*}
\widetilde{\mathsf{U}} | \phi \rangle |0 \rangle_{\text{E}} = c_{11} |
   \uparrow \uparrow \rangle | \frac{\alpha^{(1)} \left( t \right) +
   \alpha^{(2)} \left( t \right)}{2} \rangle_{\text{E}} + c_{00} | \downarrow
   \downarrow \rangle | \frac{- \alpha^{(1)} \left( t \right) - \alpha^{(2)}
   \left( t \right)}{2} \rangle_{\text{E}} \;,
\end{equation*}
and 
\begin{equation*}
\widetilde{\mathsf{U}} | \psi \rangle |0 \rangle_{\text{E}} = c_{10} |
\uparrow \downarrow \rangle | \frac{\alpha^{(1)} \left( t \right) -
  \alpha^{(2)} \left( t \right)}{2} \rangle_{\text{E}} + c_{01} | \downarrow
\uparrow \rangle | \frac{- \alpha^{(1)} \left( t \right) + \alpha^{(2)}
  \left( t \right)}{2} \rangle_{\text{E}}\;, 
\end{equation*} 
where the $\alpha^{(1)} \left( t \right)$ and $\alpha^{(2)} \left( t \right)$
are the field displacements (\ref{KHalphat}) due to the first and the second
qubit.

If the couplings to the environment are equal for both qubits, say, because
they are all sitting in the same place and seeing the same field, we have
${\alpha^{(1)} \left( t \right) = \alpha^{(2)} \left( t \right) \equiv \alpha
 \left( t \right)} .$ In this case, states of the form $| \phi \rangle$ are
decohered once the factor $\langle \alpha \left( t \right) | \um \alpha \left(
 t \right) \rangle_{\text{E}} = \exp (- 2 \left| \alpha \left( t \right)
\right|^2)$ is approximately zero. States of the form $| \psi \rangle$, on the other hand,
are not affected at all, and one says that the $\left\{ | \psi \rangle
\right\}$ span a (two-dimensional) {\tmem{decoherence-free
   subspace}}\index{decoherence-free subspace}\index{decoherence-free subspace!-- in the dephasing of $N$
 qubits}. It shows up because the environment cannot tell the difference
between the states $| \uparrow \downarrow \rangle$ and $| \downarrow \uparrow
\rangle$ if it couples only to the sum of the excitations.

For an arbitrary number of qubits, using an $N$-digit binary notation, e.g. $| \uparrow \downarrow \uparrow
\rangle \equiv |101_2 \rangle = |5 \rangle$, one has
\begin{eqnarray}
  \langle m| \tilde{\rho} (t) |n \rangle & = & \langle m| \tilde{\rho} (0) |n
  \rangle \nonumber \\
  &  & \times \tmop{tr} \!\left(\! \exp \! \left[ \sum_{j = 0}^{N - 1} (m_j - n_j)
  \sum_k \left( \alpha_k^{(j)} (t) \mathsf{b}_k^{\dag} - [\alpha^{(j)}_k
  (t)]^{\ast}  \mathsf{b}_k \right) \right] \rho_{\tmop{th}} \right) \;,
  \nonumber \\
\end{eqnarray}
where $m_j \in \left\{ 0, 1 \right\}$ indicates the $j \text{-th}$ digit in
the binary representation of the number $m$.

We can distinguish different limiting cases:

\subsubsection{Qubits Feel the Same Reservoir} 

If the separation of the qubits is small compared to the wave lengths of the
field modes they are effectively interacting with the same reservoir,
$\alpha_k^{(j)} = \alpha_k$. One can push the $j$-summation to the $\alpha$'s
in this case, so that, compared to the single qubit, one merely has to replace
$\alpha_k$ by $\sum (m_j - n_j) \alpha_k \bignone $. We find
\begin{eqnarray}
  \chi_{mn} (t) = \exp \left( - \left| \sum_{j = 0}^{N - 1} (m_j - n_j)
  \bignone \right|^2  \left( F_{\tmop{vac}} (t) + F_{\tmop{th}} (t) \right)
  \right) 
\end{eqnarray}
with $F_{\tmop{vac}} (t)$ and $F_{\tmop{th}} (t)$ given by (\ref{KHGva}) and
(\ref{KHGth}).

Hence, in the worst case, one observes an increase of the decay rate by $N^2$
compared to the single qubit rate. This is the case for the coherence between
the states $|0 \rangle$ and $|2^N - 1 \rangle$, which have the maximum
difference in the number of excitations. On the other hand, the states with an
equal number of excitations form a {\tmem{decoherence-free
   subspace}}\index{decoherence-free subspace!-- in the dephasing of $N$
 qubits} in the present model, with a maximal dimension of $\binom{N}{N / 2}$.

\subsubsection{Qubits See Different Reservoirs}

In the other extreme, the qubits are so far apart from each other that each
field mode couples only to a single qubit. This suggests a re-numbering of the
field modes,
\begin{equation*}
\alpha_k^{(j)} \longrightarrow \alpha_{k_j} \;, 
\end{equation*}
and leads, after transforming the $j$-summation into a tensor-product, to
\begin{eqnarray}
  \chi_{mn} (t) & = & \prod_{j = 0}^{N - 1} \tmop{tr} \left( \bigotimes_{k_j}
  \mathsf{D}_{k_j} \left( \left( m_j - n_j \right) \alpha_{k_j} (t) \right)
  \rho_{\tmop{th}}^{(k_j)} \bignone \right) \bignone \nonumber\\
  & = & \prod_{j = 0}^{N - 1} \bignone \exp \Bigl( - \underset{= \left| m_j -
  n_j \bignone \right|}{\underbrace{\left| m_j - n_j \bignone \right|^2}} 
  \left( F_{\tmop{vac}} (t) + F_{\tmop{th}} (t) \right) \Bigr) \nonumber\\
  & = & \exp \Biggl( - \underbrace{\sum_{j = 0}^{N - 1} \bignone \left| m_j -
  n_j \bignone \right|}_{\text{Hamming distance}}  \left( F_{\tmop{vac}} (t) +
  F_{\tmop{th}} (t) \right) \Biggr) . 
\end{eqnarray}
Hence, the decay of coherence is the same for all pairs of states with the
same Hamming distance. In the worst case, we have an increase by a factor of
$N$ compared to the single qubit case, and there are no decoherence-free
subspaces.

An intermediate case is obtained if the coupling depends on the position
$\tmmathbf{r}_j$ of the qubits. A reasonable model, corresponding to point
scatterings of fields with wave vector $\tmmathbf{k}$, is given by
$g_k^{\left( j \right)} = g_k \exp \left( \mathi \tmmathbf{k} \cdot
\tmmathbf{r}_j \right)$, and its implications are studied in
{\cite{Doll2006a}}.

The model for decoherence discussed in this section is rather exceptional in
that the dynamics of the system can be calculated exactly for some
choices of the environmental spectral density. In general, one has to resort
to approximate descriptions for the dynamical effect of the environment; we
turn to this problem in the following section.

\section{Markovian Dynamics of Open Quantum Systems}\label{KHsec:Markov}

Isolated systems evolve, in the Schr\"odinger picture and for the general case
of mixed states, according to the von Neumann equation,
\begin{equation}
  \partial_t \rho = \frac{1}{\mathi \hbar} [ \mathsf{H}, \rho] \;.
\end{equation}
One would like to have a similar differential equation for the reduced
dynamics of an ``open'' quantum system, which is in contact with its
environment. If we extend the description to include the entire environment
$\mathcal{H}_{\text{E}}$ and its coupling to the system, then the total state
in $\mathcal{H}_{\tmop{tot}} =\mathcal{H} \otimes \mathcal{H}_{\text{E}}$
evolves unitarily. The partial trace over $\mathcal{H}_{\text{E}}$ gives the
evolved system state, and its time derivative reads
\begin{equation}
  \partial_t \rho = \frac{\mathd}{\mathd t} \tmop{tr}_{\text{E}}  \left(
  \mathsf{U}_{\tmop{tot}} (t) \rho_{\tmop{tot}} (0)
  \mathsf{U}_{\tmop{tot}}^{\dag} (t) \right) = \frac{1}{\mathi \hbar}
  \tmop{tr}_{\text{E}}  \left( [ \mathsf{H}_{\tmop{tot}}, \rho_{\tmop{tot}}]
  \right) \;.
\end{equation}
This exact equation is not closed and therefore not particularly helpful as it
stands. However, it can be used as the starting point to derive approximate
time evolution equations for $\rho$, in particular, if it is permissible to
take the initial system state to be uncorrelated with the environment.

These equations are often non-local in time, though, in agreement with
causality, the change of the state at each point in time depends only on the
state evolution in the past. In this case, the evolution equation is called a
{\tmem{generalized master equation}}\index{master equation!--
 generalized}\index{generalized master equation|see{master equation}}. It
can be specified in terms of superoperator-functionals\index{superoperator}, i.e. linear operators
which take the density operator $\rho$ with its past time-evolution until time
$t$ and map it to the differential change of the operator at that time,
\begin{equation}
  \partial_t \rho = \mathcal{K} \left[ \left\{ \rho_{\tau} : \tau < t \right\}
  \right] \;. \label{KHnonmarkform}
\end{equation}
An interpretation of this dependence on the system's past is that the
environment has a memory, since it affects the system in a way which
depends on the history of the system-environment interaction. One may hope
that on a coarse-grained time-scale, which is large compared to the
inter-environmental correlation times, these memory effects might become
irrelevant. In this case, a proper master equation might be appropriate, where
the infinitesimal change of $\rho$ depends only on the instantaneous system
state, through a {\tmem{Liouville}} super-operator $\mathcal{L}$,
\begin{equation}
  \partial_t \rho =\mathcal{L} \rho \;. \label{KHmastereqform}
\end{equation}
Master equations of this type are also called {\tmem{Markovian}}\index{master
 equation!-- Markovian|see{Markovian master equation}}, because of their resemblance to the differential
Chapman-Kolmogorov equation for a classical Markov process. However, since a
number of approximations are involved in their derivation, it is not clear
whether the corresponding time evolution respects important properties of a
quantum state, such as its positivity. We will see that these constraints
restrict the possible form of $\mathcal{L}$ rather strongly.

\subsection{Quantum Dynamical Semigroups}\label{KHsemigroups}\index{dynamical semigroup!-- quantum}

The notion a of {\tmem{quantum dynamical semigroup}} permits a rigorous
formulation of the Markov assumption in quantum theory. To introduce it we
first need a number of concepts from the theory of open quantum systems
{\cite{Holevo2001a,Davies1976a,Spohn1980a,Alicki1987a}}.

\subsubsection{Dynamical Maps}\index{map!-- dynamical}

A dynamical map is a one-parameter family of trace-preserving, convex linear,
and completely positive maps (CPM) 
\begin{eqnarray}
  \mathcal{W}_t : \rho_0 & \mapsto & \rho_t, \hspace{1em} \hspace{1em}
  \hspace{1em} \text{$\tmop{for} t \in \mathbbm{R}_0^+$} \;,
\end{eqnarray}
satisfying $\mathcal{W}_0 = \tmop{id}$. As such, it yields the most general
description of a time evolution which maps an arbitrary initial state $\rho_0$
to valid states at later times.

Specifically, the condition of trace preservation guarantees the normalization
of the state,
\begin{eqnarray}
  \tmop{tr} \left( \rho_t \right) & = & 1\;, \nonumber
\end{eqnarray}
and the convex linearity, i.e.
\begin{eqnarray*}
  \mathcal{W}_t \left( \lambda \rho_0 + \left( 1 - \lambda \right) \rho'_0
  \right) & = & \lambda \mathcal{W}_t \left( \rho_0 \right) + \left( 1 -
  \lambda \right) \mathcal{W}_t \left( \rho'_0 \right) \hspace{2em} \text{for
  all $0 \leqslant \lambda \leqslant 1$} \;,
\end{eqnarray*}
ensures that the transformation of mixed states is consistent with the
classical notion of ignorance. The final requirement of {\tmem{complete
   positivity}}\index{completely positive!-- map}\index{map!-- completely positive} is stronger than
mere positivity of $\mathcal{W}_t \left( \rho_0 \right)$. It means that in
addition all the tensor product extensions of $\mathcal{W}_t$ to spaces of
higher dimension, defined with the identity map $\tmop{id}_{\tmop{ext}}$, are
positive,
\begin{eqnarray*}
  \mathcal{W}_t \otimes \tmop{id}_{\tmop{ext}} & > & 0 \;,
\end{eqnarray*}
that is, the image of any positive operator in the higher dimensional space is
again a positive operator. This guarantees that the system state remains
positive even if it is the reduced part of a non-separable state evolving in a
higher dimensional space.

\subsubsection{Kraus Representation}\index{Kraus!-- representation}

Any dynamical map admits an {\tmem{operator-sum
   representation}}\index{map!-- dynamical@-- -- operator-sum representation}
of the form (\ref{KHops}) {\cite{Alicki1987a}},
\begin{equation}
  \mathcal{W}_t (\rho) = \sum_{k = 1}^N \mathsf{W}_k (t) \rho
  \mathsf{W}_k^{\dag} (t) \bignone  \label{KHsupop}
\end{equation}
with the completeness relation{\footnote{\label{KHtracedecreasing}In case of a
   {\tmem{trace-decreasing}}, convex linear, completely positive map the
   condition (\ref{KHeq:completeness}) is replaced by $\sum_k
   \mathsf{W}^{\dag}_k (t) \mathsf{W}_k (t) \bignone <\mathbbm{I}$, i.e. the
   operator $\mathbbm{I}- \sum_k \mathsf{W}^{\dag}_k (t) \mathsf{W}_k (t)
   \bignone$ must be positive.}}
\begin{eqnarray}
  \sum_{k = 1}^N \mathsf{W}^{\dag}_k (t) \mathsf{W}_k (t) \bignone
  =\mathbbm{I} \;. &  &  \label{KHeq:completeness}
\end{eqnarray}
The number of the required Kraus operators $\mathsf{W}_k (t)$ is limited by
the dimension of the system Hilbert space, $N \leqslant \dim \left(
 \mathcal{H} \right)^2$ (and confined to a countable set in case of an
infinite-dimensional, separable Hilbert space), but their choice is not
unique.

\subsubsection{Semigroup Assumption}

We can now formulate the assumption that the $\{ \mathcal{W}_t : t \in
\mathbbm{R}_0^+ \}$ form a continuous {\tmem{dynamical
   semigroup}}{\footnote{The inverse element required for a {\tmem{group}}
   structure is missing for general, irreversible CPMs.}}
{\cite{Davies1976a,Alicki1987a}}:
\begin{eqnarray}
  \mathcal{W}_{t_2} \left( \mathcal{W}_{t_1} (\cdot) \right) & \overset{!}{=}
  & \mathcal{W}_{t_1 + t_2}(\cdot)  \hspace{1em} \hspace{1em} \hspace{1em} \text{for
  all $t_1, t_2 > 0$} \;
\end{eqnarray}
and  $\mathcal{W}_{0}=\textrm{id}$.
This statement is rather strong, and it is certainly violated for truly
microscopic times. But it seems not unreasonable on the level of a
coarse-grained time scale, which is long compared to the time it takes for the
environment to ``forget'' the past interactions with the system due to the
dispersion of correlations into the many environmental degrees of freedom.

For a given dynamical semigroup there exists, under rather weak conditions, a
generator\index{dynamical semigroup!-- generator}, i.e. a superoperator
$\mathcal{L}$ satisfying
\begin{equation}
  \mathcal{W}_t = \mathe^{\mathcal{L}t}  \hspace{1em} \hspace{1em}
  \hspace{1em} \text{for $t > 0$} \;. \label{KHeq:formalsolution}
\end{equation}
In this case $\mathcal{W}_t (\rho)$ is the formal solution of the Markovian
master equation (\ref{KHmastereqform}).

\subsubsection{Dual Maps}\index{map!-- dual}

So far we used the Schr\"odinger picture, i.e. the notion that the state of
an open quantum system evolves in time, ${\rho_t = \mathcal{W}_t (\rho_0)}$.
Like in the description of closed quantum systems, one can also take the
Heisenberg point of view, where the state does not evolve, while the operators
$\mathsf{A}$ describing observables acquire a time dependence. The
corresponding map $\mathcal{W}_t^{\sharp} : \mathsf{A_0} \mapsto \mathsf{A}_t$
is called the {\tmem{dual map}}, and it is related to $\mathcal{W}_t$ by the
requirement $\tmop{tr} ( \mathsf{A} \mathcal{W}_t (\rho)) = \tmop{tr} (\rho
\mathcal{W}_t^{\sharp} ( \mathsf{A}))$. In case of a dynamical semigroup,
$\mathcal{W}_t^{\sharp} = \exp \left( \mathcal{L}^{\sharp} t
\right)$\index{dynamical semigroup!-- dual generator}, the
equation of motion takes the form $\partial_t \mathsf{A} =
\mathcal{L}^{\sharp} \mathsf{A}$, with the dual Liouville operator determined
by $\tmop{tr} ( \mathsf{A} \mathcal{L} \left( \rho \right)) = \tmop{tr} (\rho
\mathcal{L}^{\sharp} \left( \mathsf{A} \right))$. From a mathematical point of
view, the Heisenberg picture is much more convenient since the observables
form an algebra, and it is therefore preferred in the mathematical literature.

\subsection{The Lindblad Form}\index{master equation!-- Lindblad
 form}\label{KHsec:lindblad} 

We can now derive the general form of the generator of a dynamical semigroup,
taking $\dim \left( \mathcal{H} \right) = d < \infty$ for simplicity
{\cite{Breuer2002a,Alicki1987a}}. The bounded operators on $\mathcal{H}$ then
form a $d^2$-dimensional vector space which turns into a Hilbert space, if
equipped with the Hilbert-Schmidt scalar product $( \mathsf{A}_{},
\mathsf{B}_{}) \assign \tmop{tr} ( \mathsf{A} ^{\dag} \mathsf{B}_{})$.

Given an orthonormal basis of operators $\{ \mathsf{E}_j : 1 \leqslant j
\leqslant d^2 \} \subset L \left( \mathcal{H} \right)$,
\begin{eqnarray}
  ( \mathsf{E}_i, \mathsf{E}_j) \assign \tmop{tr} ( \mathsf{E}_i ^{\dag}
  \mathsf{E}_j) = \delta_{i j}\;, &&
\end{eqnarray}
any Hilbert-Schmidt operator $\mathsf{W}_k$ can be expanded as
\begin{eqnarray}
  \mathsf{W}_k = \sum_{j = 1}^{d^2} ( \mathsf{E}_j, \mathsf{W}_k) \mathsf{E}_j
  \;. &  & 
\end{eqnarray}
We can choose one of the basis operators, say the $d^2$-th, to be proportional
to the identity operator,
\begin{eqnarray}
  \mathsf{E}_{d^2} = \frac{1}{\sqrt{d}} \mathbbm{I} \;, &  & 
\end{eqnarray}
so that all other basis elements are traceless,
\begin{equation}
  \tmop{tr} ( \mathsf{E}_j) = \begin{cases}
    0 & \text{for $j = 1, \ldots, d^2 - 1$} \;,\\
    \sqrt{d} & \text{for $j = d^2$}\;.
  \end{cases}
\end{equation}
Representing the superoperator of the dynamical map (\ref{KHsupop}) in the $\{
\mathsf{E}_j \}$ basis we have
\begin{equation}
  \mathcal{W}_t (\rho) = \sum_{i, j = 1}^{d^2} c_{i j} (t) \mathsf{E}_i \rho
  \mathsf{E}_j^{\dag}
\end{equation}
with a time dependent, hermitian and positive coefficient matrix,
\begin{equation}
  c_{i j} (t) = \sum_{k = 1}^N ( \mathsf{E}_i, \mathsf{W}_k (t)) (
  \mathsf{E}_j, \mathsf{W}_k (t))^*
\end{equation}
(positivity can be checked in a small calculation). We can now calculate the
semigroup generator\index{dynamical semigroup!-- generator} in terms of the differential quotient by writing the terms
including the element $\mathsf{E}_{d^2}$ separately:
\begin{eqnarray}
  \mathcal{L} \rho & = & \lim_{\tau \rightarrow 0}  \frac{\mathcal{W}_{\tau}
  (\rho) - \rho}{\tau} \nonumber\\
  & = & \underbrace{\lim_{\tau \rightarrow 0}  \frac{\frac{1}{d} c_{d^2 d^2}
  (\tau) - 1}{\tau}}_{c_0 \in \mathbbm{R}} \rho + \underbrace{\lim_{\tau
  \rightarrow 0}  \sum_{j = 1}^{d^2 - 1} \frac{c_{j d^2} (\tau)}{\sqrt{d}
  \tau}  \mathsf{E}_j}_{\mathsf{B} \in L (\mathcal{H})} \rho + \rho
  \underbrace{\lim_{\tau \rightarrow 0}  \sum_{j = 1}^{d^2 - 1} \frac{c_{d^2
  j} (\tau)}{\sqrt{d} \tau}  \mathsf{E}_j^{\dag}}_{\mathsf{B}^{\dag} \in L
  (\mathcal{H})} \nonumber\\
  &  & + \sum_{i, j = 1}^{d^2 - 1} \underbrace{\lim_{\tau \rightarrow 0} 
  \frac{c_{i j} (\tau)}{\tau} }_{\alpha_{i j} \in \mathbbm{R}} \mathsf{E}_i
  \rho \mathsf{E}_j^{\dag} \nonumber\\
  & = & c_0 \rho + \mathsf{B} \rho + \rho \mathsf{B}^{\dag} + \sum_{i, j =
  1}^{d^2 - 1} \alpha_{i j}  \mathsf{E}_i \rho \mathsf{E}_j^{\dag} 
  \nonumber\\
  & = & \frac{1}{\mathi \hbar} [ \mathsf{H}, \rho] + \frac{1}{\hbar} (
  \mathsf{G} \rho + \rho \mathsf{G}) + \sum_{i, j = 1}^{d^2 - 1} \alpha_{i j} 
  \mathsf{E}_i \rho \mathsf{E}_j^{\dag} \;.
\end{eqnarray}
In the last equality the following hermitian operators with the dimension of
an energy were introduced:
\begin{eqnarray*}
  \mathsf{G} & = & \frac{\hbar}{2} ( \mathsf{B} + \mathsf{B}^{\dag} + c_0) \;,\\
  \mathsf{H} & = & \frac{\hbar}{2 \mathi} ( \mathsf{B} - \mathsf{B}^{\dag}) \;.
\end{eqnarray*}
By observing that the conservation of the trace implies $\tmop{tr}
(\mathcal{L} \rho) = 0$, one can relate the operator $\mathsf{G}$ to the matrix
$\tmmathbf{\alpha}= \left( \alpha_{i j} \right)$, since
\[ 0 = \tmop{tr} (\mathcal{L} \rho) = 0 + \tmop{tr} \left[ \left( \frac{2
      \mathsf{G}}{\hbar} + \sum_{i, j = 1}^{d^2 - 1} \alpha_{i j}
    \mathsf{E}_j^{\dag} \mathsf{E}_i \right) \rho \right] \] 
must hold for all $\rho$. It follows that
\[ \mathsf{G} = \um \frac{\hbar}{2}  \sum_{i, j = 1}^{d^2 - 1} \alpha_{i j} 
\mathsf{E}_j^{\dag} \mathsf{E}_i \;. \] 
This leads to the {\tmem{first standard form}}\index{dynamical semigroup!-- generator@-- -- first standard form} for the generator of a dynamical semigroup: 
\begin{equation}
  \mathcal{L} \rho = \underbrace{\frac{1}{\mathi \hbar} [ \mathsf{H},
  \rho]}_{\text{unitary part}} + \underset{\tmop{incoherent}
  \tmop{part}}{\underbrace{\sum_{i, j = 1}^{d^2 - 1} \alpha_{i j}  \left(
  \mathsf{E}_i \rho \mathsf{E}_j^{\dag} - \frac{1}{2}  \mathsf{E}_j^{\dag} 
  \mathsf{E}_i \rho - \frac{1}{2} \rho \mathsf{E}_j^{\dag}  \mathsf{E}_i
  \right)}} \;. \label{KHfirstLindblad}
\end{equation}
The complex coefficients $\alpha_{i j}$ have dimensions of frequency and
constitute a positive matrix $\tmmathbf{\alpha}$.

The \emph{second standard form}\index{dynamical semigroup!-- generator@-- --
 second standard form}\index{master equation!-- second standard form}
or {\tmem{Lindblad form}}\index{master equation!--  Lindblad form} is obtained
by diagonalizing the coefficient matrix $\tmmathbf{\alpha}$. The corresponding
unitary matrix $\tmmathbf{U}$ satisfying
$\tmmathbf{U}\tmmathbf{\alpha}\tmmathbf{U}^{\dag} = {\tmop{diag} (\gamma_1,
 \ldots, \gamma_{d^2 - 1})}$ allows to define the dimensionless operators
$\mathsf{L}_k \assign \sum_{j = 1}^{d^2 - 1} \mathsf{E}_j U_{j k}^{\dag}$ so
that ${\mathsf{E}_j = \sum_{k = 1}^{d^2 - 1} \mathsf{L}_k U_{k j}}$ and
therefore{\footnote{It is easy to see that the dual Liouville operator
   discussed in Sect.~\ref{KHsemigroups} reads, in Lindblad form,
\begin{eqnarray} \label{KHsecondLindblad}
  \mathcal{L}^{\sharp} \left( \mathsf{A} \right) & = & \frac{1}{\mathi \hbar}
  [ \mathsf{A}, \mathsf{H}] + \sum_{k = 1}^N \gamma_k  \left(
  \mathsf{L}^{\dag}_k  \mathsf{A}  \mathsf{L}_k - \frac{1}{2} 
  \mathsf{L}_k^{\dag}  \mathsf{L}_k  \mathsf{A} - \frac{1}{2} \mathsf{A} 
  \mathsf{L}^{\dag}_k  \mathsf{L}_k  \right) \;. \nonumber
\end{eqnarray}
Note that this implies $\mathcal{L^{\sharp}} \left( \mathbbm{I} \right) = 0$,
while $\mathcal{L} \left( \mathbbm{I} \right) = \sum_k \gamma_k [
\mathsf{L}_k, \mathsf{L}^{\dag}_k]$, and $\tmop{tr} \left( \mathcal{L}
 \mathsf{X} \right) = 0$.}}
\begin{equation}
  \mathcal{L} \rho = \frac{1}{\mathi \hbar} [ \mathsf{H}, \rho] + \sum_{k =
  1}^N \gamma_k  \left( \mathsf{L}_k \rho \mathsf{L}_k^{\dag} - \frac{1}{2} 
  \mathsf{L}_k^{\dag}  \mathsf{L}_k \rho - \frac{1}{2} \rho
  \mathsf{L}^{\dag}_k  \mathsf{L}_k  \right) \label{KHLindblad}
\end{equation}
with $N \leqslant d^2 - 1.$ This shows that the general form of a generator of
a dynamical semigroup is specified by a single hermitian operator
$\mathsf{H}$, which is not necessarily equal to the Hamiltonian of the
isolated system, see below, and at most $d^2 - 1$ arbitrary operators
$\mathsf{L}_k$ with attributed positive rates $\gamma_k$. These are called
{\tmem{Lindblad operators}}\index{Lindblad operators}{\footnote{Lindblad
   showed in 1976 that the form (\ref{KHLindblad}) is obtained even for
   infinite-dimensional systems provided the generator $\mathcal{L}$ is
   bounded (which is usually not the case).}} or {\tmem{jump
   operators}}\index{jump operators}, a name motivated in the following
section.

It is important to note that a given generator $\mathcal{L}$ does not
determine the jump operators uniquely. In fact, the equation is invariant
under the transformation
\begin{eqnarray}
  \mathsf{L}_k & \rightarrow & \mathsf{L}_k + c_k \;, \label{KHeq:Lshift}\\
  \mathsf{H} & \rightarrow & \mathsf{H} + \frac{\hbar}{2 \mathi}  \sum_j
  \gamma_j \bignone \left( c_j^{\ast}  \mathsf{L}_j - c_j  \mathsf{L}^{\dag}_j
  \right)  \;, 
\end{eqnarray}
with $c_k\in\mathbb{C}$, so that the $\mathsf{L}_k$ can be chosen traceless. In this case, the only
remaining freedom is a unitary mixing,
\begin{equation}
  \sqrt{\gamma_i}  \mathsf{L}_i \longrightarrow \sum_j U_{i j}' 
  \sqrt{\gamma_j}  \mathsf{L}_j \;. \bignone \bignone \label{KHeq:Lmix}
\end{equation}
If $\mathcal{L}$ shows an additional invariance, e.g. with respect to
rotations or translations, the form of the Lindblad operators will be further
restricted, see e.g.~{\cite{Petruccione2005a}}.

\subsection{Quantum Trajectories}
\label{KHsec:unravelling}

Generally, if we write the Liouville superoperator\index{Liouville!--
 superoperator}\index{superoperator!-- Liouville} $\mathcal{L}$ as the sum of two parts, $\mathcal{L}_0$ and
$\mathcal{S}$, then the formal solution (\ref{KHeq:formalsolution}) of the
master equation (\ref{KHmastereqform}) can be expressed as
\begin{eqnarray}
  \mathcal{W}_t & = & \mathe^{\left( \mathcal{L}_0 + \mathcal{S} \right) t}
  \hspace{0.6em} = \hspace{0.6em} \sum_{n = 0}^{\infty} \frac{t^n}{n!} 
  \bignone \left( \mathcal{L}_0 + \mathcal{S} \right)^n \nonumber\\
  & = & \sum_{n = 0}^{\infty} \sum_{k_0, \ldots, k_n = 0}^{\infty} \bignone
  \frac{t^{n + \sum_j k_j \bignone}}{\left( n + \sum_j k_j \right) !} 
  \underset{n \text{ times}}{\underbrace{\mathcal{L}_0^{k_n} \mathcal{S} 
  \mathcal{L}_0^{k_{n - 1}} \mathcal{S} \cdots \mathcal{S} 
  \mathcal{L}_0^{k_1} \mathcal{S}  \mathcal{L}_0^{k_0}}} \nonumber\\
  & = & \sum_{n = 0}^{\infty} \int_0^t \mathd t_n \bignone \int_0^{t_n}
  \mathd t_{n - 1} \bignone \cdots \bignone \int^{t_2}_0 \mathd t_1
  \nonumber\\
  &  & \times \sum_{k_0, \ldots, k_n = 0}^{\infty} \frac{\left( t - t_n
  \right)^{k_n}}{k_n !}  \frac{\left( t_n - t_{n - 1} \right)^{k_{n -
  1}}}{k_{n - 1} !} \cdots \frac{\left( t_1 - 0 \right)^{k_0}}{k_0 !}
  \nonumber\\
  &  & \times \mathcal{L}_0^{k_n} \mathcal{S}  \mathcal{L}_0^{k_{n - 1}}
  \mathcal{S} \cdots \mathcal{S}  \mathcal{L}_0^{k_1} \mathcal{S} 
  \mathcal{L}_0^{k_0} \nonumber\\
  & = & \mathe^{\mathcal{L}_0 t} + \sum_{n = 1}^{\infty} \int_0^t \mathd t_n
  \bignone \int_0^{t_n} \mathd t_{n - 1} \bignone \cdots \bignone \int^{t_2}_0
  \mathd t_1 \nonumber\\
  &  & \times \mathe^{\mathcal{L}_0 \left( t - t_n \right)} \mathcal{S}
  \mathe^{\mathcal{L}_0 \left( t_n - t_{n - 1} \right)} \mathcal{S} \cdots
  \mathe^{\mathcal{L}_0 \left( t_2 - t_1 \right)} \mathcal{S}
  \mathe^{\mathcal{L}_0 t_1} \;.  \label{KHeq:gendyson}
\end{eqnarray}
The step from the second to the third line, where $\bignone t^{n + \sum_j k_j
 \bignone} / (n + \sum_j k_j) !$ is replaced by $n$ time integrals, can be
checked by induction.

The form (\ref{KHeq:gendyson}) is a generalized Dyson expansion, and the
comparison with the Dyson series\index{Dyson series} for unitary evolutions
suggests to view $\exp \left( \mathcal{L}_0 \tau \right)$ as an
``unperturbed'' evolution and $\mathcal{S}$ as a ``perturbation'', such that
the exact time evolution $\mathcal{W}_t$ is obtained by an integration over
all iterations of the perturbation, separated by the unperturbed evolutions.

The particular Lindblad form (\ref{KHLindblad}) of the generator suggests to
introduce the completely positive {\tmem{jump superoperators}} 
\index{completely positive!-- jump superoperators}\index{superoperator!-- jump}
\begin{eqnarray}
  \mathcal{L}_k \rho & = & \gamma_k  \mathsf{L}_k \rho \mathsf{L}_k^{\dag} \;, 
\end{eqnarray}
along with the {\tmem{non-hermitian}} operator\index{operator!-- non-hermitian}
\begin{eqnarray}
  \mathsf{H}_{\mathbbm{C}} & = & \mathsf{H} - \frac{\mathi \hbar}{2} \sum_{k =
  1}^N \gamma_k  \mathsf{L}_k^{\dag} \mathsf{L}_k \;.  \label{KHeq:Hdamp}
\end{eqnarray}
The latter has a negative imaginary part, $\tmop{Im} \left(
 \mathsf{H}_{\mathbbm{C}} \right) < 0$, and can be used to construct
\begin{eqnarray}
  \mathcal{L}_0 \rho & = & \frac{1}{\mathi \hbar}  \left(
  \mathsf{H}_{\mathbbm{C}} \rho - \rho \mathsf{H}_{\mathbbm{C}}^{\dag}
  \right) \;. 
\end{eqnarray}
It follows that the sum of these superoperators yields the Liouville operator
\index{Liouville!-- operator}
(\ref{KHLindblad})
\begin{eqnarray}
  \mathcal{L} & = & \hspace{0.6em} \mathcal{L}_0 + \sum_{k = 1}^N
  \mathcal{L}_k \;. 
\end{eqnarray}
Of course, neither $\mathcal{L}_0$ nor $\mathcal{S} = \sum_{k = 1}^N
\mathcal{L}_k$ generate a dynamical semigroup. Nonetheless, they are useful
since the interpretation of (\ref{KHeq:gendyson}) can now be taken one
step further. We can take the point of view that the $\mathcal{L}_k$ with $k
\geqslant 1$ describe elementary transformation events due to the environment
(``jumps''), which occur at random times with a rate $\gamma_k$. A particular
realization of $n$ such events is specified by a sequence of the form 
\begin{eqnarray}
  R^t_n & = & \left( t_1, k_1 ; t_2, k_2 ; \ldots . ; t_n, k_n) \;. 
  \label{KHeq:records} \right.
\end{eqnarray}
The attributed times satisfy $0 < t_1 < \ldots . < t_n < t$, and the $k_j \in
\left\{ 1, \ldots, N \right\}$ indicate which kind of event ``took place''. We
call $R^t_n$ a {\tmem{record}} of length $n$.

The general time evolution $\mathcal{W}_t$ can thus be written as an
integration over all possible realizations of the jumps, with the ``free''
evolution $\exp \left( \mathcal{L}_0 \tau \right)$ in between,
\begin{eqnarray}
  \mathcal{W}_t & = & \mathe^{\mathcal{L}_0 t_{}} + \sum_{n = 1}^{\infty}
  \int_0^t \mathd t_n \bignone \int_0^{t_n} \mathd t_{n - 1} \bignone \cdots
  \bignone \int^{t_2}_0 \mathd t_1 \bignone   \label{KHeq:jumpexp} \nonumber\\
  &  & \phantom{\mathe^{\mathcal{L}_0 t_{}} +} \times \sum_{\left\{ R_n
  \right\}} \underset{\mathcal{K}_{R_n^t}}{\underbrace{\mathe^{\mathcal{L}_0
  \left( t - t_n \right)} \mathcal{L}_{k_n} \mathe^{\mathcal{L}_0 \left( t_n -
  t_{n - 1} \right)} \mathcal{L}_{k_{n - 1}} \cdots \mathe^{\mathcal{L}_0
  \left( t_2 - t_1 \right)} \mathcal{L}_{k_1} \mathe^{\mathcal{L}_0 t_1}}}
  \nonumber \\
\end{eqnarray}
As a result of the negative imaginary part in (\ref{KHeq:Hdamp}) the $\exp
\left( \mathcal{L}_0 \tau \right)$ are {\tmem{trace decreasing}}\footnote{See
the note \ref{KHtracedecreasing}.} completely positive maps,
\begin{eqnarray}
  \mathe^{\mathcal{L}_0 \tau_{}} \rho & = & \exp \left( - \frac{i \tau}{\hbar}
  \mathsf{H}_{\mathbbm{C}}  \right) \rho \exp \left( \frac{i \tau}{\hbar}
  \mathsf{H}^{\dag}_{\mathbbm{C}}  \right) > 0 \,,\\
  \frac{\mathd}{\mathd \tau} \tmop{tr} \left( \mathe^{\mathcal{L}_0 \tau_{}}
  \rho \right) & = & \tmop{tr} \left( \mathcal{L}_0 \mathe^{\mathcal{L}_0
  \tau_{}} \rho \right) \hspace{0.6em} = \hspace{0.6em} - \sum_{k = 1}^N
  \bignone \tmop{tr} ( \underset{> 0}{\underbrace{\mathcal{L}_k
  \mathe^{\mathcal{L}_0 \tau_{}}}} \rho) < 0 \;.
\end{eqnarray}
It is now natural to interpret $\tmop{tr} \left( \mathe^{\mathcal{L}_0 t_{}}
 \rho \right)$ as the probability that no jump occurs during the
time interval $t$,
\begin{eqnarray}
  \tmop{Prob} \left( R_0^t | \rho \right) & \assign & \tmop{tr} \left(
  \mathe^{\mathcal{L}_0 t_{}} \rho \right) \;.  \label{KHeq:probnull}
\end{eqnarray}
To see that this makes sense, we attribute to each record of length $n$ a
$n$-time probability density. For a given record $R_n^t$ we define
\begin{eqnarray}
  \tmop{prob} \left( R_n^t | \rho \right) & \assign & \tmop{tr} \left(
  \mathcal{K}_{R^t_n} \rho \right) \;,  \label{KHeq:probrecord}
\end{eqnarray}
in terms of the superoperators from the second line in (\ref{KHeq:jumpexp}),
\begin{eqnarray}
  \mathcal{K}_{R_n^t} & \assign & \mathe^{\mathcal{L}_0 \left( t - t_n
  \right)} \mathcal{L}_{k_n} \mathe^{\mathcal{L}_0 \left( t_n - t_{n - 1}
  \right)} \mathcal{L}_{k_{n - 1}} \cdots \mathe^{\mathcal{L}_0 \left( t_2 -
  t_1 \right)} \mathcal{L}_{k_1} \mathe^{\mathcal{L}_0 t_1} \;. 
\end{eqnarray}
This is reasonable since the $\mathcal{K}_{R_n^t}$ are completely positive
maps that do not preserve the trace. Indeed, the probability density for a
record is thus determined both by the corresponding jump operators, which
involve the rates $\gamma_k$, and by the $\mathe^{\mathcal{L}_0 \tau_{}}
\rho$, which account for the fact that the likelihood for the absence of a
jump decreases with the length of the time interval.

This notion of probabilities is consistent, as can be seen by adding the
probability (\ref{KHeq:probnull}) for no jump to occur during the interval
$\left( 0 ; t \right)$ to the integral over the probability densities
(\ref{KHeq:probrecord}) of all possible jump sequences. As required, the
result is unity,
\begin{eqnarray}
  \tmop{Prob} \left( R_0^t | \rho \right) + \sum_{n = 1}^{\infty} \int_0^t
  \mathd t_n \bignone \int_0^{t_n} \mathd t_{n - 1} \bignone \cdots \bignone
  \int^{t_2}_0 \mathd t_1 \sum_{\left\{ R_n^t \right\}} \tmop{prob} \left(
  R^t_n | \rho \right) \bignone & = & 1 \nonumber
\end{eqnarray}
for all $\rho$ and $t \geqslant 0$. This follows immediately from the trace
preservation of the map (\ref{KHeq:jumpexp}).

It is now natural to normalize the transformation defined by the
$\mathcal{K}_{R^t_n}$. Formally, this yields the state transformation
conditioned to a certain record $R_n^t$. It is called a
{\tmem{quantum trajectory}}\index{quantum!-- trajectory},
\begin{eqnarray}
  \mathcal{T} \left( \rho |R_n^t \right) & \assign & \frac{\mathcal{K}_{R^t_n}
  \rho}{\tmop{tr} \left( \mathcal{K}_{R^t_n} \rho \right)} \;. 
  \label{KHeq:condtrafo}
\end{eqnarray}
Note that this definition comprises the trajectory corresponding to a
null-record $R_0^t$, where $\mathcal{K}_{R_0^t} = \exp \left( \mathcal{L}_0 t
\right)$. These completely positive, trace-pre\-serving, {\tmem{non}}linear
maps $\rho \mapsto \mathcal{T} \left( \rho |R_n^t \right)$ are defined for all
states $\rho$ that yield a finite probability (density) for the given record
$R_n^t$, i.e. if the denominator in (\ref{KHeq:condtrafo}) does not vanish.

Using these notions, the exact solution of a general Lindblad master equation
(\ref{KHeq:gendyson}) may thus be rewritten in the form
\begin{eqnarray}
  \rho_t & = & \tmop{Prob} \left( R^t_0 \right) \bignone \mathcal{T} \left(
  \rho |R_0^t \right) \nonumber\\
  &  & + \sum_{n = 1}^{\infty}\! \int_0^t \! \mathd t_n \! \int_0^{t_n}
  \! \mathd t_{n - 1} \cdots\! \int^{t_2}_0 \! \mathd t_1
  \!\sum_{\left\{ R_n \right\}} \tmop{prob} \left( R^t_n | \rho \right)
  \mathcal{T} \!\left( \rho |R_n^t \right) \;.
\nonumber\\
&&  \label{KHeq:unravel}
\end{eqnarray}
It shows that the general Markovian quantum dynamics can be understood as a
summation over all quantum trajectories $\mathcal{T} \left( \rho |R_n^t
\right)$ weighted by their probability (density). This is called a
{\tmem{stochastic unravelling}} of the master equation.\index{master equation!-- unravelling} The set of trajectories and their weights are
labeled by the possible records (\ref{KHeq:records}), and determined by the
Lindblad operators of the master equation (\ref{KHLindblad}).

The semigroup property described by the master equation shows up if a record
is formed by joining the records of adjoining time intervals, $\left( 0 ; t'
\right)$ and $\left( t' ; t \right)$,
\begin{eqnarray}
  R_{n + m}^{\left( 0 ; t \right)} & \assign & (R_n^{\left( 0 ; t' \right)} ;
  R_m^{\left( t' ; t \right)}) \;. 
\end{eqnarray}
As one expects, the probabilities and trajectories satisfy
\begin{eqnarray}
  \tmop{prob} (R_{n + m}^{\left( 0 ; t \right)} | \rho) & = & \tmop{prob}
  (R_n^{\left( 0 ; t' \right)} | \rho) \tmop{prob} (R_m^{\left( t' ; t
  \right)} | \mathcal{T} (\rho |R_n^{\left( 0 ; t' \right)})) \;, 
\end{eqnarray}
and
\begin{eqnarray}
  \mathcal{T} (\rho |R_{n + m}^{\left( 0 ; t \right)}) & = & \mathcal{T} (
  \mathcal{T} (\rho |R_n^{\left( 0 ; t' \right)}) |R_m^{\left( t' ; t
  \right)}) \;. 
\end{eqnarray}

Note finally, that the concept of quantum trajectories fits seamlessly into
the framework of generalized measurements discussed in
Sect.~\ref{KHsec:monitoring}. In particular, the conditioned state
transformation $\mathcal{T} \left( \rho |R_n^t \right)$ has the form
(\ref{KHeq:efficientm}) of an efficient measurement transformation,
\begin{eqnarray}
  \mathcal{T} \left( \rho |R_n^t \right) & = & \frac{\mathsf{M}_{R_n^t} \rho
  \mathsf{M}_{R_n^t}^{\dag}}{\tmop{tr} \left( \mathsf{M}_{R_n^t}^{\dag}
  \mathsf{M}_{R_n^t} \rho \right)} 
\end{eqnarray}
with compound measurement operators
\begin{eqnarray}
  \mathsf{M}_{R_n^t} & \assign & \mathe^{- \mathi \mathsf{H}_{\mathbbm{C}}
  \left( t - t_n \right) / \hbar} \mathsf{L}_{k_n} \cdots \mathsf{L}_{k_2}
  \mathe^{- \mathi \mathsf{H}_{\mathbbm{C}} \left( t_2 - t_1 \right) / \hbar}
  \mathsf{L}_{k_1} \mathe^{- \mathi \mathsf{H}_{\mathbbm{C}} t_1 / \hbar} \;. 
\end{eqnarray}
This shows that we can legitimately view the open quantum dynamics generated
by $\mathcal{L}$ as due to the {\tmem{continuous monitoring}}\index{continuous
 monitoring!-- master equation|see{Markovian master equation}}\index{Markovian master
 equation!-- continuous monitoring} of the system by the environment. We just
have to identify the (aptly named) record $R_n^t$ with the total outcome of a
hypothetical, continuous measurement during the interval $\left( 0 ; t
\right)$. The jump operators $\mathcal{L}_k$ then describe the effects of the
corresponding elementary measurement events{\footnote{For all these appealing
   notions, it should be kept in mind that the $\mathcal{L}_k$ are not
   uniquely specified by a given generator $\mathcal{L}$, see
   (\ref{KHeq:Lshift})--(\ref{KHeq:Lmix}). Different choices of the Lindblad
   operators lead to different unravellings of the master equation, so that
   these hypothetical measurement events must not be viewed as ``real''
   processes. }}
(``clicks of counter $k$''). Since the absence of any click during the
``waiting time'' $\tau$ may also confer information about the system, this
lack of an event constitutes a measurement as well, which is described by the
non-unitary operators $\exp \left( - \mathi \mathsf{H}_{\mathbbm{C}} \tau /
 \hbar \right)$. A hypothetical demon, who has the full record $R^t_n$
available, would then be able to predict the final state \ $\mathcal{T} \left(
 \rho |R_n^t \right)$. In the absence of this information we have to resort
to the probabilistic description (\ref{KHeq:unravel}) weighting each quantum
trajectory with its (Bayesian) probability.

We can thus conclude that the dynamics of open quantum dynamics, and therefore
decoherence, can in principle be understood in terms of an information
transfer to the environment. Apart from this conceptual insight, the
unravelling of a master equation provides also an efficient stochastic
simulation method for its numerical integration. In these {\tmem{quantum
   jump}} approaches
{\cite{Carmichael1993a,Molmer1993a,Plenio1998a}}, which are based on the
observation that the quantum trajectory (\ref{KHeq:condtrafo}) of a pure state
remains pure, one generates a finite ensemble of trajectories such that the
ensemble mean approximates the solution of the master equation.

\subsection{Exemplary Master Equations}
\label{KHsec:examples}

Let us take a look at a number of very simple Markovian master
equations{\footnote{See also Sect. 6 in Cord M\"uller's contribution for a
   discussion of the master equation describing spin relaxation.}}, which
are characterized by a single Lindblad operator $\mathsf{L}$ (together with a
hermitian operator $\mathsf{H}$). The first example gives a general
description of dephasing, while the others are empirically known to describe
dissipative phenomena realistically. We may then ask what they predict about
decoherence.

\subsubsection{Dephasing}\index{dephasing!-- master equation}\index{master
 equation!-- for dephasing}

The simplest choice is to take the Lindblad operator to be
proportional{\footnote{As an exception, $\gamma$ does not have the
   dimensions of a rate here (to avoid clumsy notation).}} to the Hamiltonian
of a discrete quantum system, i.e. to the generator of the unitary dynamics,
$\mathsf{L} = \sqrt{\gamma} \mathsf{H}$. The Lindblad equation
\begin{equation}
  \partial_t \rho_t = \frac{1}{\mathi \hbar} [ \mathsf{H}, \rho_t] + \gamma
  \left( \mathsf{H} \rho_t  \mathsf{H} - \frac{1}{2}  \mathsf{H}^2 \rho_t -
  \frac{1}{2} \rho_t  \mathsf{H}^2 \right)
\end{equation}
is immediately solved in the energy eigenbasis, $\mathsf{H} = \sum_m E_m |m
\rangle \langle m| \bignone$:
\begin{equation}
  \rho_{m n} (t) \equiv \langle m| \rho_t |n \rangle = \rho_{m n} (0) \exp
  \left( \um \frac{\mathi}{\hbar} (E_m - E_n)t - \frac{\gamma}{2} (E_m - E_n)^2
  t \right) \;.
\end{equation}
As we expect from the discussion of qubit dephasing in
Sect.~\ref{KHsec:qubit}, the energy eigenstates are unaffected by the
non-unitary dynamics if the environmental effect commutes with $\mathsf{H}$.
The coherences show the exponential decay that we found in the ``thermal
regime'' (of times $t$ which are long compared to the inverse Matsubara
frequency). The comparison with (\ref{KHeq:thermalregime}) indicates that
$\gamma$ should be proportional to the temperature of the environment.

\subsubsection{Amplitude Damping of the Harmonic Oscillator}

Next, we choose $\mathsf{H}$ to be the Hamiltonian of a harmonic oscillator,
$\mathsf{H} = \hbar \omega \mathsf{a}^{\dag} \mathsf{a}$, and take as Lindblad
operator the ladder operator, $\mathsf{L} = \mathsf{a}$. The resulting
Lindblad equation is known empirically to describe the quantum dynamics of a
damped harmonic oscillator.\index{damped harmonic oscillator}

Choosing as initial state a coherent state\index{damped harmonic
 oscillator!-- coherent states}\index{coherent states!-- for the damped
 harmonic oscillator}, see (\ref{KHeq:displacementoperator}) and (\ref{KHeq:alphadef}) below,
\begin{eqnarray}
  \rho_0 = | \alpha_0 \rangle \langle \alpha_0 | & \equiv & \mathsf{D}
  (\alpha_0) |0 \rangle \langle 0| \mathsf{D}^{\dag} (\alpha_0) \nonumber\\
  & = & \mathe^{- \left| \alpha_0 \right|^2} \mathe \tmop{xp} (\alpha_0 
  \mathsf{a}^{\dag}) |0 \rangle \langle 0| \exp \left( \alpha_0^{\ast} 
  \mathsf{a} \right) \;, \label{KHrho0}
\end{eqnarray}
we are faced with the exceptional fact that the state {\tmem{remains pure}}
during the Lindblad time evolution. Indeed, the solution of the Lindblad
equation reads
\begin{equation}
  \rho_t = | \alpha_t \rangle \langle \alpha_t | \label{KHrohtcoh}
\end{equation}
with
\begin{equation}
  \hspace{2em} \alpha_t = \alpha_0 \exp \left( \um \mathi \omega t -
  \frac{\gamma}{2} t \right) \,,\label{KHalphadamp}
\end{equation}
as can be verified easily using (\ref{KHrho0}). It describes how the coherent
state spirals in phase space towards the origin, approaching the ground state
as $t \rightarrow \infty$. The rate $\gamma$ is the {\tmem{dissipation
   rate}}\index{dissipation!-- rate}\index{damped harmonic oscillator!-- dissipation rate} since it quantifies the energy loss, as
shown by the time dependence of the energy expectation value,
\begin{equation}
  \langle \alpha_t | \mathsf{H} | \alpha_t \rangle = \mathe^{\um \gamma t}
  \langle \alpha_0 | \mathsf{H} | \alpha_0 \rangle \;.
\end{equation}

\paragraph{Superposition of Coherent States}\index{coherent states!--
 superposition of}

What happens if we start out with a superposition of coherent states,
\begin{eqnarray}
  | \psi_0 \rangle & = & \frac{1}{\sqrt{\mathcal{N}}}  \left( | \alpha_0
  \rangle + | \beta_0 \rangle \right) \hspace{1em} \hspace{1em} \hspace{1em} 
  \label{KHpsicat}
\end{eqnarray}
with $\mathcal{N} = 2 + 2 \tmop{Re} \langle \alpha_0 | \beta_0 \rangle$, in
particular, if the separation in phase space is large compared to the quantum
uncertainties, $\left| \alpha_0 - \beta_0 \right| \gg 1$ ? The initial density
operator corresponding to (\ref{KHpsicat}) reads
\begin{equation}
  \rho_0 = \frac{1}{\mathcal{N}}  \left( | \alpha_0 \rangle \langle \alpha_0 |
  + | \beta_0 \rangle \langle \beta_0 | + c_0 | \alpha_0 \rangle \langle
  \beta_0 | + c_0^{\ast} | \beta_0 \rangle \langle \alpha_0 | \right)
\end{equation}
with $c_0 = 1$. One finds that the ansatz
\begin{equation}
  \rho_t = \frac{1}{\mathcal{N}}  \left( | \alpha_t \rangle \langle \alpha_t |
  + | \beta_t \rangle \langle \beta_t | + c_t | \alpha_t \rangle \langle
  \beta_t | + c_t^{\ast} | \beta_t \rangle \langle \alpha_t | \right)
\end{equation}
solves the Lindblad equation with (\ref{KHalphadamp}), provided
\begin{equation}
  c_t = c_0 \exp \left( \left[ \um \frac{1}{2}  \left| \alpha_0 - \beta_0
  \right|^2 + \mathi \tmop{Im} (\alpha_0 \beta_0^{\ast}) \right] \left( 1 -
  \mathe^{\um \gamma t} \right) \right) \;.
\end{equation}
That is, while the coherent ``basis'' states have the same time evolution as
in (\ref{KHrohtcoh}), the initial coherence $c_0$ gets additionally
suppressed. For times that are short compared to the dissipative time scale,
{$t \ll \gamma^{\um 1}$}, we have an exponential decay
\begin{equation}
  \left. |c_t | = |c_0 | \exp \Big( -
  \underset{\gamma_{\tmop{deco}}}{\underbrace{\frac{\gamma}{2}  \left|
  \alpha_0 - \beta_0 \right|^2 }} t \Big) \right.
\end{equation}
with a rate $\gamma_{\tmop{deco}}$. For macroscopically distinct
superpositions, where the phase space distance of the quantum states is much
larger than their uncertainties, $\left| \alpha_0 - \beta_0 \right| \gg 1$, the decoherence rate
$\gamma_{\tmop{deco}}$ can be much greater than the dissipation rate,
\begin{equation}
  \frac{\gamma_{\tmop{deco}}}{\gamma} = \frac{1}{2}  \left| \alpha_0 - \beta_0
  \right|^2 \gg 1 \;.
\end{equation}
This quadratic increase of the decoherence rate with the separation between
the coherent states has been confirmed experimentally in a series of beautiful
cavity QED experiments in Paris, using field states with an average of 5--9
photons {\cite{Raimond2001a,Haroche2004a}}.

Given this empiric support we can ask about the prediction for a material,
macroscopic oscillator. As an example, we take a pendulum with a mass of $m
=$100\,g and a period of {$2 \pi / \omega = 1\,\text{s}$}, and assume that we
can prepare it in a superposition of coherent states with a separation of $x
=$1\,cm. The mode variable $\alpha$ is related to position and momentum by
\begin{eqnarray}
  \alpha & = & \sqrt{\frac{m \omega}{2 \hbar}} \left( x + \mathi \frac{p}{m
  \omega} \right) \;, \label{KHeq:alphadef}
\end{eqnarray}
so that we get the prediction
\begin{eqnarray*}
  \gamma_{\tmop{deco}} & \simeq & 10^{30} \gamma \;.
\end{eqnarray*}
This purports that even with an oscillator of enormously low friction
corresponding to a dissipation rate of $\gamma =$1/year the coherence is lost
on a timescale of $10^{- 22}$\,s{\emdash}in which light travels the distance
of about a nuclear diameter.

This observation is often evoked to explain the absence of macroscopic
superpositions. However, it seems unreasonable to assume that anything
physically relevant takes place on a timescale at which a signal travels at
most by the diameter of an atomic nucleus. Rather, one expects that the
decoherence rate should saturate at a finite value if one increases the phase
space distance between the superposed states.

\subsubsection{Quantum Brownian Motion}

Next, let us consider a particle in one dimension. A possible choice for the
Lindblad operator is a linear combination of its position and momentum
operators,
\begin{equation}
  \mathsf{L} = \frac{p_{\tmop{th}}}{\hbar}  \mathsf{x} +
  \frac{\mathi}{p_{\tmop{th}}}  \mathsf{p} \;. \label{KHLqbm}
\end{equation}
Here $p_{\tmop{th}}$ is a momentum scale, which will be related to the
temperature of the environment below. The hermitian operator is taken to be
the Hamiltonian of a particle in a potential $V \left( x \right)$, plus a term
due to the environmental coupling,
\begin{equation}
  \mathsf{H} = \frac{\mathsf{p}^2}{2 m} + V ( \mathsf{x}) + \frac{\gamma}{2} (
  \mathsf{x}  \mathsf{p} + \mathsf{p}  \mathsf{x}) \;. \label{KHHqbm}
\end{equation}
This additional term will be justified by the fact that the resulting Lindblad
equation is almost equal to the {\tmem{Caldeira--Leggett master
    equation}}. The latter is the high-temperature limit of the exact
evolution equation following from a harmonic bath model of the environment
{\cite{Caldeira1983a,Unruh1989a}}, see Sect.~\ref{KHsec:weakcoup}.  It is
empirically known to describe the frictional quantum dynamics of a Brownian
particle, and, in particular, for $t\to\infty$ it leads to the canonical Gibbs
state in case of quadratic potentials.

The choices (\ref{KHLqbm}) and (\ref{KHHqbm}) yield the following Lindblad
equation:
\begin{eqnarray}
   \partial_t \rho_t & = & \overset{\tmop{Caldeira--Leggett}
   \tmop{master} \tmop{equation}}{\overbrace{\frac{1}{\mathi \hbar} [
   \frac{\mathsf{p}^2}{2 m} + V ( \mathsf{x}), \rho_t] +
   \underset{\tmop{dissipation}}{\underbrace{\frac{\gamma}{\mathi \hbar} 
   \left[ \mathsf{x}, \mathsf{p} \rho_t + \rho_t  \mathsf{p} \right]}} 
   \underset{\tmop{position} \tmop{localization}}{\underbrace{-
   \frac{\gamma}{2}  \frac{p^2_{\tmop{th}}}{\hbar^2}  \left[ \mathsf{x}, \left[
   \mathsf{x}, \rho_t \right] \right]}}}}  \nonumber\\
   &  & - \frac{\gamma}{2}  \frac{1}{p_{\tmop{th}}^2}  \left[ \mathsf{p},
   \left[ \mathsf{p}, \rho_t \right] \right]  \;. \label{KHqbmme}
\end{eqnarray}
The three terms in the upper line (with $p_{\tmop{th}}$ from (\ref{KHpth}))
constitute the Caldeira--Leggett master equation\index{Caldeira--Leggett master
  equation}\index{master equation!-- Caldeira--Leggett}. It is a Markovian, but not a completely positive master equation.
In a sense, the last term in (\ref{KHqbmme}) adds the minimal modification
required to bring the Caldeira--Leggett master equation into Lindblad form
{\cite{Breuer2002a,Diosi1993a}}.

To see the most important properties of (\ref{KHqbmme}) let us take a look at
the time evolution of the relevant observables in the Heisenberg picture. As
discussed in Sect.~\ref{KHsemigroups}, the Heisenberg equations of motion
are determined by the dual Liouville operator $\mathcal{L}^{\sharp}$. In the
present case, it takes the form
\begin{eqnarray}
  \mathcal{L^{\sharp}} \left( \mathsf{A} \right) & = & \frac{1}{\mathi \hbar}
  [ \mathsf{A}, \frac{\mathsf{p}^2}{2 m} + V ( \mathsf{x})] -
  \frac{\gamma}{\mathi \hbar}  \left( \mathsf{p}  \left[ \mathsf{x},
  \mathsf{A} \right] + \left[ \mathsf{x}, \mathsf{A} \right] \mathsf{p}
  \right) - \frac{\gamma}{2}  \frac{p^2_{\tmop{th}}}{\hbar^2}  \left[
  \mathsf{x}, \left[ \mathsf{x}, \mathsf{A} \right] \right]  \nonumber\\
  &  & - \frac{\gamma}{2}  \frac{1}{p_{\tmop{th}}^2}  \left[ \mathsf{p},
  \left[ \mathsf{p}, \mathsf{A} \right] \right] \;. 
\end{eqnarray}
It is now easy to see that
\begin{eqnarray}
  \mathcal{L^{\sharp}} \left( \mathsf{x} \right) & = & \frac{\mathsf{p}}{m}\;,
  \nonumber\\
  \mathcal{L^{\sharp}} \left( \mathsf{p} \right) & = & \um V' ( \mathsf{x}) -
  2 \gamma \mathsf{p} \;.  \label{KHfriction}
\end{eqnarray}
Hence, the force arising from the potential is complemented by a frictional
force which will drive the particle into thermal equilibrium. The fact that
this frictional component stems from the second term in (\ref{KHqbmme})
indicates that the latter describes the {\tmem{dissipative}} effect of the
environment.

In the absence of an external potential, $V = 0$, the time evolution
determined by (\ref{KHfriction}) is easily obtained, since $\left(
 \mathcal{L^{\sharp}} \right)^n \left( \mathsf{p} \right) = \left( - 2 \gamma
\right)^n \mathsf{p}$ for $n \in \mathbbm{N}$:
%% \begin{eqnarray}
%%    \mathsf{p}_t & = & \mathe^{\mathcal{L}^{\sharp} t} \mathsf{p} \hspace{0.6em}
%%    = \hspace{0.6em} \sum_{n = 0}^{\infty} \frac{\bignone \left( - 2 \gamma t
%%    \right)^n}{n!} \mathsf{p} \hspace{0.6em} = \hspace{0.6em} \mathe^{- 2 \gamma
%%    t} \mathsf{p} \hspace{1em} \text{[for $V = 0$]} \\
%%    \mathsf{x}_t & = & \mathe^{\mathcal{L}^{\sharp} t} \mathsf{x} \hspace{0.6em}
%%    = \hspace{0.6em} \mathsf{x} + \frac{1}{m} \sum_{n = 1}^{\infty}
%%    \frac{\bignone t^n}{n!} \left( \mathcal{L^{\sharp}} \right)^{n - 1} \left(
%%    \mathsf{p} \right) \nonumber\\
%%    & = & \mathsf{x} + \frac{\mathsf{p} - \mathsf{p}_t}{2 \gamma m}
%%    \hspace{0.6em} \hspace{1em} \text{[for $V = 0$]} 
%% \end{eqnarray}
\begin{equation}
 \begin{aligned}
  \mathsf{p}_t &= \mathe^{\mathcal{L}^{\sharp} t} \mathsf{p} = \sum_{n =
  0}^{\infty} \frac{\left( - 2 \gamma t \right)^n}{n!} \mathsf{p} = \mathe^{- 2 \gamma
  t} \mathsf{p} \\
  \mathsf{x}_t & = \mathe^{\mathcal{L}^{\sharp} t} \mathsf{x} = \mathsf{x} + \frac{1}{m} \sum_{n = 1}^{\infty}
  \frac{\bignone t^n}{n!} \left( \mathcal{L^{\sharp}} \right)^{n - 1} \left(
  \mathsf{p} \right) = \mathsf{x} + \frac{\mathsf{p} - \mathsf{p}_t}{2 \gamma m}
 \end{aligned}
\qquad \text{[for $V = 0$] \;.} 
\end{equation}
Note that, unlike in closed systems, the Heisenberg operators do not retain
their commutator, $[ \mathsf{x}_t, \mathsf{p}_t] \neq i \hbar$ for $t > 0$
(since the map $\mathcal{W}_t^{\sharp} = \exp \left( \mathcal{L}^{\sharp} t
\right)$ is non-unitary). Similarly, $( \mathsf{p}^2)_t \neq \left(
 \mathsf{p}_t \right)^2$ for $t > 0$, so that the kinetic energy operator
$\mathsf{T} = \mathsf{p}^2 / 2 m$ has to be calculated separately. Noting
\begin{eqnarray}
  \mathcal{L^{\sharp}} \left( \mathsf{T} \right) & = & \gamma
  \frac{p_{\tmop{th}}^2}{2 m} - 4 \gamma \mathsf{T}  \hspace{1em} \text{[for
  $V = 0$]\;,} 
\end{eqnarray}
we find
\begin{eqnarray}
  \mathsf{T}_t & = & \frac{p_{\tmop{th}}^2}{8 m} + \left( \mathsf{T} -
  \frac{p_{\tmop{th}}^2}{8 m} \right) \mathe^{- 4 \gamma t} \hspace{1em}
  \text{[for $V = 0$]} \;. 
\end{eqnarray}
This shows how the kinetic energy approaches a constant determined by the
momentum scale $p_{\tmop{th}}$. We can now relate $p_{\tmop{th}}$ to a
temperature by equating the stationary expectation value $\tmop{tr} \left(
 \rho \mathsf{T}_{\infty} \right) = p_{\tmop{th}}^2 / 8 m$ with the the
average kinetic energy $\frac{1}{2} k_{\rm B}T$ in a one-dimensional thermal distribution. This leads
to
\begin{eqnarray}
  p_{\tmop{th}} & = & 2 \sqrt{mk_B T}  \label{KHpth}
\end{eqnarray}
for the momentum scale in (\ref{KHLqbm}). Usually, one is not able to state
the operator evolution in closed form. In those cases it may be helpful to
take a look at the Ehrenfest equations for their expectation values. For
example, given $\langle \mathsf{p}^2 \rangle_t = 2 m \langle \mathsf{T}
\rangle_t$, the other second moments, $\langle \mathsf{x}^2 \rangle_t$ and
$\langle \mathsf{p} \mathsf{x} + \mathsf{x} \mathsf{p} \rangle_t$ form a
closed set of differential equations. Their solutions, given in
{\cite{Breuer2002a}}, yield the time evolution of the position variance
$\sigma_x^2 \left( t \right) = {\langle \mathsf{x}^2 \rangle_t - \langle
 \mathsf{x} \rangle^2_t}$. It has the asymptotic form
\begin{eqnarray}
  \text{$\sigma_x^2 \left( t \right)$} & \sim & \frac{k_{\text{B}} T}{m
  \gamma} t \hspace{1em} \text{as $t \rightarrow \infty$} \;, 
\end{eqnarray}
which shows the {\tmem{diffusive}} behavior expected of a (classical) Brownian
particle{\footnote{Note that the definition of $\gamma$ differs by a factor of
   2 in part of the literature.}}.

Let us finally take a closer look at the physical meaning of the third term in
(\ref{KHqbmme}), which is dominant if the state is in a superposition of
spatially separated states. Back in the Schr\"odinger picture we have in
position representation, {$\rho_t (x, x') = \langle x| \rho_t |x' \rangle$},
\begin{equation}
  \partial_t \rho_t (x, x') = -
  \underset{\gamma_{\tmop{deco}}}{\underbrace{\frac{\gamma}{2} 
  \frac{p_{\tmop{th}}^2}{\hbar^2} (x - x')^2}} \rho_t (x, x') \hspace{1em} +
  \text{[the other terms] \;.}
\end{equation}
The ``diagonal elements'' $\rho \left( x, x \right)$ are unaffected by this
term, so that it leaves the particle density invariant. The coherences in
position representation, on the other hand, get exponentially suppressed,
\begin{equation}
  \rho_t (x, x') = \exp \left( \um \gamma_{\tmop{deco}} t \right) \rho_0 (x,
  x') \;. \label{KHeq:quadinc}
\end{equation}
Again the decoherence rate is determined by the square of the relevant
distance $|x - x' |$,
\begin{equation}
  \frac{\gamma_{\tmop{deco}}}{\gamma} = 4 \mathpi \frac{(x -
  x')^2}{\Lambda_{\tmop{th}}^2} \;. \label{KHCLquad}
\end{equation}
Like in Sect.~\ref{KHsec:examples}, the rate $\gamma_{\tmop{deco}}$ will be
much larger than the dissipative rate provided the distance is large on the
quantum scale, here given by the {\tmem{thermal de Broglie wavelength}}
\begin{eqnarray}
  \Lambda_{\tmop{th}}^2 & = &  \frac{2 \mathpi \hbar^2}{mk_B T} \;. 
  \label{KHthermdeBroglie}
\end{eqnarray}
In particular, one finds $\gamma_{\tmop{deco}} \ggg \gamma$ if the separation
is truly macroscopic. Again, it seems unphysical that the decoherence rate
does not saturate as $|x - x' | \rightarrow \infty$, but grows above all
bounds. One might conclude from this that non-Markovian master equations are
more appropriate on these short timescales. However, I will argue that (unless
the environment has very special properties) Markovian master equations are
well suited to study decoherence processes, provided they involve an
appropriate description of the microscopic dynamics.

\section{Microscopic Derivations}

In this section we discuss two important and rather different strategies to
obtain Markovian master equations based on microscopic considerations.

\subsection{The Weak Coupling Formulation}\index{master equation!-- weak coupling
formulation}
\label{KHsec:weakcoup}

The most widely used form of incorporating the environment is the weak
coupling approach. Here one assumes that the total Hamiltonian is ``known''
microscopically, usually in terms of a simplified model,
\begin{eqnarray*}
  \mathsf{H}_{\tmop{tot}} & = & \mathsf{H} + \mathsf{H}_{\text{E}} +
  \mathsf{H}_{\text{int}}
\end{eqnarray*}
and takes the interaction part $\mathsf{H}_{\text{int}}$ to be ``weak'' so
that a perturbative treatment of the interaction is permissible.

The main assumption, called the {\tmem{Born approximation}}\index{Born!--
 approximation}, states that $\mathsf{H}_{\text{int}}$ is sufficiently small
so that we can take the total state as factorized, both initially,
$\rho_{\tmop{tot}} (0) = \rho (0) \otimes \rho_{\text{E}}$, and also at $t >
0$ in those terms which involve $\mathsf{H}_{\text{int}}$ to second order.
\begin{eqnarray}
  \text{Assumption 1} : \hspace{1em} & \rho_{\tmop{tot}} (t) \simeq \rho (t)
  \otimes \rho_{\text{E}} \!  & \hspace{1em} \text{[to 2nd order in
  $\mathsf{H}_{\text{int}}$]} \; . \label{KHeq:A1}
\end{eqnarray}
Here $\rho_{\text{E}}$ is the stationary state of the environment, $[
\mathsf{H}_{\text{E}}, \rho_{\text{E}}] = 0$. Like above, the use of the
interaction picture is indicated with a tilde, cf.~(\ref{KHintpic}), so
that the von Neumann equation for the total system reads
\begin{eqnarray}
  \partial_t  \tilde{\rho}_{\tmop{tot}} & = & \frac{1}{\mathi \hbar} [
  \widetilde{\mathsf{H}}_{\text{int}} (t), \tilde{\rho}_{\tmop{tot}} (t)]
  \nonumber\\
  & = & \frac{1}{\mathi \hbar} [ \widetilde{\mathsf{H}}_{\text{int}} (t),
  \tilde{\rho}_{\tmop{tot}} (0)] + \frac{1}{(\mathi \hbar)^2} \! \int_0^t \!\mathd
  s \bignone [ \widetilde{\mathsf{H}}_{\text{int}} (t), [
  \widetilde{\mathsf{H}}_{\text{int}} (s), \tilde{\rho}_{\tmop{tot}} (s)]]
  \;. \nonumber \\ 
  \label{KHvN2}
\end{eqnarray}
In the second equation, which is still exact, the von Neumann equation in its
integral version was inserted into the differential equation version.  Using a
basis of Hilbert-Schmidt operators of the product Hilbert space, see Sect.
\ref{KHsec:lindblad}, one can decompose the general
$\widetilde{\mathsf{H}}_{\text{int}}$ into the form
\begin{equation}
  \widetilde{\mathsf{H}}_{\text{int}} (t) = \sum_k \widetilde{\mathsf{A}}_k
  (t) \otimes \widetilde{\mathsf{B}}_k (t)  
\end{equation}
with $\widetilde{\mathsf{A}}_k = \widetilde{\mathsf{A}}_k^{\dag},
\widetilde{\mathsf{B}}_k = \widetilde{\mathsf{B}}_k^{\dag}$. The first
approximation is now to replace $\tilde{\rho}_{\tmop{tot}} (s)$ by
$\tilde{\rho} (s) \otimes \rho_{\text{E}}$ in the double commutator of
(\ref{KHvN2}), where $\widetilde{\mathsf{H}}_{\text{int}}$ appears to second
order. Performing the trace over the environment one gets\index{master
 equation!-- generalized}
\begin{eqnarray}
  \partial_t  \tilde{\rho} (t) & = & \tmop{tr}_{\text{E}} (\partial_t 
  \tilde{\rho}_{\tmop{tot}}) \nonumber\\
  & \cong & \frac{1}{\mathi \hbar}  \sum_k \langle \widetilde{\mathsf{B}}_k
  (t) \rangle_{\rho_{\text{E}}} [ \widetilde{\mathsf{A}}_k (t), \tilde{\rho}
  (0)] \nonumber\\
  &  & + \frac{1}{(\mathi \hbar)^2} \bignone \sum_{k \ell} \biggl\{ \int_0^t \!
  \mathd s \,\underset{C_{k \ell} (t - s)}{\underbrace{\langle
  \widetilde{\mathsf{B}}_k (t) \widetilde{\mathsf{B}}_{\ell} (s)
  \rangle_{\rho_{\text{E}}}}} \{ \widetilde{\mathsf{A}}_k (t)
  \widetilde{\mathsf{A}}_{\ell} (s) \tilde{\rho} (s) -
  \widetilde{\mathsf{A}}_{\ell} (s) \tilde{\rho} (s) \widetilde{\mathsf{A}}_k
  (t)\} \nonumber\\
  &  & \phantom{+ \frac{1}{(\mathi \hbar)^2} \bignone \sum_{k \ell}} +
  \text{h.c.} \biggr\}\;.  \label{KHme1}
\end{eqnarray}
All the relevant properties of the environment are now expressed in terms of
the (complex) {\tmem{bath correlation functions}} $C_{k \ell} (t - s)$. Since
$[ \mathsf{H}_{\text{E}},\rho_{\text{E}}] = 0$, they depend only on the time
difference $t - s$,
\begin{equation}
   C_{k \ell} (\tau) = \tmop{tr} \left( \mathe^{\mathi \mathsf{H}_{\text{E}}
   \tau} \mathsf{B}_k  \mathe^{\um \mathi \mathsf{H}_{\text{E}} \tau}
   \mathsf{B}_{\ell} \rho_{\text{E}} \right) \equiv \left\langle \mathe^{\mathi
   \mathsf{H}_{\text{E}} \tau} \mathsf{B}_k  \mathe^{\um \mathi
   \mathsf{H}_{\text{E}} \tau} \mathsf{B}_{\ell}
   \right\rangle_{\rho_{\text{E}}} \; .
\end{equation}
This function is determined by the environmental state alone, and it is
typically appreciable only for a small range of $\tau$ around $\tau = 0$.

Equation (\ref{KHme1}) has the closed form of a generalized master equation,
but it is non-local in time, i.e. non-Markovian. Viewing the second term as a
superoperator $\mathcal{K}$, which depends essentially on $t - s$ we have
\begin{equation}
  \partial_t  \tilde{\rho} (t) = \underbrace{\frac{1}{\mathi \hbar} [\langle
  \widetilde{\mathsf{H}}_{\text{int}} (t) \rangle_{\rho_{\text{E}}}, \tilde{\rho}
  (0)]}_{\text{disregarded}} + \int_0^t \mathd s\mathcal{K}(t - s)
  \tilde{\rho} (s) \bignone \label{KHeq:nonloc} \;,
\end{equation}
where $\mathcal{K}$ is a superoperator memory kernel of the form
(\ref{KHnonmarkform}). We may disregard the first term since the model
Hamiltonian $\mathsf{H}_{\text{E}}$ can always be reformulated such that
$\langle \widetilde{\mathsf{B}}_k (t) \rangle_{\rho_{\text{E}}} = 0$.

A naive application of second order of perturbation theory would now replace
$\tilde{\rho} (s)$ by the initial $\tilde{\rho} (0)$. However, since the
memory kernel is dominant at the origin it is much more reasonable to replace
$\tilde{\rho} (s)$ by $\tilde{\rho} (t)$. The resulting master equation is
local in time,
\begin{equation} 
\partial_t  \tilde{\rho} (t) \cong 0 + \left( \int_0^t \mathd
 s\mathcal{K}(t - s) \right) \tilde{\rho} (t) \;. 
\end{equation}
It is called the {\tmem{Redfield equation}}\index{Redfield equation} and it is
{\tmem{not}} Markovian, because the integrated superoperator still depends on
time. Since the kernel is appreciable only at the origin it is reasonable to
replace $t$ in the upper integration limit by $\infty$.

These steps are summarized by the {\tmem{Born-Markov
   approximation}}\index{Born-Markov approximation}:
\begin{eqnarray}
  \text{Assumption 2} : &  & \int_0^t \mathd s\mathcal{K}(t - s) \tilde{\rho}
  \bignone (s) \cong \int_0^{\infty} \mathd s\mathcal{K}(s) \tilde{\rho}
  \bignone (t) \;. \label{KHeq:A2}
\end{eqnarray}
It leads from (\ref{KHeq:nonloc}) to a Markovian master equation provided
$\left\langle \mathsf{H}_{\text{int}} \right\rangle_{\rho_{\text{E}}} = 0$.

However, by no means is such a master equation guaranteed to be completely
positive. An example is the Caldeira--Leggett master equation discussed in
Sect. \ref{KHsec:examples}. It can be derived by taking the environment to be
a bath of bosonic field modes whose field amplitude is coupled linearly to the
particle's position operator. A model assumption on the spectral density of
the coupling then leads to the frictional behavior of (\ref{KHfriction})
{\cite{Caldeira1983a,Weiss1999a}}.

A {\tmem{completely positive}} master equation can be obtained by a further
simplification, the ``{\tmem{secular}}'' approximation\index{secular approximation@\emph{secular}
 approximation}, which is applicable if the system
Hamiltonian $\mathsf{H}$ has a discrete, non-degenerate spectrum.  The system
operators $\mathsf{A}_k$ can then be decomposed in the system energy
eigenbasis. Combining the contributions with equal energy differences
\begin{eqnarray}
  \mathsf{A}_k (\omega) & = & \sum_{E' - E = \hbar \omega} \langle E|
  \mathsf{A}_k |E' \rangle |E \rangle \langle E' | \bignone \bignone =
  \mathsf{A}_k^{\dag} (\omega) \;,
\end{eqnarray}
we have
\begin{eqnarray}
  \mathsf{A}_k & = & \sum_{\omega} \mathsf{A}_k (\omega) \;. 
\end{eqnarray}
The time dependence of the operators in the interaction picture is now
particularly simple,
\begin{equation}
  \widetilde{\mathsf{A}}_k (t) = \sum_{\omega} \mathe^{\um \mathi \omega t} 
  \mathsf{A}_k (\omega) \bignone \;.
\end{equation}
Inserting this decomposition we find
\begin{eqnarray}
  \partial_t  \tilde{\rho} (t) & = & \sum_{k \ell} \sum_{\omega \omega'}
  \mathe^{\mathi (\omega - \omega') t} \Gamma_{k \ell} (\omega')\{
  \mathsf{A}_{\ell} (\omega') \tilde{\rho} (t) \mathsf{A}_k^{\dag} (\omega) -
  \mathsf{A}_k^{\dag} (\omega) \mathsf{A}_{\ell} (\omega') \tilde{\rho} (t)\}+ \text{h.c.}\nonumber\\
  \label{KHme3}
\end{eqnarray}
with
\begin{equation}
  \Gamma_{k \ell} (\omega) = \frac{1}{\hbar^2}  \int_0^{\infty} \mathd s
  \mathe^{\mathi \omega s} \langle \widetilde{\mathsf{B}}_k (s)
  \mathsf{B}_{\ell} (0) \rangle_{\rho_{\text{E}}} \bignone \;.
\end{equation}
For times $t$ which are large compared to the time scale given by the smallest
system energy spacings it is reasonable to expect that only equal pairs of
frequencies $\omega$, $\omega'$ contribute appreciably to the sum in
(\ref{KHme3}), since all other contributions are averaged out by the wildly
oscillating phase factor. This constitutes the {\tmem{rotating wave
   approximation}}\index{rottating wave approximation}, our third assumption
\begin{eqnarray}
  \text{Assumption 3} : \hspace{1em} \sum_{\omega \omega'} \mathe^{\mathi
  (\omega - \omega') t} f (\omega, \omega') & \simeq & \sum_{\omega} f
  (\omega, \omega) \;.  \label{KHeq:A3}
\end{eqnarray}
It is now useful to rewrite
\begin{equation}
  \Gamma_{k \ell} (\omega) = \frac{1}{2} \gamma_{k \ell} (\omega) + \mathi
  S_{k \ell} (\omega)
\end{equation}
with $\gamma_{k \ell} (\omega)$ given by the full Fourier transform of the
bath correlation function,
\begin{equation}
  \gamma_{k \ell} (\omega) = \Gamma_{k \ell} (\omega) + \Gamma_{\ell k}^{\ast}
  (\omega) = \frac{1}{\hbar^2}  \int_{\um \infty}^{\infty} \mathd t \,
  \mathe^{\mathi \omega t} \bignone \left\langle \widetilde{\mathsf{B}}_k (t)
  \mathsf{B}_{\ell} (0) \right\rangle_{\rho_{\text{E}}} \;,
\end{equation}
and the hermitian matrix $S_{k \ell} (\omega)$ defined by
\begin{equation}
  \text{$S_{k \ell} (\omega) = \frac{1}{2 \mathi} (\Gamma_{k \ell} (\omega) -
  \Gamma_{\ell k}^{\ast} (\omega))$} \;.
\end{equation}
The matrix $\gamma_{k \ell} (\omega)$ is {\tmem{positive}}{\footnote{To see
   that the matrix $\left(\tmmathbf{\gamma} \left( \omega \right)\right)_{k,
     \ell} \equiv \gamma_{k \ell} (\omega)$ is positive we write
\begin{eqnarray}
  \left( \vec{v}, \tmmathbf{\gamma} \vec{v} \right) & = & \sum_{k \ell}
  v_k^{\ast} \bignone \gamma_{k \ell} (\omega) v_{\ell}\nonumber\\
  & = & \frac{1}{\hbar^2}
  \int \mathd t \mathe^{\mathi \omega t} \bignone \sum_{k \ell}  \left\langle
  \bignone \mathe^{\mathi \mathsf{H}_{\text{E}} t / \hbar}  \mathsf{B}_k (0)
  v_k^{\ast} \mathe^{\um \mathi \mathsf{H}_{\text{E}} t / \hbar} 
  \mathsf{B}_{\ell} (0) v_{\ell} \right\rangle_{\rho_{\text{E}}} \nonumber\\
  & = & \int \mathd t \mathe^{\mathi \omega t} \bignone  \left\langle
  \mathe^{\mathi \mathsf{H}_{\text{E}} t / \hbar}  \mathsf{C}^{\dag}
  \mathe^{\um \mathi \mathsf{H}_{\text{E} } t / \hbar}  \mathsf{C}
  \right\rangle_{\rho_{\text{E}}}  \label{KHeq:gampos}
\end{eqnarray}
with $\mathsf{C} \assign \hbar^{- 1} \sum_{\ell} \bignone v_{\ell}
\mathsf{B}_{\ell} (0)$. One can now check that due to its particular form the
correlation function
\begin{eqnarray}
  f \left( t \right) & = & \left\langle \mathe^{\mathi \mathsf{H}_{\text{E}} t
  / \hbar}  \mathsf{C}^{\dag} \mathe^{\um \mathi \mathsf{H}_{\text{E} } t /
  \hbar}  \mathsf{C} \right\rangle_{\rho_{\text{E}}} \nonumber
\end{eqnarray}
appearing in (\ref{KHeq:gampos}) is {\tmem{of positive type}}, meaning that
the $n \times n$-matrices $\left( f \left( t_i - t_j \right) \right)_{ij}$
defined by an arbitrary choice of $t_1, \ldots ., t_n$ and $n \in \mathbbm{N}$
are positive. According to Bochner's theorem {\cite{Lukacs1966a}} the Fourier
transform of a function which is of positive type is positive, which proves
the positivity of (\ref{KHeq:gampos}).}} so that we end up with a master
equation of the first Lindblad form (\ref{KHfirstLindblad}),
\begin{eqnarray}
  \partial_t \tilde{\rho} (t) & = & \frac{1}{\mathi \hbar}  \left[
  \mathsf{H}_{\tmop{Lamb}}, \tilde{\rho} (t) \right] + \sum_{k \ell \omega}
  \gamma_{k \ell} (\omega) \biggl( \mathsf{A}_{\ell}
  (\omega) \tilde{\rho} (t) \mathsf{A}_k^{\dag} (\omega) \nonumber\\
  & & \phantom{\frac{1}{\mathi \hbar}  \left[
  \mathsf{H}_{\tmop{Lamb}}, \tilde{\rho} (t) \right]} - \frac{1}{2} 
  \mathsf{A}_k^{\dag} (\omega) \mathsf{A}_{\ell} (\omega) \tilde{\rho} (t) -
  \frac{1}{2}  \tilde{\rho} (t) \mathsf{A}_k^{\dag} (\omega) \mathsf{A}_{\ell}
  (\omega) \biggr) \; . \nonumber\\ 
\end{eqnarray}
The hermitian operator
\begin{eqnarray}
  \mathsf{H}_{\tmop{Lamb}} & = & \hbar \sum_{k \ell \omega} S_{k \ell}
  (\omega) \mathsf{A}_k^{\dag} (\omega) \mathsf{A}_{\ell} (\omega) \bignone 
  \label{KHeq:Hlamb}
\end{eqnarray}
describes a renormalization of the system energies due to the coupling with
the environment, the {\tmem{Lamb shift}}\index{Lamb shift}. Indeed, one finds
$\left[ \mathsf{H}, \mathsf{H}_{\tmop{Lamb}} \right] = 0$.

Reviewing the three approximations
(\ref{KHeq:A1}), (\ref{KHeq:A2}), (\ref{KHeq:A3}) in view of the decoherence
problem one comes to the conclusion that they all seem to be well justified if
the environment is generic and the coupling is sufficiently weak. Hence, the
master equation should be alright for times beyond the short-time transient
which is introduced due to the choice of a product state as initial state.
Evidently, the problem of non-saturating decoherence rates encountered in
Sect. \ref{KHsec:examples} is rather due to
the {\tmem{linear}} coupling assumption, corresponding to a ``dipole
approximation'', which is clearly invalid once the system states are separated
by a larger distance than the wavelength of the environmental field modes.

This shows the need to incorporate realistic, nonlinear environmental
couplings with a finite range. A convenient way of deriving such master
equations is discussed in the next section.

\subsection{The Monitoring Approach}\index{continuous monitoring}\index{master equation!-- monitoring approach}

The following method to derive microscopic master equations differs
considerably from the weak coupling treatment discussed above. It is
not based on postulating an approximate ``total'' Hamiltonian of
system plus environment, but on two operators, which can be characterized
individually in an operational sense. This permits to describe the
environmental coupling in a non-perturbative fashion, and to incorporate the
Markov assumption right from the beginning, rather than introducing it in the
course of the calculation.

The approach may be motivated by the observation made in
Sects.~\ref{KHsec:monitoring} and \ref{KHsec:unravelling} that
environmental decoherence can be understood as due to the information transfer
from the system to the environment occurring in a sequence of indirect
measurements. In accordance with this, we will picture the environment as
monitoring the system continuously by sending probe particles which scatter
off the system at random times. This notion will be applicable whenever the
interaction with the environment can reasonably be described in terms of
individual interaction events or ``collisions'', and it suggests a formulation
in terms of scattering theory, like in Sect.~\ref{KHsec:scattering}. The
Markov assumption is then easily incorporated by disregarding the change of
the environmental state after each collision {\cite{Hornberger2007b}}.

\begin{figure}[h]
\centering
 \includegraphics[scale=.47]{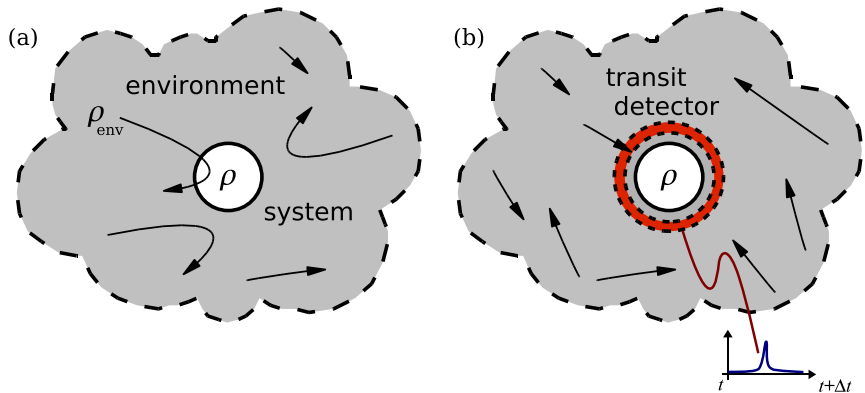}
% \resizebox{117mm}{!}{\epsfig{file=monitoring.eps}}
  \caption{(a) In the monitoring approach the system is taken to interact at
  most with one environmental {(quasi-)}particle at a time, so that three-body
  collisions are excluded. Moreover, in agreement with the Markov assumption,
  it is assumed that the environmental particles disperse their correlation
  with the system before scattering again. (b) In order to consistently
  incorporate the state-dependence of the collision rate into the dynamic
  description of the scattering process, we imagine that the system is
  monitored continuously by a transit detector, which tells at a temporal
  resolution $\Delta t$ whether a particle is going to scatter off the system,
  or not.\label{KHfig:monitoring}}
\end{figure}

When setting up a differential equation, one would like to write the temporal
change of the system as the rate of collisions multiplied by the state
transformation due to an individual scattering. However, in general not only
the transformed state will depend on the original system state, but also the
collision rate, so that such a naive ansatz would yield a nonlinear equation, 
violating the basic principles of quantum mechanics.
To account for this state dependence of the collision rate in a proper way we
will apply the concept of generalized measurements discussed in
Sect.~\ref{KHsec:monitoring}. Specifically, we shall assume that the system is
surrounded by a hypothetical, minimally invasive detector, which tells at any
instant whether a probe particle has passed by and is going to scatter off the
system, see Fig.~\ref{KHfig:monitoring}.

The rate of collisions is then described by a positive operator
$\mathsf{\Gamma}$ acting in the system-probe Hilbert space. Given the
uncorrelated state $\varrho_{\tmop{tot}} = \rho \otimes \rho_E$, it determines
the probability of a collision to occur in a small time interval $\Delta t$,
\begin{eqnarray}
  \tmop{Prob} \left( \text{C}_{\Delta t} | \rho \otimes \rho_E \right) & = &
  \Delta t \tmop{tr} \left( \mathsf{\Gamma} \left[ \rho \otimes \rho_E \right]
  \right) \;.  \label{KHeq:probscat}
\end{eqnarray}
Here, $\rho_E$ is the stationary reduced single particle state of the
environment. The microscopic definition of $\mathsf{\Gamma}$ will in general
involve the current density operator of the relative motion and a total
scattering cross section, see below.

The important point to note is that the information that a collision will take place changes our knowledge about the state, as described by 
the generalized measurement transformation
(\ref{KHeq:transformation-state-measurement}). At the same time, we have to
keep in mind that the measurement is not real, but is introduced here only for
enabling us to account for the state dependence of the collision probability.
It is therefore reasonable to take the detection process as
{\tmem{efficient}}\index{detection process!-- efficient}, see
Sect.~\ref{KHsec:monitoring}, and {\tmem{minimally-invasive}}\index{detection
 process!-- minimally-invasive}, i.e. $\mathsf{U}_{\alpha} = \mathbbm{I}$ in
(\ref{KHeq:effectivemeastrafo}), so that neither unnecessary uncertainty nor a
reversible back-action is introduced. This implies that after a (hypothetical)
detector click, but prior to scattering, the system-probe state will have the
form
\begin{eqnarray}
  \mathcal{M} \left( \varrho_{\tmop{tot}} | \text{C}_{\Delta t} \right) & = &
  \frac{ \mathsf{\Gamma}^{1 / 2} \varrho_{\tmop{tot}}  \mathsf{\Gamma}^{1 /
  2}}{\tmop{tr} \left( \mathsf{\Gamma} \varrho_{\tmop{tot}} \right)} \;. 
  \label{KHeq:mestra}
\end{eqnarray}
This measurement transformation reflects our improved knowledge about the
incoming two-particle wave packet, and it may be viewed as enhancing those
parts which are heading towards a collision. Similarly, the absence of a detection
event during $\Delta t$ changes the state, and  this occurs with the probability
$\tmop{Prob} \left( \overline{\text{C}}_{\Delta t} \right) = 1 - \tmop{Prob}
\left( \text{C}_{\Delta t} \right)$.
%, and the state conditioned on a null-event is given by{\footnote{Although more general transformations are conceivable, this one is distinguished by the fact that it introduces no further     (time-dependent) operators.}}
%\begin{eqnarray}
%   \mathcal{M} \left( \varrho_{\tmop{tot}} | \overline{\text{C}}_{\Delta t}
%   \right) & = & \frac{\varrho_{\tmop{tot}} - \Delta t \mathsf{\Gamma}^{1 / 2}
%   \varrho_{\tmop{tot}}  \mathsf{\Gamma}^{1 / 2}}{1 - \Delta t \tmop{tr} \left(
%   \mathsf{\Gamma} \varrho_{\tmop{tot}} \right)} \;.  \label{KHeq:compmestra}
%\end{eqnarray}

Using the state transformation (\ref{KHeq:mestra}) we can now formulate the unconditioned system-probe state after a coarse grained time $\Delta t$ as the {{mixture}} of the
colliding state transformed by the S-matrix and the untransformed
non-colliding one, weighted with their respective probabilities,
\begin{eqnarray}
  \varrho'_{\tmop{tot}}\! \left( \Delta t \right) & = & \tmop{Prob} \left(
  \text{C}_{\Delta t} | \varrho_{\tmop{tot}} \right) \mathsf{S} \mathcal{M}
  \left( \varrho_{\tmop{tot}} | \text{C}_{\Delta t} \right) \mathsf{S}^{\dag}\!
  + \tmop{Prob} \left( \overline{\text{C}}_{\Delta t} | \varrho_{\tmop{tot}}
  \right)  \mathcal{M} \left( \varrho_{\tmop{tot}} |
  \overline{\text{C}}_{\Delta t} \right) \nonumber\\
  & = & \mathsf{S} \mathsf{\Gamma}^{1 / 2} \varrho_{\tmop{tot}} 
  \mathsf{\Gamma}^{1 / 2} \mathsf{S}^{\dag} \Delta t + \varrho_{\tmop{tot}} -
  \mathsf{\Gamma}^{1 / 2} \varrho_{\tmop{tot}}  \mathsf{\Gamma}^{1 / 2}
  \Delta t \;. \label{KHeq:rhoinf}
\end{eqnarray}
Here, the complementary map $\mathcal{M} \left( \cdot |\overline{\text{C}}_{\Delta t} \right)$ is fixed by the requirement that the state $ \varrho_{\tmop{tot}}$ should  remain unchanged both if the collision probability vanishes, $\mathsf{\Gamma}=0$, and if the scattering has no effect,  $\mathsf{S}=\mathbbm{I}$.

Focusing on the nontrivial part $\mathsf{T}$ of the two-particle S-matrix $\mathsf{S} =
\mathbbm{I} + \mathi \mathsf{T}$ one finds that the unitarity of $\mathsf{S}$
implies that
\begin{eqnarray}
 \mathrm{Im}(\mathsf{T})\equiv \frac{1}{2\mathi} \left( \mathsf{T} - \mathsf{T}^{\dag} \right) & = & \frac{1}{2}
  \mathsf{T}^{\dag} \mathsf{T} \;. 
\end{eqnarray}
Using this relation we can write the differential quotient as
\begin{eqnarray}
  \frac{\varrho_{\tmop{tot}}' \left( \Delta t \right) -
  \varrho_{\tmop{tot}}}{\Delta t} & = & \mathsf{T} \mathsf{\Gamma}^{1 / 2}
  \varrho_{\tmop{tot}} \mathsf{\Gamma}^{1 / 2} \mathsf{T}^{\dag} - \frac{1}{2}
  \mathsf{T}^{\dag} \mathsf{T} \mathsf{\Gamma}^{1 / 2} \varrho_{\tmop{tot}}
  \mathsf{\Gamma}^{1 / 2} \\
  &  & - \frac{1}{2} \mathsf{\Gamma}^{1 / 2} \varrho_{\tmop{tot}}
  \mathsf{\Gamma}^{1 / 2} \mathsf{T}^{\dag} \mathsf{T} + \frac{\mathi}{2}
  \left[ \mathsf{T} + \mathsf{T}^{\dag}, \mathsf{\Gamma}^{1 / 2}
  \varrho_{\tmop{tot}} \mathsf{\Gamma}^{1 / 2} \right] \;. \nonumber
\end{eqnarray}
It is now easy to arrive at a closed differential equation. We trace out the
environment, assuming, in accordance with the Markov approximation, that the
factorization $\varrho_{\tmop{tot}} = \rho \otimes \rho_E$ is valid prior to
each monitoring interval $\Delta t$. Taking the limit of continuous monitoring
$\Delta t \rightarrow 0$, approximating $ \tmop{Tr}_E \left( \left[ \tmop{Re}(\mathsf{T} ), \mathsf{\Gamma}^{1 / 2} \varrho_{\tmop{tot}} \mathsf{\Gamma}^{1 / 2} \right] \right)\simeq \tmop{Tr}_E \left( \left[\mathsf{\Gamma}^{1 / 2} \tmop{Re}(\mathsf{T} )\mathsf{\Gamma}^{1 / 2}, \varrho_{\tmop{tot}}\right] \right)$, and adding the generator $\mathsf{H}$ of the free
system evolution we arrive at {\cite{Hornberger2007b}}
\begin{eqnarray}
  \frac{\mathd}{\mathd t} \rho & = & \frac{1}{\mathi \hbar} \left[ \mathsf{H},
  \rho \right] + \mathi \tmop{Tr}_E \left( \left[  \mathsf{\Gamma}^{1 / 2}\tmop{Re}(\mathsf{T}) \mathsf{\Gamma}^{1 / 2},  \rho \otimes \rho_E  \right] \right) \nonumber\\
  &  & + \tmop{Tr}_E \left( \mathsf{T} \mathsf{\Gamma}^{1 / 2} \left[ \rho
  \otimes \rho_E \right] \mathsf{\Gamma}^{1 / 2} \mathsf{T}^{\dag} \right)
  \nonumber\\
  &  & - \frac{1}{2} \tmop{Tr}_E \left( \mathsf{\Gamma}^{1 / 2}
  \mathsf{T}^{\dag} \mathsf{T}  \mathsf{\Gamma}^{1 / 2} \left[ \rho \otimes
  \rho_E \right] \right) \nonumber\\
  &  & - \frac{1}{2} \tmop{Tr}_E \left( \left[ \rho \otimes \rho_E \right] 
  \mathsf{\Gamma}^{1 / 2} \mathsf{T}^{\dag} \mathsf{T} \mathsf{\Gamma}^{1 / 2}
  \right) \;.  \label{KHeq:me1}
\end{eqnarray}
This general monitoring master equation, entirely specified by the rate operator $\mathsf{\Gamma}$, the scattering operator $\mathsf{S} = \mathbbm{I} + \mathi \mathsf{T}$, and the environmental state $ \rho_E$, is non-perturbative in the sense that the collisional interaction is nowhere assumed to be weak. It is manifestly markovian even before the environmental trace is carried out, and one finds, by doing the trace in the eigenbasis of $ \rho_E$,  that is has the general Lindblad structure (\ref{KHsecondLindblad}) of the generator of a quantum dynamical semigroup.
The second term in (\ref{KHeq:me1}), which involves a commutator,
accounts for the renormalization of the system energies due to the coupling to
the environment, just like (\ref{KHeq:Hlamb}), while the last three lines describe the incoherent effect of the coupling to the environment.

So far, the discussion was very general. To obtain concrete master equations
one has to specify system and environment, along with the operators
$\mathsf{\Gamma}$ and $\mathsf{S}$ describing their interaction. In the
following applications, we will assume the environment to be an ideal Maxwell
gas, whose single particle state
\begin{eqnarray}
  \rho_{\tmop{gas}} & = & \frac{\Lambda_{\tmop{th}}^3}{\Omega} \exp \left( -
  \beta \frac{\mathsf{p}^2}{2 m} \right)  \label{KHeq:gas}
\end{eqnarray}
is characterized by the inverse temperature $\beta$, the normalization volume
$\Omega$, and the thermal de Broglie wave length $\Lambda_{\tmop{th}}$ defined
in (\ref{KHthermdeBroglie}).

\subsection{Collisional Decoherence of a Brownian
 Particle}\index{Brownian particle!-- decoherence}
\label{KHsec:colldeco}

As a first application of the monitoring approach, let us consider the
``localization'' of a mesoscopic particle by a gaseous environment.
Specifically, we will assume that the mass $M$ of this Brownian particle is
much greater than the mass $m$ of the gas particles. In the limit $m/M\to 0$ the energy
exchange during an elastic collision vanishes, so that the mesoscopic particle
will not thermalize in our description, but we expect that the off-diagonal elements  of its position representation will get reduced, as discussed in
Sect.~\ref{KHsec:examples}.

This can be seen by considering the effect of the S-matrix in the limit $m / M \rightarrow 0$. In general, a collision keeps the center-of-mass invariant, and only the relative
coordinates are affected. Writing $\mathsf{S}_0$ for the S-matrix in the
center of mass frame and denoting the momentum eigenstates of the Brownian and
the gas particle by $|\tmmathbf{P} \rangle$ and $|\tmmathbf{p} \rangle$,
respectively, we have {\cite{Taylor1972a}}
\begin{equation}
\mathsf{S} |\tmmathbf{P} \rangle |\tmmathbf{p} \rangle_{} = \int \mathd^3
\tmmathbf{Q}|\tmmathbf{P}-\tmmathbf{Q} \rangle |\tmmathbf{p}+\tmmathbf{Q}
\rangle_{} \langle \frac{m_{\ast}}{m} \tmmathbf{p}- \frac{m_{\ast}}{M}
\tmmathbf{P}+\tmmathbf{Q}| \mathsf{S}_0 | \frac{m_{\ast}}{m} \tmmathbf{p}-
\frac{m_{\ast}}{M} \tmmathbf{P} \rangle \;, 
\end{equation} 
where $m_{\ast} = Mm / \left( M + m \right)$ is the reduced mass and
$\tmmathbf{Q}$ is the transfered momentum (and thus the change of the relative
momentum). In the limit of a large Brownian mass we have $m_{\ast} / m
\rightarrow 1$ and $m_{\ast} / M \rightarrow 0$, so that
\begin{equation}
  \mathsf{S} |\tmmathbf{P} \rangle |\tmmathbf{p} \rangle_{} \rightarrow \int
  \mathd^3 \tmmathbf{Q}|\tmmathbf{P}-\tmmathbf{Q} \rangle
  |\tmmathbf{p}+\tmmathbf{Q} \rangle \langle \tmmathbf{p}+\tmmathbf{Q}|
  \mathsf{S}_0 |\tmmathbf{p} \rangle \hspace{2em} \text{[for $M \gg m$]} \;.
  \label{KHeq:Sinf}
\end{equation}
It follows that a position eigenstate $|\tmmathbf{X} \rangle$ of the Brownian
particle remains unaffected by a collision,
\begin{equation}
  \mathsf{S}_{} |\tmmathbf{X} \rangle | \psi_{\tmop{in}} \rangle_{\text{E}} =
  |\tmmathbf{X} \rangle \underbrace{\left( \mathe^{\um \mathi \mathsf{p} \cdot
  \tmmathbf{X}/ \hbar}  \mathsf{S}_0  \mathe^{\mathi \mathsf{p} \cdot
  \tmmathbf{X}/ \hbar} \right) | \psi_{\tmop{in}} \rangle_{\text{E}}}_{|
  \psi^{\left( \tmmathbf{X} \right)}_{\tmop{out}} \rangle_{\text{E}}}\;,
  \label{KHeq:Sinftrans}
\end{equation}
as can be seen by inserting identities in terms of the momentum eigenstates.
Here, $| \psi_{\mathi n} \rangle_E$ denotes an arbitrary single-particle wave
packet state of a gas atom. The exponentials in (\ref{KHeq:Sinftrans}) effect
a translation of $\mathsf{S}_0$ from the origin to the position
$\tmmathbf{X}$, so that the scattered state of the gas particle $|
\psi^{\left( \tmmathbf{X} \right)}_{\tmop{out}} \rangle_{\text{E}}$ depends on
the location of the Brownian particle.

Just like in Sect.~\ref{KHsec:nutshell}, a single collision will thus reduce
the spatial coherences $\rho \left( \tmmathbf{X}, \tmmathbf{X}' \right) =
\langle \tmmathbf{X}| \rho |\tmmathbf{X}' \rangle$ by the overlap of the gas
states scattered at positions $\tmmathbf{X}$ and $\tmmathbf{X}'$,
\begin{eqnarray}
  \rho' \left( \tmmathbf{X}, \tmmathbf{X}' \right) & = & \rho \left(
  \tmmathbf{X}, \tmmathbf{X}' \right) \langle
  \psi^{(\tmmathbf{X}')}_{\tmop{out}} | \psi^{\left( \tmmathbf{X}
  \right)}_{\tmop{out}} \rangle_{\text{E}} \;. 
\end{eqnarray}
The reduction factor will be the smaller in magnitude the better the scattered
state of the gas particle can ``resolve'' between the positions $\tmmathbf{X}$
and $\tmmathbf{X}'$.

In order to obtain the dynamic equation we need to specify the rate operator.
Classically, the collision rate is determined by the product of the current
density $j = n_{\tmop{gas}} v_{\tmop{rel}}$ and the total cross section
$\sigma \left( p_{\tmop{rel}} \right)$, and therefore $\mathsf{\Gamma}$ should
be expressed in terms of the corresponding operators. This is particularly
simple in the large mass limit $M \rightarrow \infty$, where $v_{\tmop{rel}} =
\left| \tmmathbf{p}/ m -\tmmathbf{P}/ M \right| \rightarrow \left|
 \tmmathbf{p} \right| / m$, so that the current density and the cross section
depend only on the momentum of the gas particle, leading to
\begin{eqnarray}
  \mathsf{\Gamma} & = &  \bignone n_{\tmop{gas}} \frac{\left| \mathsf{p}
  \right|}{m} \sigma \left( \mathsf{p} \right) \;.  \label{KHeq:Gamma1}
\end{eqnarray}
If the gas particle moves in a normalized wave packet heading towards the
origin then the expectation value of this operator will indeed determine the
collision probability. However, this expression depends only on the modulus of
the velocity so that it will yield a finite collision probability even if the
particle is heading away form the origin. Hence, for (\ref{KHeq:me1}) to make
sense either the S-matrix should be modified to keep such a non-colliding state unaffected,
or $\mathsf{\Gamma}$ should contain in addition a projection to the subset of
incoming states, see the discussion below.

In momentum representation, $\rho \left( \tmmathbf{P}, \tmmathbf{P}' \right) =
\langle \tmmathbf{P}| \rho |\tmmathbf{P}' \rangle$, equation (\ref{KHeq:me1})
assumes the general structure{\footnote{The second term in (\ref{KHeq:me1})
   describes forward scattering and vanishes for momentum diagonal
   $\rho_E$.}}\index{master equation!-- monitoring approach@-- -- for a Brownian particle}
\begin{eqnarray}
  \partial_t \rho \left( \tmmathbf{P}, \tmmathbf{P}' \right) & = &
  \frac{1}{\mathi \hbar} \frac{P^2 - \left( P' \right)^2}{2 M} \rho \left(
  \tmmathbf{P}, \tmmathbf{P}' \right) \nonumber\\
  &  & + \int \mathd \tmmathbf{P}_0 \mathd \tmmathbf{P}_0'  \bignone \rho
  \left( \tmmathbf{P}_0, \tmmathbf{P}'_0 \right) M \left( \tmmathbf{P},
  \tmmathbf{P}' ; \tmmathbf{P}_0, \tmmathbf{P}_0' \right) \nonumber\\
  &  & - \frac{1}{2} \int \mathd \tmmathbf{P}_0 \rho \left( \tmmathbf{P}_0,
  \tmmathbf{P}' \right) \int \mathd \tmmathbf{P}_f \bignone M \left(
  \tmmathbf{P}_f, \tmmathbf{P}_f ; \tmmathbf{P}_0, \tmmathbf{P} \right)
  \nonumber\\
  &  & - \frac{1}{2} \int \mathd \tmmathbf{P}_0' \rho \left( \tmmathbf{P},
  \tmmathbf{P}_0' \right) \int \mathd \tmmathbf{P}_f \bignone M \left(
  \tmmathbf{P}_f, \tmmathbf{P}_f ; \tmmathbf{P}', \tmmathbf{P}_0' \right) \;. 
\end{eqnarray}
The dynamics is therefore characterized by a single complex function
\begin{eqnarray}
  M \left( \tmmathbf{P}, \tmmathbf{P}' ; \tmmathbf{P}_0, \tmmathbf{P}_0'
  \right) & = & \langle \tmmathbf{P}| \tmop{tr}_{\tmop{gas}} \left( 
  \mathsf{T} \mathsf{\Gamma}_{}^{1 / 2}  \left[ |\tmmathbf{P}_0 \rangle
  \langle \tmmathbf{P}'_0 | \otimes \rho_{\tmop{gas}} \right]
  \mathsf{\Gamma}_{}^{1 / 2} \mathsf{T^{\dag}} \right) |\tmmathbf{P}'
  \rangle\;,\nonumber \\
\end{eqnarray}
which has to be evaluated. Inserting the diagonal representation of the gas
state (\ref{KHeq:gas})
\begin{eqnarray}
  \rho_{\tmop{gas}} & = & \frac{(2 \pi \hbar)^3}{\Omega} \int \mathd
  \tmmathbf{p}_0 \mu \left( \tmmathbf{p}_0 \right) \bignone |\tmmathbf{p}_0
  \rangle \langle \tmmathbf{p}_0 |  \label{KHeq:rhodiag}
\end{eqnarray}
it reads, with the choices (\ref{KHeq:Sinf}) and (\ref{KHeq:Gamma1}) for
$\mathsf{S}$ and $\mathsf{\Gamma}$,
\begin{eqnarray}
  M \! \left( \tmmathbf{P}, \tmmathbf{P}';\! \tmmathbf{P}\!-\!\tmmathbf{Q},
  \tmmathbf{P}'\! -\!\tmmathbf{Q}' \right)\! &=& \! \int\! \mathd \tmmathbf{p}_1 \mathd
  \tmmathbf{p}_0 \mu \left( \tmmathbf{p}_0 \right) \delta \left(
  \tmmathbf{Q}+\tmmathbf{p}_1 -\tmmathbf{p}_0 \right) \delta \left(
  \tmmathbf{Q}' +\tmmathbf{p}_0 -\tmmathbf{p}_1 \right) \nonumber\\
  &  &\! \times \frac{n_{\tmop{gas}}}{m}  \left| \tmmathbf{p}_0 \right| \sigma
  \left( \tmmathbf{p}_0 \right) \frac{\left( 2 \mathpi \hbar
  \right)^3}{\Omega} | \langle \tmmathbf{p}_1 | \mathsf{T}_0 |\tmmathbf{p}_0
  \rangle |^2  \nonumber\\
  & = & \delta \left( \tmmathbf{Q}-\tmmathbf{Q}' \right) \int \mathd
  \tmmathbf{p}_0 \,\mu \left( \tmmathbf{p}_0 \right)  \frac{n_{\tmop{gas}}}{m}
  \left| \tmmathbf{p}_0 \right| \sigma \left( \tmmathbf{p}_0 \right)
  \nonumber\\
  &  & \times \frac{\left( 2 \mathpi \hbar \right)^3}{\Omega} | \langle
  \tmmathbf{p}_0 -\tmmathbf{Q}| \mathsf{T}_0 |\tmmathbf{p}_0 \rangle |^2
  \nonumber\\
  & =: & \delta \left( \tmmathbf{Q}-\tmmathbf{Q}' \right) M_{\tmop{in}}
  \left( \tmmathbf{Q} \right) \;. 
\end{eqnarray}
This shows that, apart form the unitary motion, the dynamics is simply
characterized by momentum exchanges described in terms of gain and loss terms,
\begin{eqnarray}
  \partial_t \rho \left( \tmmathbf{P}, \tmmathbf{P}' \right) \text{} & \text{}
  = & \frac{1}{\mathi \hbar} \frac{P^2 - \left( P' \right)^2}{2 M} \rho \left(
  \tmmathbf{P}, \tmmathbf{P}' \right) + \int \mathd \tmmathbf{Q} \bignone \rho
  \left( \tmmathbf{P}-\tmmathbf{Q}, \tmmathbf{P}' -\tmmathbf{Q} \right)
  M_{\tmop{in}} \left( \tmmathbf{Q} \right) \nonumber\\
  &  & - \rho \left( \tmmathbf{P}, \tmmathbf{P}' \right)  \int \mathd
  \tmmathbf{Q}M_{\tmop{in}} \left( \tmmathbf{Q} \right) \;.  \label{KHeq:mepr}
\end{eqnarray}
We still have to evaluate the function $M_{\tmop{in}} \left( \tmmathbf{Q}
\right)$, which can be clearly interpreted as the rate of collisions leading
to a momentum gain $\tmmathbf{Q}$ of the Brownian particle,
\begin{eqnarray}
  M_{\tmop{in}} \left( \tmmathbf{Q} \right) & = & \frac{n_{\tmop{gas}}}{m}
  \int \mathd \tmmathbf{p}_0 \mu \left( \tmmathbf{p}_0 \right)  \left|
  \tmmathbf{p}_0 \right| \sigma \left( \tmmathbf{p}_0 \right) \frac{\left( 2
  \mathpi \hbar \right)^3}{\Omega} | \langle \tmmathbf{p}_0 -\tmmathbf{Q}|
  \mathsf{T}_0 |\tmmathbf{p}_0 \rangle |^2 \;. 
\nonumber\\ \label{KHeq:Min1}
\end{eqnarray}
It involves the momentum matrix element of the on-shell $\mathsf{T}_0$-matrix,
$\mathsf{S}_0 = 1 + i \mathsf{T}_0$, which, according to elastic scattering
theory {\cite{Taylor1972a}}, is proportional to the scattering amplitude $f$,
\begin{eqnarray}
  \langle \tmmathbf{p}_f | \mathsf{T}_0 |\tmmathbf{p}_i \rangle & = & \frac{f
  (\tmmathbf{p}_f, \tmmathbf{p}_i)}{2 \pi \hbar} \delta \left( \frac{p_f^2}{2}
  - \frac{p_i^2}{2} \right) \;.  \label{KHTme}
\end{eqnarray}
The delta function ensures the conservation of energy during the collision. At
first sight, this leads to an ill-defined expression since the matrix element
(\ref{KHTme}) appears as a squared modulus in (\ref{KHeq:Min1}), so that the
tree-dimensional integration is over a squared delta function.

The appearance of this problem can be traced back to our disregard of the
projection to the subset of incoming states in the definition
(\ref{KHeq:Gamma1}) of $\mathsf{\Gamma}$. When evaluating $M_{\tmop{in}}$ we
used the diagonal representation (\ref{KHeq:rhodiag}) for $\rho_{\tmop{gas}}$
in terms of (improper) momentum eigenstates, which comprise both incoming and
outgoing characteristics if viewed as the limiting form of a wave packet. One
way of implementing the missing projection to incoming states would be to use
a different convex decomposition of $\rho_{\text{gas}}$, which admits a
separation into incoming and outgoing contributions {\cite{Hornberger2003b}}.
This way, $M_{\tmop{in}}$ can indeed be calculated properly, albeit in a
somewhat lengthy calculation. A shorter route to the same result sticks to the
diagonal representation, but modifies the definition of $\mathsf{S}$ in a
formal sense so that it keeps all outgoing state invariant.{\footnote{In
    general, even a purely outgoing state gets transformed by $\mathsf{S}$,
    since the definition of the S-matrix involves a backward time-evolution
    {\cite{Taylor1972a}}.}} The conservation of the probability current,
which must still be guaranteed by any such modification, then implies a simple
rule how to deal with the squared matrix element {\cite{Hornberger2003b}},
\begin{equation}
  \frac{\left( 2 \mathpi \hbar \right)^3}{\Omega} \left| \langle
  \tmmathbf{p}_f | \mathsf{T}_0 |\tmmathbf{p}_i \rangle \right|^2
  \longrightarrow \frac{\left| f (\tmmathbf{p}_f, \tmmathbf{p}_i)
  \right|^2}{p_i \sigma (p_i)} \delta \left( \frac{p_f^2}{2} - \frac{p_i^2}{2}
  \right) \;. \label{KHeq:rr}
\end{equation}
Here $\sigma (p) = {\int \mathd \Omega' \bignone \left| f
   (p\tmmathbf{n}', p\tmmathbf{n}) \right|}^2$ is the total elastic cross
section. With this replacement we obtain immediately
\begin{eqnarray} \label{KEeq:Minfinal}
  M_{\tmop{in}} \left( \tmmathbf{Q} \right) & = & \frac{n_{\tmop{gas}}}{m} \!
  \int\! \mathd \tmmathbf{p}_0 \,\mu \left( \tmmathbf{p}_0 \right)  \left| f
  (\tmmathbf{p}_0 -\tmmathbf{Q}, \tmmathbf{p}_0) \right|^2 \delta \left(
  \frac{\tmmathbf{p}^2_0}{2} - \frac{\left( \tmmathbf{p}_0 -\tmmathbf{Q}
  \right)^2}{2} \right) \;. \nonumber \\ 
\end{eqnarray}
As one would expect, the rate of momentum changing collisions is determined
by a thermal average over the differential cross section $\mathd \sigma /
\mathd \Omega = \left| f \right|^2$.

Also for finite mass ratios $m / M$ a master equation can be obtained this
way, although the calculation is more complicated {\cite{Hornberger2006b,Hornberger2008a}}.
The resulting linear quantum Boltzmann equation then describes on equal
footing the decoherence and dissipation effects of a gas on the quantum motion
of a particle.

The ``localizing'' effect of a gas on the Brownian particle can now be seen,
after going into the interaction picture in order to remove the unitary part
of the evolution, and by stating the master equation in position
representation.  From Eqs.~(\ref{KHeq:mepr}) and (\ref{KEeq:Minfinal}) one obtains
\begin{eqnarray}
  \partial_t  \tilde{\rho} (\tmmathbf{X}, \tmmathbf{X}') & = & \um F
  (\tmmathbf{X}-\tmmathbf{X}') \tilde{\rho} (\tmmathbf{X}, \tmmathbf{X}') 
  \label{KHeq:lamr}
\end{eqnarray}
with {\tmem{localization rate}} {\cite{Hornberger2003b}}
\begin{eqnarray}
  F (\tmmathbf{x}) & = & \int_0^{\infty} \mathd v \bignone \nu (v) \bignone
  \bignone n_{\tmop{gas}} \bignone v \int \frac{\mathd \Omega_1 \mathd
  \Omega_2}{4 \mathpi}  \left( 1 - \mathe^{\mathi mv \left( \tmmathbf{n}_1
  -\tmmathbf{n}_2 \right) \cdot \tmmathbf{x}/ \hbar} \right)  \nonumber\\
  &  & \times {\left| f (mv\tmmathbf{n}_2, mv\tmmathbf{n}_1) \right|}^2 \;. 
  \label{KHeq:locrate}
\end{eqnarray}
Here, the unit vectors $\tmmathbf{n}_1, \tmmathbf{n}_2$ are the directions of
incoming and outgoing gas particles associated to the elements of solid angle
$\mathd \Omega_1$ and $\mathd \Omega_2$ and $\nu \left( v \right)$ is the
velocity distribution in the gas. Clearly, $F \left( \tmmathbf{x} \right)$
determines how fast the spatial coherences corresponding to the distance
$\tmmathbf{x}$ decay.

One angular integral in (\ref{KHeq:locrate}) can be performed in the case of
isotropic scattering, $f (\tmmathbf{p}_f, \tmmathbf{p}_i) = f \left( \cos
 \left( \tmmathbf{p}_f, \tmmathbf{p}_i \right) ; E = p_i^2 / 2 m \right)$. In
this case,
\begin{eqnarray}
  F (\tmmathbf{x}) & = & \int_0^{\infty} \bignone \mathd v \bignone \nu (v)
  \bignone \bignone n \bignone _{\tmop{gas}} v \left\{ \sigma (mv) - 2 \pi
  \int_{- 1}^1 \mathd \left( \cos \theta \right) \left| f \left( \cos \theta ;
  E = \frac{m}{2} v^2 \right) \right|^2 \right. \nonumber\\
  &  & \left. \phantom{\int_0^{\infty} \bignone \mathd v \bignone \nu (v)
  \bignone \bignone n \bignone _{\tmop{gas}} v \left\{ \sigma_{\tmop{tot}} (v)
  - \right.} \times \bignone \tmop{sinc} \left( 2 \sin \left( \frac{\theta}{2}
  \right)  \frac{mv \left| \tmmathbf{x} \right|}{\hbar} \right)  \right\}\;, 
  \label{KHeq:locrate2}
\end{eqnarray}
with $\tmop{sinc} (x) = \sin (x) / x$ and $\theta$ the (polar) scattering
angle. 
\begin{figure}[tb]
\centering 
%\resizebox{10cm}{!}{\epsfig{file=locrate.eps}}
\includegraphics[scale=.5]{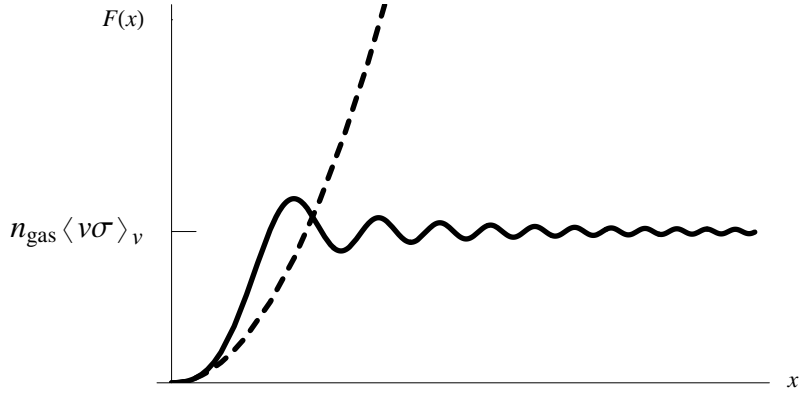}
  \caption{The localization rate (\ref{KHeq:locrate2}) describing the loss of
  wave-like behavior in a Brownian particle state saturates for large
  distances at the average collision rate. In contrast, the Caldeira--Leggett
  model predicts a quadratic increase beyond all bounds (dashed line), see
  (\ref{KHeq:quadinc}). This indicates that linear coupling models should
  be taken with care if time scales are involved that differ strongly from the
  dissipation time scale.\label{KHfig:locrate}}
\end{figure}

The argument of the sinc function is equal to the momentum exchange during the
collision times the distance in units of $\hbar$. As $\left|
 \tmmathbf{X}-\tmmathbf{X}' \right| \longrightarrow 0$ the sinc approaches
unity and the angular integral yields the total cross section $\sigma$ so that
the localization rate vanishes, as required. At very small distances, a second
order expansion in the distance $\tmmathbf{x}$ is permissible and one obtains
a quadratic dependence {\cite{Joos1985a}}, such as predicted by the
Caldeira--Leggett model, see Eq.~(\ref{KHCLquad}). However, once the distance  $\left|
 \tmmathbf{X}-\tmmathbf{X}' \right|$
is sufficiently large so that the scattered state can resolve whether the
collision took place at position $\tmmathbf{X}$ or $\tmmathbf{X}'$ the sinc
function in (\ref{KHeq:locrate2}) suppresses the integrand. It follows that in
the limit of large distances the localization rate saturates, at a
value given by the average collision rate $F (\infty) = \left\langle \sigma
 vn_{\tmop{gas}} \right\rangle$, see Figure \ref{KHfig:locrate}.

Decoherence in this saturated regime of large separations has been observed,
in good agreement with this theory, in molecular interference experiments in
the presence of various gases {\cite{Hornberger2003a}}. The intermediate
regime between quadratic increase and saturation was also seen in such
experiments on momentum-exchange mediated decoherence, by studying the
influence of the heat radiation emitted by fullerene molecules on the
visibility of their interference pattern {\cite{Hackermuller2004a}}.

As a conclusion of this section, we see that the scattering approach permits
to incorporate realistic microscopic interactions transparently and without
approximation in the interaction strength. The results show clearly that
linear coupling models, which imply that decoherence rates grow above all
bounds, have a limited range of validity. They cannot be judged by their
success in describing dissipative phenomena. Frequent claims of
``universality'' in decoherence behavior, which are based on these linear
coupling models, are therefore to be treated with care.

\subsection{Decoherence of a Quantum Dot}\index{quantum
 dot!-- decoherence}\index{decoherence!-- in a quantum dot}

As a second application of the monitoring approach, let us see how the
dynamics of an immobile object with discrete internal structure, such as an
implementation of a quantum dot, gets affected by an environment of ideal gas
particles. For simplicity, we take the gas again in the Maxwell state
(\ref{KHeq:gas}), though different dispersion relations, e.g. in the case of
phonon quasi-particles, could be easily incorporated. The interaction between
system and gas will be described in terms of the in general inelastic
scattering amplitudes determined by the interaction potential.

In the language of scattering theory the energy eigenstates of the
non-motional degrees of freedom are called channels. In our case of a
structureless gas they form a discrete basis of the system Hilbert space. In
the following, the notation $| \alpha \rangle$, not to be confused with the
coherent states of Sect.~\ref{KHsec:examples}, will be used to indicate the
system eigenstates of energy $E_{\alpha}$. In this channel basis,
$\rho_{\alpha \beta} = \langle \alpha | \rho | \beta \rangle$, the equation of
motion (\ref{KHeq:me1}) takes on the form of a general discrete master
equation of Lindblad type\index{master equation!-- monitoring approach@-- -- for a quantum dot},
\begin{eqnarray}
  \partial_t \rho_{\alpha \beta} & = & \frac{E_{\alpha} + \varepsilon_{\alpha}
  - E_{\beta} - \varepsilon_{\beta}}{\mathi \hbar} \rho_{\alpha \beta} +
  \sum_{\alpha_0 \beta_0} \rho_{\alpha_0 \beta_0} \bignone M_{\alpha
  \beta}^{\alpha_0 \beta_0} \nonumber\\
  &  & - \frac{1}{2}  \sum_{\alpha_0 } \rho_{\alpha_0 \beta_{}} \bignone
  \bignone \sum_{\gamma} M_{\gamma \gamma}^{\alpha_0 \alpha} - \frac{1}{2} 
  \sum_{\beta_0} \rho_{\alpha \beta_0} \bignone \sum_{\gamma} M_{\gamma
  \gamma}^{\beta \beta_0} \;.  \label{KHeq:master}
\end{eqnarray}
The real energy shifts $\varepsilon_{\alpha}$ given below describe the
coherent modification of the system energies due to the presence of the
environment. They are due to the second term in \ (\ref{KHeq:me1}) and are the
analogue of the Lamb shift (\ref{KHeq:Hlamb}) encountered in the weak coupling
calculation. The incoherent effect of the environment, on the other hand, is
described by the set of complex rate coefficients
\begin{eqnarray}
  M_{\alpha \beta}^{\alpha_0 \beta_0} & = & \langle \alpha | \tmop{Tr}_E
  \left( \mathsf{T} \mathsf{\Gamma}^{1 / 2} \left[ | \alpha_0 \rangle \langle
  \beta_0 | \otimes \rho_{\tmop{gas}} \right] \mathsf{\Gamma}^{1 / 2}
  \mathsf{T}^{\dag} \right) | \beta \rangle \;.  \label{KHeq:M}
\end{eqnarray}
In order to calculate these quantities we need again to specify the rate
operator $\mathsf{\Gamma}$. In the present case, it is naturally given in
terms of the current density operator $\mathsf{j} = n_{\tmop{gas}} \mathsf{p}
/ m$ of the impinging gas particles multiplied by the channel-specific total
scattering cross sections $\sigma \left( \tmmathbf{p}, \alpha \right)$,
\begin{eqnarray}
  \mathsf{\Gamma} & = & \sum_{\alpha} | \alpha \rangle \langle \alpha |
  \otimes n_{\tmop{gas}}  \frac{\left| \mathsf{p} \right|}{m} \sigma \left(
  \mathsf{p}, \alpha \right) \;.  \label{KHeq:Gamma}
\end{eqnarray}
Like in Sect.~\ref{KHsec:colldeco}, this operator should in principle contain
a projection to the subset of incoming states of the gas particle. Again, this
can be accounted for in two different ways in the calculation of the rates
(\ref{KHeq:M}). By using a non-diagonal decomposition of $\rho_{\tmop{gas}}$,
which permits to disregard the outgoing states, one obtains{\footnote{For the
   special case of factorizing interactions, $\mathsf{H}_{\tmop{int}} =
   \mathsf{A} \otimes \mathsf{B}_E$, and for times large compared to all
   system time scales this result can be obtained rigorously in a standard
   approach {\cite{Dumcke1985a}}, by means of the ``low density limit''
   scaling method {\cite{Alicki1987a,Breuer2002a}}.}}
\begin{eqnarray}
  M_{\alpha \beta}^{\alpha_0 \beta_0} & = & \chi_{\alpha \beta}^{\alpha_0
  \beta_0}  \frac{n_{\tmop{gas}}}{m^2} \int \mathd \tmmathbf{p} \bignone
  \bignone \mathd \tmmathbf{p}_0 \mu \left( \tmmathbf{p}_0 \right) f_{\alpha
  \alpha_0} \left( \tmmathbf{p}, \tmmathbf{p}_0  \right) \nonumber\\
  &  & \times f_{\beta \beta_0}^{\ast} \left( \tmmathbf{p}, \tmmathbf{p}_0
  \right) \delta \left( \frac{\tmmathbf{p}^2 -\tmmathbf{p}_0^2}{2 m} +
  E_{\alpha} - E_{\alpha_0} \right)\;,  \label{KHeq:M3}
\end{eqnarray}
with the Kronecker-like factor
\begin{eqnarray}
  \chi_{\alpha \beta}^{\alpha_0 \beta_0} & \assign & \left\{ \begin{array}{ll}
    1 & \text{if $E_{\alpha} - E_{\alpha_0} = E_{\beta} - E_{\beta_0}$}\\
    0 & \text{otherwise} \;.
  \end{array} \right. 
\end{eqnarray}
The energy shifts are determined the real parts of the forward scattering
amplitude,
\begin{eqnarray}
  \varepsilon_{\alpha} & = & - 2 \pi \hbar^2 \frac{n_{\tmop{gas}}}{m} \int
  \mathd \tmmathbf{p}_0 \mu \left( \tmmathbf{p}_0 \right) \tmop{Re} \left[
  f_{\alpha \alpha} \left( \tmmathbf{p}_0, \tmmathbf{p}_0 \right) \right] . 
\end{eqnarray}
Some details of this calculation can be found in {\cite{Hornberger2007b}}.
Rather than repeating them here we note that the result (\ref{KHeq:M3}) can be
obtained directly by using the diagonal representation (\ref{KHeq:rhodiag}) of
$\rho_{\tmop{gas}}$ and the multichannel-generalization of the replacement
rule (\ref{KHeq:rr}),
\begin{eqnarray}
  \frac{\left( 2 \pi \hbar \right)^3}{\Omega} \langle \alpha \tmmathbf{p}|
  \mathsf{T} | \alpha_0 \tmmathbf{p}_0 \rangle \langle \beta_0 \tmmathbf{p}_0
  | \mathsf{T}^{\dag} | \beta \tmmathbf{p} \rangle & \rightarrow & 
  \frac{\chi_{\alpha \beta}^{\alpha_0 \beta_0}}{p_0 m} \frac{f_{\alpha
  \alpha_0} \left( \tmmathbf{p}, \tmmathbf{p}_0 \right) f^{\ast}_{\beta
  \beta_0} \left( \tmmathbf{p}, \tmmathbf{p}_0 \right)}{\sqrt{\sigma \left(
  p_0, \alpha_0 \right) \sigma \left( p_0, \beta_0 \right)} }\;\;\;\;  \nonumber\\
  &  & \times \delta \left( \frac{\tmmathbf{p}^2 -\tmmathbf{p}_0^2}{2 m} +
  E_{\alpha} - E_{\alpha_0} \right)\; . 
\end{eqnarray}
The expression for the complex rates simplifies further if the scattering
amplitudes are rotationally invariant, $f_{\alpha \alpha_0} \left( \cos \left(
   \tmmathbf{p}, \tmmathbf{p}_0 \right) ; E = p_0^2 / 2 m \right)$. In this
case we have
\begin{eqnarray}
  M_{\alpha \beta}^{\alpha_0 \beta_0} & = & \chi_{\alpha \beta}^{\alpha_0
  \beta_0} \int_0^{\infty} \mathd v \nu \left( v \right) n_{\tmop{gas}}
  v_{\tmop{out}} \left( v \right) 2 \pi \int_{- 1}^1 \mathd \left( \cos \theta
  \right) \nonumber\\
  &  & \times f_{\alpha \alpha_0} \left( \cos \theta ; E = \frac{m}{2} v^2
  \right) f_{\beta \beta_0}^{\ast} \left( \cos \theta ; E = \frac{m}{2} v^2
  \right)  \label{KHeq:M4}
\end{eqnarray}
with $\nu \left( v \right) $the velocity distribution like in
(\ref{KHeq:locrate2}), and
\begin{eqnarray}
  v_{\tmop{out}} \left( v \right) & = & \sqrt{v^2 - \frac{2}{m} \left(
  E_{\alpha} - E_{\alpha_0} \right)} 
\end{eqnarray}
the velocity of a gas particle after a possibly inelastic collision.

This shows that limiting cases of (\ref{KHeq:master}) display the expected
dynamics. For the populations $\rho_{\alpha \alpha}$ it reduces to a rate
equation, where the cross sections $\sigma_{\alpha \alpha_0} \left( E \right)
= 2 \pi \int \mathd \left( \cos \theta \right) \left| f_{\alpha \alpha_0}
 \left( \cos \theta ; E \right) \right|^2$ for scattering from channel
$\alpha_0$ to $\alpha$ determine the transition rates,
\begin{eqnarray}
  M_{\alpha \alpha}^{\alpha_0 \alpha_0} & = & \int \mathd v \nu \left( v
  \right) n_{\tmop{gas}} v_{\tmop{out}} \left( v \right) \sigma_{\alpha
  \alpha_0} \left( \frac{m}{2} v^2 \right) \;. 
\end{eqnarray}
In the case of purely elastic scattering, on the other hand, i.e. for
$M_{\alpha \beta}^{\alpha_0 \beta_0} = M_{\alpha \beta}^{\alpha \beta}
\delta_{\alpha \alpha_0} \delta_{\beta \beta_0}$, the coherences are found to
decay exponentially,
\begin{eqnarray}
  \partial_t \left| \rho_{\alpha \beta} \right| & = & - \gamma_{\alpha
  \beta}^{\tmop{elastic}} \left| \rho_{\alpha \beta} \right| \;. 
\end{eqnarray}
The corresponding pure dephasing rates\index{dephasing!-- rate} are determined by the
difference of the scattering amplitudes,
\begin{eqnarray}
  \gamma_{\alpha \beta}^{\tmop{elastic}} & = & \pi \int \mathd v \nu \left( v
  \right) n_{\tmop{gas}} v_{\tmop{out}} \left( v \right) \int_{- 1}^1 \mathd
  \left( \cos \theta \right) \nonumber\\
  &  & \times \left| f_{\alpha \alpha} \left( \cos \theta ; \frac{m}{2} v^2
  \right) - f_{\beta \beta} \left( \cos \theta ; \frac{m}{2} v^2 \right)
  \right|^2 \;.  \label{KHeq:gammaelastic}
\end{eqnarray}
As one expects in this case, the better the scattering environment can
distinguish between system states $| \alpha \rangle$ and $| \beta \rangle$ the
more coherence is lost in this elastic process.

In the general case, the decay of off-diagonal elements will be due to a
combination of elastic and inelastic processes. Although little can be said
without specifying the interaction, it is clear that the integral over $\left|
 f_{\alpha \alpha} - f_{\beta \beta} \right|^2$ in (\ref{KHeq:gammaelastic}),
a ``decoherence cross section'' without classical interpretation, is not
related to the inelastic cross sections characterizing the population
transfer, and may be much larger. In this case, the resulting decoherence will
be again much faster than the corresponding relaxation time scales.

\section{Robust States and the Pointer Basis}

We have seen that, even though the decoherence predictions of linear coupling
models has to be taken with great care, the general observation remains valid
that the loss of coherence may occur on a time scale
$\gamma_{\tmop{deco}}^{\um 1}$ that is much shorter the relaxation time
$\gamma^{- 1}$. Let us therefore return to the general description of open
systems in terms of a semigroup generator $\mathcal{L}$, and ask what we can
say about a general state after a time $t$ which is still small compared to
the relaxation time, but much larger than the decoherence time scale. From a
classical point of view, which knows only about relaxation, the state has
barely changed, but in the quantum description it may now be well approximated
by a mixture determined by particular projectors $\mathsf{P}_{\ell}$,
\begin{equation}
  \mathe^{\mathcal{L}t} : \rho \xrightarrow{\gamma_{\tmop{deco}}^{\um
  1} \ll t \ll \gamma^{- 1}}
  \rho_t \simeq \rho' = \sum_{\ell} \tmop{tr} (\rho \mathsf{P}_{\ell})
  \bignone  \mathsf{P}_{\ell} .
\end{equation}
This set of projectors \{$\mathsf{P}_{\ell}$\}, which depend at most weakly on
$t$, is called {\tmem{pointer basis}}\index{pointer!-- basis} {\cite{Zurek1981a}}
or set of {\tmem{robust states}}\index{robust states|see{pointer states}}\index{state!-- robust|see{pointer states}}\index{pointer!-- states} {\cite{Diosi2000a}}. It
is distinguished by the fact that a system prepared in such a state is hardly
affected by the environment, while a superposition of two distinct pointer
states decoheres so rapidly that it is never observed in practice.

We encountered this behavior with the damped harmonic oscillator discussed in
Sect. \ref{KHsec:examples}. There the coherent oscillator states remained pure
under Markovian dynamics, while superpositions between (macroscopically
distinct) coherent states decayed rapidly. Hence, in this case the coherent
states\index{coherent states!-- and robust states} $\mathsf{P}_{\alpha} = |
\alpha \rangle \langle \alpha |$ can be said to form an (over-complete) set of
robust states\index{pointer states!-- and coherent states}, leading to the
mixture
\begin{equation}
  \rho' = \int \mathd \mu \left( \alpha \right) \tmop{tr} (\rho
  \mathsf{P}_{\alpha}) \mathsf{P}_{\alpha} \;,
\end{equation}
with appropriate measure $\mu$.

The name {\tmem{pointer basis}} is well-fitting because the existence of such
robust states is a prerequisite for the description of an ideal measurement
device in a quantum framework. A macroscopic---and therefore
decohering---apparatus implementing the measurement of an observable
$\mathsf{A}$ is ideally constructed in such a way that macroscopically
distinct positions of the ``pointer'' are obtained for the different
eigenstates of $\mathsf{A}$. Provided these pointer positions of the device are
robust, the correct values are observed with certainty if the quantum
system is in an eigenstate of the observable. Conversely, if the quantum
system is not in an eigenstate of $\mathsf{A}$, the apparatus will
{\tmem{not}} end up in a superposition of pointer positions, but be found at a
definite position, albeit probabilistically, with a probability given by the
Born rule.

The main question regarding pointer states is, given the environmental
coupling or the generator $\mathcal{L}$, what determines whether a state is
robust or not, and how can we determine the set of pointer states without
solving the master equation for all initial states. It is fair to say that
this issue is not fully understood, except for very simple model environments,
nor is it even clear how to quantify robustness.

An obvious ansatz, due to Zurek {\cite{Zurek2003a,Zurek1993a}}, is to sort all
pure states in the Hilbert space according to their (linear) entropy
production rate\index{entropy!-- production rate}, or rate of loss of
purity\index{purity!-- rate of loss},
\begin{equation}
  \partial_t \mathsf{S}_{\tmop{lin}} [\rho] = \um 2 \tmop{tr} \left( \rho
  \mathcal{L}(\rho) \right) \;.
\end{equation}
It has been called ``predictability sieve'' since the least entropy producing
and therefore most predictable states are candidate pointer states
{\cite{Zurek2003a}}. 

In the following, a related approach will be described, following the
presentation in {\cite{Strunz2002a,Gisin1995a}}. It is based on a time
evolution equation for robust states. Since such an equation must distinguish
particular states from their linear superpositions, it is necessarily
nonlinear.

\subsection{Nonlinear Equation for Robust States}

We seek a nonlinear time evolution equation for robust pure states
$\mathsf{P}_t$ which, on the one hand, preserves their purity, and on the
other, keeps them as close as possible to the evolved state following the
master equation.

A simple nonlinear equation keeping a pure state pure is given by the following
extension of the Heisenberg form for the infinitesimal time step,
\begin{equation}
  \mathsf{P}_{t + \delta t} = \mathsf{P}_t + \delta t \left( \frac{1}{\mathi}
  [ \mathsf{A}_t, \mathsf{P}_t] + [ \mathsf{P}_t, [ \mathsf{P}_t,
  \mathsf{B}_t]] \right) \;,
\end{equation}
where $\mathsf{A} \text{} \mathsf{}$ and $\mathsf{B}$ are hermitian operators.
In fact, the unitary part can be absorbed into the nonlinear part by
introducing the hermitian operator $\mathsf{X}_t = - \mathi [ \mathsf{A}_t,
\mathsf{P}_t] + \mathsf{B}_t$. It ``generates'' the infinitesimal time
translation of the projectors (and may be a function of $\mathsf{P}_t$),
\begin{equation}
  \left. \mathsf{P}_{t + \delta t} = \mathsf{P}_t + \delta t [ \mathsf{P}_t, [
  \mathsf{P}_t, \mathsf{X}_t] \right] \;.
\end{equation}
With this choice one confirms easily that the evolved operator has indeed the
properties of a projector, to leading order in $\delta t$,
\begin{equation}
  \mathsf{P}_{t + \delta t}^{\dag} = \mathsf{P}_{t + \delta t}
\end{equation}
and
\begin{eqnarray}
  \left( \mathsf{P}_{t + \delta t} \right)^2 & = & \mathsf{P}_{t + \delta t} +
  O (\delta t^2) \;. 
\end{eqnarray}
The corresponding differential equation reads
\begin{equation}
  \partial_t \mathsf{P}_t = \frac{\mathsf{P}_{t + \delta t} -
  \mathsf{P}_t}{\delta t} = \left[ \mathsf{P}_t, \left[ \mathsf{P}_t,
  \mathsf{X}_t \right] \right] \;.
\end{equation}
To determine the operator $\mathsf{X}_t$ one minimizes the distance between
the time derivatives of the truly evolved state and the projector. If we
visualize the pure states as lying on the boundary of the convex set of mixed
states, then a pure state will in general dive into the interior under the
time evolution generated by $\mathcal{L}$. The minimization chooses the
operator $\mathsf{X}_t$ in such a way that $\mathsf{P}_t$ sticks to the
boundary, while remaining as close as possible to the truly evolved state.

The (Hilbert-Schmidt) distance between the time derivatives can be calculated
as
\begin{eqnarray}
  \| \underbrace{\mathcal{L}( \mathsf{P}_t)}_{\equiv \mathsf{Z}} - \partial_t
  \mathsf{P}_t \|^2_{\tmop{HS} \mathsf{}} & = & \tmop{tr} \left[ \left(
  \mathsf{Z} - \left[ \mathsf{P}_t, \left[ \mathsf{P}_t, \mathsf{X}_t \right]
  \right] \right)^2 \right] \nonumber\\
  & = & \tmop{tr} \left( \mathsf{Z}^2 - 2 ( \mathsf{Z}^2 \mathsf{P}_t -
  \left( \mathsf{Z}  \mathsf{P}_t \right)^2) \right) \nonumber\\
  &  & + 2 \tmop{tr} \left( ( \mathsf{Z} - \mathsf{X})^2  \mathsf{P}_t -
  \left( \left( \mathsf{Z} - \mathsf{X} \right)  \mathsf{P}_t \right)^2
  \right) \;. 
\end{eqnarray}
We note that the first term is independent of $\mathsf{X}$, whereas the second
one is non-negative. With the obvious solution $\mathsf{X}_t = \mathsf{Z} \equiv
\mathcal{L} ( \mathsf{P}_t)$ one gets a nonlinear evolution equation for
robust states $\mathsf{P}_t$, which is trace and purity preserving
{\cite{Gisin1995a}},
\begin{equation}
  \partial_t \mathsf{P}_t = [ \mathsf{P}_t, [ \mathsf{P}_t, \mathcal{L}(
  \mathsf{P}_t)]] \;.
\end{equation}
It is useful to write down the equation in terms of the vectors $| \xi
\rangle$ which correspond to the pure state $\mathsf{P}_t = | \xi \rangle
\langle \xi |$,
\begin{equation}
  \partial_t | \xi \rangle = [\mathcal{L}(| \xi \rangle \langle \xi |) -
  \underbrace{\langle \xi |\mathcal{L}(| \xi \rangle \langle \xi |) | \xi
  \rangle}_{\text{``decay rate''}}] | \xi \rangle \;.
\end{equation}
If we take $\mathcal{L}$ to be of the Lindblad form (\ref{KHLindblad}) the
equation reads
\begin{eqnarray}
  \partial_t | \xi \rangle & = & \frac{1}{\mathi \hbar}  \mathsf{H} | \xi
  \rangle + \sum_k \gamma_k \left[ \langle \mathsf{L}_k^{\dag} \rangle_{\xi}
  \left( \mathsf{L}_k - \left\langle \mathsf{L}_k \right\rangle_{\xi} \right)
  - \frac{1}{2}  \left( \mathsf{L}_k^{\dag}  \mathsf{L} - \langle
  \mathsf{L}_k^{\dag}  \mathsf{L}_k \rangle \right) \right] \bignone | \xi
  \rangle \nonumber\\
  &  & - \frac{1}{\mathi \hbar} \langle \mathsf{H} \rangle_{\xi} | \xi
  \rangle \;.  \label{KHnl3}
\end{eqnarray}
Its last term is usually disregarded because it gives rise only to an
additional phase if $\langle \mathsf{H} \rangle_{\xi}$ is constant. The
meaning of the nonlinear equation (\ref{KHnl3}) is best studied in terms of
concrete examples.

\subsection{Applications}

\subsubsection{Damped Harmonic Oscillator}\index{damped harmonic oscillator!--
 robust states}\index{pointer states!-- of the damped harmonic oscillator}

Let us start with the damped harmonic oscillator discussed in
Sect.~\ref{KHsec:examples}. By setting $\mathsf{H} = \hbar \omega
\mathsf{a}^{\dag} \mathsf{a}$ and $\mathsf{L} = \mathsf{a}$ (\ref{KHnl3})
turns into
\begin{equation}
  \partial_t | \xi \rangle = \um \mathi \omega \mathsf{a}^{\dag}  \mathsf{a} |
  \xi \rangle + \gamma \left( \langle \mathsf{a}^{\dag} \rangle_{\xi} (
  \mathsf{a} - \left\langle \mathsf{a} \right\rangle_{\xi}) - \frac{1}{2} 
  \left( \mathsf{a}^{\dag}  \mathsf{a} - \langle \mathsf{a}^{\dag}  \mathsf{a}
  \rangle_{\xi} \right) \right) | \xi \rangle \;.
\end{equation}
Note that the first term of the non-unitary part vanishes if $| \xi \rangle$
is a coherent state\index{coherent states!-- definition}, i.e. an eigenstate of
$\mathsf{a}$. This suggests the ansatz $| \xi \rangle = | \alpha \rangle$
which leads to
\begin{equation}
  \partial_t | \alpha \rangle = \left[ \left( \um \mathi \omega -
  \frac{\gamma}{2} \right) \alpha \mathsf{a}^{\dag} + \frac{\gamma}{2} |
  \alpha |^2 \right] | \alpha \rangle \;.
\end{equation}
It is easy to convince oneself that this equation is solved by
\begin{equation}
  | \alpha_t \rangle = | \alpha_0  \mathe^{\um \mathi \omega t - \gamma t / 2}
  \rangle = \mathe^{- | \alpha_t |^2 / 2} \mathe^{\alpha_t  \mathsf{a}^{\dag}}
  |0 \rangle 
\end{equation}
with $\alpha_t = \alpha_0 \exp \left( - \mathi \omega t - \gamma t / 2
\right)$. It shows that the predicted robust states\index{pointer states!-- and coherent states} are indeed given by the slowly decaying coherent
states\index{coherent states!-- and robust states} encountered in
Sect.~\ref{KHsec:examples}.

\subsubsection{Quantum Brownian Motion}

A second example is given by the Brownian motion of a quantum particle. The
choice
\begin{eqnarray}
  \mathsf{H} = \frac{\mathsf{p}^2}{2 m}  \hspace{1em} & \tmop{and} &
  \hspace{1em} \mathsf{L} = \frac{\sqrt{8 \mathpi}}{\Lambda_{\tmop{th}}} 
  \mathsf{x} 
\end{eqnarray}
yields a master equation of the form (\ref{KHqbmme}) but without the
dissipation term. Inserting these operators into (\ref{KHnl3}) leads to
\begin{equation}
  \partial_t | \xi \rangle = \frac{\mathsf{p}^2}{2 m \mathi \hbar} | \xi
  \rangle - \gamma \frac{4 \mathpi}{\Lambda_{\tmop{th}}^2} [( \mathsf{x} -
  \left\langle \mathsf{x} \right\rangle_{\xi})^2 - \underbrace{\langle (
  \mathsf{x} - \left\langle \mathsf{x} \right\rangle_{\xi})^2
  \rangle_{\xi}}_{\sigma^2_{\xi} ( \mathsf{x})}] | \xi \rangle \;.
  \label{KHqbmpointer}
\end{equation}
The action of the non-unitary term is apparent in the position representation,
$\xi (x) = \langle x| \xi \rangle$. At positions $x$ which are distant from
mean position $\left\langle \mathsf{x} \right\rangle_{\xi}$ as compared to the
dispersion $\sigma_{\xi} ( \mathsf{x}) = \big\langle \left( \mathsf{x} -
 \langle \mathsf{x} \rangle_{\xi} \right)^2 \big\rangle_{\xi}^{1 / 2}$
the term is negative and the value $\xi (x)$ gets suppressed. Conversely, the
part of the wave function close to the mean position gets enhanced,
\begin{equation}
\langle x| \xi \rangle = \left\{ \begin{array}{ll}
     \text{suppressed} & \text{if } |x - \left\langle \mathsf{x}
     \right\rangle_{\xi} | > \sigma_{\xi} ( \mathsf{x})\\
     \text{enhanced } & \text{if } |x - \left\langle \mathsf{x}
     \right\rangle_{\xi} | < \sigma_{\xi} ( \mathsf{x})\;.
   \end{array} \right. 
\end{equation}
This localizing effect is countered by the first term in (\ref{KHqbmpointer})
which causes the dispersive broadening of the wave function. Since both
effects compete we expect stationary, soliton-like solutions of the equation.

Indeed, a Gaussian ansatz for $| \xi \rangle$ with ballistic motion, i.e.
$\left\langle \mathsf{p} \right\rangle_{\xi} = p_0$, $\left\langle \mathsf{x}
\right\rangle_{\xi} = x_0 + p_0 t / m$, and a {\tmem{fixed}} width
$\sigma_{\xi} ( \mathsf{x}) = \sigma_0 $ solves (\ref{KHqbmpointer}) provided \cite{Diosi2000a}
\begin{equation}
  \sigma_0^2 = \frac{1}{4 \mathpi}  \sqrt{\frac{k_B T}{2 \hbar \gamma}}
  \Lambda_{\tmop{th}}^2 = \left( \frac{\hbar^3}{8 \gamma m^2 k_B T} \right)^{1
  / 2}\;,
\end{equation}
see (\ref{KHthermdeBroglie}). As an example, let us consider a dust particle
with a mass of 10\,$\mu$g in the interstellar medium interacting only with the
microwave background of $T = 2.7\,\mathrm{K}$. Even if we take a very small
relaxation rate of $\gamma=1/(13.7\times 10^{9}\,{\rm y})$, corresponding to the inverse
age of the universe, the width of the solitonic wave packet describing the center of mass
is as small as $2$\,pm. This sub-atomic value demonstrates again the remarkable efficiency of the decoherence mechanism to induce classical behavior in the quantum state of macroscopic objects.

\subsection*{Acknowledgments}

Many thanks to \'Alvaro Tejero Cantero who provided me with his notes, typed
with the lovely {\TeXmacs} program during the lecture. The present text is
based on his valuable input. I am also grateful to Marc Busse and Bassano
Vacchini for helpful comments on the manuscript.

This work was supported by the Emmy Noether program of the DFG.

%\setcounter{page}{2}
%\thispagestyle{empty}
%\noindent
%\textsf{\large Klaus Hornberger}
%\\[\baselineskip]
%\textsf{\Large Introduction to Decoherence Theory}
%\\[3\baselineskip]
%This text corresponds more or less to a chapter\\ 
%soon to be published version in: 
%\\[0.5\baselineskip]
%A. Buchleitner, C. Viviescas, and M. Tiersch (Eds.),\\
%Entanglement and Decoherence. Foundations and Modern Trends,\\
%Lecture Notes in Physics \textbf{768}, Springer, Berlin (2009)\\
%DOI 10.1007/978-3-540-88169-8\_5 
%\\[0.5\baselineskip]
%Short reference: K. Hornberger, Lect. Notes Phys. \textbf{768}, 223-278 (2009)
\vspace*{-1\baselineskip}
\tableofcontents

\end{document}